\documentclass{interact}
\usepackage[numbers,sort&compress,merge]{natbib}
\bibpunct[, ]{[}{]}{,}{n}{,}{,}

\usepackage{graphicx}

\usepackage{color}
\usepackage{textcomp}
\usepackage{subscript}
\usepackage{babel}
\usepackage{amsmath,amssymb,amsfonts,hyperref}
\usepackage{slashed,placeins}
\usepackage{braket,latexsym,epstopdf,rotating}
\usepackage[export]{adjustbox}
\usepackage[svgnames]{xcolor}
\usepackage{enumitem}
\usepackage{comment}
\usepackage{caption}
\usepackage{ragged2e}
\usepackage{subcaption}
\usepackage{siunitx}
\usepackage{float}

\usepackage[normalem]{ulem}
\usepackage{cancel} 
\usepackage{array}
\usepackage{calc}

\definecolor{ao(english)}{rgb}{0.0, 0.5, 0.0}
\definecolor{alizarin}{rgb}{0.82, 0.1, 0.26}
\definecolor{lightskyblue}{rgb}{0.53, 0.81, 0.98}

\newcommand{\Htwoplus}[1]{{H$_2^+$}}
\newcommand{\HDplus}[1]{{HD$^+$}}
\newcommand{\antiHtwoplus}[1]{{$\overline{\mathrm{H}}_2^-$}}

\newcommand{\PMT}[1]{{PT}}
\newcommand{\precisiontrap}[1]{{precision trap}}

\begin{document}

\title{On the potential for high-accuracy spectroscopy of \Htwoplus{} and \antiHtwoplus{}  in Penning traps for a test of CPT invariance}

\author{
\name{Stephan Schiller\textsuperscript{a}\thanks{CONTACT S.~Schiller. Email: step.schiller@hhu.de}, Juan M. Cornejo\textsuperscript{b}, Nikita Poljakov\textsuperscript{c}, Christian Ospelkaus\textsuperscript{c}, Stefan Ulmer\textsuperscript{a}, Dimitar Bakalov\textsuperscript{d}
}
\affil{\textsuperscript{a}Institut für Experimentalphysik, Heinrich-Heine-Universität Düsseldorf, 40225 Düsseldorf, Germany;
\\ {\textsuperscript{b}Departamento de F{\'i}sica de la Materia Condensada, Universidad de C{\'a}diz, Puerto Real, 11519, Spain};
\\ {\textsuperscript{c}Institut für Quantenoptik, Leibniz-Universität Hannover, Germany};
\\ {\textsuperscript{d}Institute for Nuclear Research and Nuclear Energy, Bulgarian Academy of Sciences, Tsarigradsko Chaussée 72, Sofia, 1784, Bulgaria}
}
}
\maketitle

\begin{abstract}
    The comparison of vibrational transition frequencies of \Htwoplus{} and \antiHtwoplus{} offers a new opportunity to test CPT invariance. 
    Myers [Phys.~Rev.~A~\textbf{98}, 010101(R) (2018)] proposed performing laser spectroscopy in a Penning trap (PT) with non-destructive read-out. Here, we provide an extensive analysis of this proposal, 
    introduce novel aspects, and discuss its implementation in \PMT{}s that incorporate either the continuous Stern-Gerlach effect or quantum-logic spectroscopy. We derive estimates for the achievable accuracy of the test. We find that a comparison of the vibrational frequencies at a fractional level of $1\times10^{-17}$ is a realistic prospect, using technology that is mostly already available.  We also analyze complementary CPT invariance tests, namely those of the g-factor of the bound electron/positron via electron-spin-resonance spectroscopy and of the magnetic moment of the proton/antiproton via radiofrequency spectroscopy.
\end{abstract}

\section{Introduction}

CPT invariance (CPTI) is a cherished principle of Physics \cite{lehnert2016cpt}. The advent of accelerators brought the capability of producing copious amounts of baryonic antimatter and therefore enabled  experimental tests of this principle on antiprotons and other antibaryons. Experiments on a variety of systems have been performed - consistently finding it to hold \cite{Navas2024}.  Among the low-energy tests one may highlight a number of experiments involving cold, often trapped particles: the measurement of the 1s-2s transition in hydrogen and anti-hydrogen \cite{Ahmadi2018characterization}, spectroscopy of anti-protonic helium \cite{Hori2016}, and charge-to-mass ratios and g factors of protons, antiprotons, leptons and antileptons \cite{VanDyck1987,Smorra2017,gurung2020precision,Aguillard2025}. 

Three decades ago, Dehmelt \cite{Dehmelt1995} proposed extending CPTI tests to molecules, specifically to  the molecular hydrogen ions (MHI) \Htwoplus{} and \antiHtwoplus{}. These are the simplest and presumably the first molecules for which the anti-matter version will become available, and which can be controlled relatively easily thanks to being charged. 
A CPTI test would consist in verifying whether a --  suitably chosen -- hyperfine or rovibrational transition frequency differs between \antiHtwoplus{}  compared to \Htwoplus{}. {According to Dehmelt, the spectroscopy would be performed on trapped ensembles of molecules or on a single molecule.} 

At the time of Dehmelt's proposal,  virtually no experimental data existed on the vibrational transitions of \Htwoplus{}, only on the hyperfine structure - Dehmelt and his colleagues having pioneered its study {and its trapping in radiofrequency (RF) traps \cite{Jefferts1968,Jefferts1969,Menasianthesis}.} 
One laser spectroscopy study of the related molecule \HDplus{} stands out \cite{Wing1976}.

In the intervening years, electron-spin-resonance (ESR) spectroscopy of \Htwoplus{} was demonstrated in a Penning trap \footnote{The name ``Penning" was given by Dehmelt, who first conceived and demonstrated the trap.} (\PMT{}) \cite{Loch1988} and vibrational spectroscopy of \HDplus{} in radio-frequency (RF) traps \cite{Koelemeij2007,Bressel2012}, achieving respectable accuracy. {In RF~traps, this was enabled by the technique of sympathetic cooling by laser-cooled, co-trapped atomic ions \cite{Blythe2005}}. Theoretical studies projected that rovibrational frequencies should be measurable with uncertainties at the $10^{-17}$ fractional uncertainty level in such traps \cite{Schiller2014,Karr2014,Karr2016}. 

Myers  subsequently performed an important analysis of the potential of CPTI tests in \PMT{}s  \cite{Myers2018}. 
He considered CPTI tests of three types: those of rovibrational transition frequencies ($f_\mathrm{vib}$), of the ratio of bound-electron spin-flip frequency ($f_\mathrm{sf}$) to cyclotron frequency, and of the magnetic moments of the (bound) proton/antiproton. He also pointed  out that it should be possible to compare the nuclear magnetization distributions of proton/antiproton via a comparison of the hyperfine transition frequencies.

The consideration of \PMT{}s was motivated by the fact that such traps are already used very successfully for storing single charged (anti-)particles and for performing magnetic resonance experiments \cite{Koehler2015,schneider2017double,Heisse2019}.
In particular, \PMT{}s are also operating at CERN's antimatter factory (AF), where CPTI tests are performed on single antiprotons \cite{Smorra2017}. 

Since Myers' study, the rotational and vibrational spectroscopy of MHI
has made further impressive progress \cite{Alighanbari2020,Patra2020,Kortunov2021,Alighanbari2023}, achieving a gain in accuracy of $10^5$ compared to the earliest studies \cite{Wing1976}.  In parallel, continuous work on ab initio theory has delivered a similar increase in theoretical accuracy \cite{Haidar2022b,Korobov2021}.
Recently, Shore \cite{Shore2025a,Shore2025b} and Vargas \cite{Vargas2025} have discussed in detail which aspects of the Standard Model a \Htwoplus{}/\antiHtwoplus{} CPTI test would probe.

An important milestone reported very recently  is the first Doppler-free spectroscopy of \Htwoplus{}, which provided an $8\times10^{-12}$-fractional uncertainty  measurement of one rovibrational transition frequency and 
enabled the deduction of a new value for the electron-proton mass ratio from it \cite{Schenkel2024,Alighanbari2025}. This level of frequency uncertainty is one that would make a \Htwoplus{}/\antiHtwoplus{} CPTI test relevant.

Another direction of work has been the theoretical study of reactions that might be employed to produce \antiHtwoplus{}, starting with the antiprotons produced at CERN in the AF facility, see \cite{Zammit2025} and references therein.

Assuming \antiHtwoplus{} will one day become available, it is useful to discuss already today the approaches for precision spectroscopy of this molecule. Obviously, the experimental approach and the systematic shifts encountered will be the same for the molecule and the antimolecule, provided they are studied under similar conditions, e.g.~in the same trap type.

Motivated by the experimental progress in single-particle manipulation in \PMT{}s and MHI laser spectroscopy both in RF~traps and in \PMT{}s  (see Sec.\,\ref{sec:CPTI tests using PTs}), as well as the distant prospects of creating \antiHtwoplus{}, here, our aim here is to theoretically investigate the systematic shifts of transition frequencies in two different types of \PMT{} in greater detail, and to determine what is required to push the accuracy of a CPTI test beyond the level considered by Myers. We consider  a \PMT{} employing the continuous Stern-Gerlach effect (CSGE) (similar to BASE \cite{Smorra2015} and ALPHATRAP  \cite{Sturm2019}) as well as a quantum-logic spectroscopy (QLS) \PMT{} (similar to BASE-QLEDS)~\cite{winelandExperimentalIssuesCoherent1998, heinzenQuantumlimitedCoolingDetection1990, 
Schmidt2005, Cornejo2021}. 

Our aim can be reached thanks to the already existing theory of \Htwoplus{} and so our study relies on various aspects of this theory, including the 
hyperfine structure \cite{Korobov2006}, 
the Zeeman effect \cite{Karr2008,Karr2021}, 
the Stark shift \cite{Schiller2014a}, 
the electric quadrupole shift \cite{Bakalov2014}, and electric quadrupole (E2) transitions \cite{Korobov2018a}.
As in all atomic and molecular systems, also in  \Htwoplus{} magnetic interactions of the diamagnetic type (proportional to magnetic field squared, but unrelated to the particle magnetic moments) exist. These were not mentioned in Myers's work \cite{Myers2018}. Some of us have recently developed a detailed treatment for HD$^+$ and \Htwoplus{} \cite{Schiller2025b}. Here we apply the findings for the latter species. 

We also provide detailed considerations on how the relativistic Doppler shift can be accounted for. Our focus is on rovibrational transitions, but we also treat lepton-spin-flip and nuclear-spin-flip transitions.

This paper is structured as follows. 
Section~\ref{sec:CPTI tests using PTs} motivates the use of \PMT{} and provides general considerations on the measurement strategy. 
Section~\ref{sec:Basics} presents basic aspects of candidate rovibrational transitions and introduces the perturbations that affect every rovibrational level.  
A detailed treatment of the perturbations resulting from the strong magnetic field present in a \PMT{} follows in 
Section~\ref{sec:Magnetic perturbations}. This section concludes with a summary. The electric perturbations are then treated in Section~\ref{sec:Electric perturbations}.
Our suggested spectroscopy is based on electric-quadrupole transitions. They have been treated before for the case of weak magnetic field. 
Therefore, in Section~\ref{sec:Electric quadrupole transitions} a treatment is given for strong magnetic field.
An important systematic shift is the quadratic Doppler shift, hence Section~\ref{sec:Quadratic Doppler shift} is devoted to it. Proposed experimental implementations of CPTI tests are contained in Sections~\ref{sec:Penning traps}-\ref{sec:QLS-PT}, in which we consider two different approaches to non-destructive spectroscopy in \PMT{}s.
A discussion of the main results and conclusions in Section~\ref{sec:Discussion} closes this work. Appendices cover  additional details.
\section{CPTI tests of \Htwoplus{}/\antiHtwoplus{} using \PMT{}s}
\label{sec:CPTI tests using PTs}
As mentioned, rovibrational spectroscopy of \Htwoplus{} has been achieved in an RF~trap, equipped with sympathetic cooling \cite{Alighanbari2025}. Recently, also nondestructive RF~spectroscopy of ortho-\Htwoplus{} has been demonstrated \cite{Holzapfel2025} in a cryogenic RF~trap.
This raises the question, why consider \PMT{}s for trapping \Htwoplus{}/\antiHtwoplus{}? 

For the sympathetic cooling of \antiHtwoplus{} by a positively charged atomic coolant (the only type available), the standard linear RF~trap geometry with ``continuous" electrodes would have to be replaced with a featuring segmented electrodes. These would prevent the coalescence  between the two oppositely charged ions, while still providing a sufficiently strong coupling to enable sympathetic cooling. While this approach might work, it has not yet been demonstrated.

By contrast, a cryogenic \PMT{}s can cool any single positive or negative ion to a low temperature by resistive cooling (although not as low as in the above case of sympathetic cooling). This is a robust, well-established technique.

An important experimental aspect is that \PMT{}s can be constructed to provide an excellent vacuum, enabling them to store antimatter ions for years \cite{sellner2017improved}, at typical consumption rates of one particle every two months. This is important in the context of \antiHtwoplus{}: these particles may turn out to be very difficult to produce, and may thus be available in only small numbers. This would require reducing (anti-) particle loss to a minimum and an excellent vacuum is a necessary condition. 

Cryogenic RF~traps have also been developed in substantial numbers world-wide and might eventually exhibit similar storage performance to that described above. However, this remains to be proven.

An additional argument in favour of \PMT{}s is that advanced \PMT{}s enable the nondestructive detection of the ion's spin state, i.e.~of its projection onto a (given) measurement axis, via the continuous Stern-Gerlach effect (CSGE) \cite{Dehmelt1986,Dehmelt1986a, ulmer2011observation}. This feature could also be used to detect rovibrational transitions, as proposed by Myers. The experimental demonstration of optical transition detection via CSGE was independently presented at nearly the same time by Egl~\textit{et al.}~\cite{Egl2019}. (In RF traps, nondestructive detection of optical transitions of molecules has already been demonstrated \cite{Wolf2016a,Sinhal2020}.)

In \PMT{}s, the magnetic field plays a dominant role, affecting not only the motion of the ion but also the internal energies. Therefore an analysis of the accuracy potential of rovibrational spectroscopy is needed. Myers provided an initial analysis. He concluded that a measurement uncertainty below the $1\times10^{-15}$ fractional level should be possible \cite{Myers2018} and in \cite{Myers2018a} stated a level below $10^{-16}$. 
He considered a case in which the single trapped molecular ion can be cooled to axial and cyclotron mode temperatures of 20\,mK. 
It should be noted that while  such low levels are routinely achieved by sub-thermal cooling of the cyclotron and magnetron modes \cite{latacz2024orders}, the axial mode - which is usually used for particle detection - is at cryostat temperature, typically $\approx 4\,$K.  Obtaining a lower axial temperature requires substantial effort. Using resonant image current coupling to laser-cooled $^9$Be$^+$-ions in a highly specialized 6-trap experiment, in \cite{Will_2022} an axial temperature of 170$\,$mK was been achieved by the BASE collaboration. 

Quite recently, important progress has been made in molecular spectroscopy in \PMT{}s. In the ALPHATRAP apparatus \cite{Sturm2019}, a measurement campaign on the heteronuclear molecule HD$^+$ has shown that it is possible to store and manipulate a single molecular ion for weeks without losing it. Its full internal state (consisting of the degrees of freedom of electron spin, proton spin, deuteron spin, rotation and vibration) can be determined non-destructively by detecting the occurrence of an electron-spin resonance transition out of or into a state \cite{Koenig2025}. 
Non-destructive, high-accuracy ESR spectroscopy has been performed \cite{Koenig2025b}. 

Finally, a rovibrational transition in \HDplus{} has been measured, representing the first time that a narrow optical transition has been measured on a non-laser-cooled particle in a \PMT{} \cite{Kortunov2026}.

We expect that the preferred CPTI test strategy will consist in performing vibrational spectroscopy of \Htwoplus{} and \antiHtwoplus{} in the \emph{same trap} and \emph{under almost the same conditions} (except for trap polarity). The two species would be measured in interleaved fashion, shuttling \Htwoplus{} and \antiHtwoplus{} into and out of the particular apparatus section where the laser excitation occurs, following conceptual ideas applied in many state-of-the-art mass spectrometers \cite{borchert202216, schussler2020detection}. 
The difference (if any) $\delta f_\mathrm{vib}= f_\mathrm{vib}(\mathrm{H}_2^+)-f_\mathrm{vib}(\bar{\mathrm{H}}_2^-)$ of their vibrational transition frequencies will be measured.  
The experimental conditions do not need to be those usually aspired to in optical clocks, i.e.~near-zero electric and magnetic fields, near-zero motional energy and near-zero black-body temperature. Instead, these parameters can be finite - even large - but should be kept sufficiently \emph{stable} when ``switching" from \Htwoplus{} to \antiHtwoplus{}, so that their effects cancel out in the frequency difference $\delta f_\mathrm{vib}$. 
The advantages of this approach are that some systematic shifts will be common-mode and that the high B-field stability of the \PMT{} magnet can be exploited. An important practical advantage could be that no atomic clock reference is required, only a reference with ultrastable - rather than ultra-accurate - frequency.

For the purpose of \emph{finding} a transition experimentally, once \Htwoplus{} or \antiHtwoplus{} is available in a \PMT{}, the theoretically predicted values of $f_\mathrm{vib}$ and $f_\mathrm{sf}$ have sufficient accuracy \cite{Haidar2022b,Korobov2021,Kullie2025,Schiller2025b}. In fact, the theory has been confirmed experimentally in a \PMT{} environment for the related HD$^+$, in the campaign mentioned above. 

In the context of a CPTI test, the absolute frequency $f_\mathrm{vib}$ is not of fundamental interest. 
Nevertheless, it may be worthwhile measuring it so that values obtained by independent research groups working on \PMT{}s can be compared. If substantially different magnetic fields are employed, an accurate comparison may necessitate pushing the ab initio theory further than the initial analysis already performed  \cite{Schiller2025b}.  
Absolute values of $f_\mathrm{vib}(\mathrm{H}_2^+)$ measured in RF traps, i.e.~at near-zero magnetic field, could also provide useful cross-checks - but their usefulness 
depends on how accurately the effects of the strong magnetic field can eventually be  computed theoretically.

\section{Basics of \Htwoplus{} }
\label{sec:Basics} 
As diatomic molecules MHI have one vibrational and one rotational degree of freedom, with quantum number $v$ and $N$, respectively, labeling the levels. Additional degrees of freedom come with the two nuclear spins and the electron spin. Small couplings lead to consideration of the total angular momentum $\mathbf{F}$ 
of the molecule, the combination of electron-spin angular momentum, nuclear angular momenta, and rotational angular momentum. To $\mathbf{F}$ are associated the  strength $F$ and projection $M_F$, and both are good quantum numbers. Every rovibrational level, denoted by $(v,N)$, is split in two or more states in a finite magnetic field, lifting the energy degeneracy associated with $M_F$. 
Depending on the rotational angular momentum $N$ of a level and the particular isotopologue, $F$ can have different values. We call the individual quantum states ``spin states",    for simplicity. They can be denoted by $(v,N,F,M_F)$, but as we are dealing here with MHI in a strong magnetic field, below we shall use a more practical notation.  A particular transition between two specific spin states of two different rovibrational levels is called a ``spin component". 

In \Htwoplus{}, the two nuclei are identical spin-1/2 particles; therefore the total nuclear spin $I$ is a good quantum number and must have the value $0$ or 1. In the following we confine the discussion to transitions between levels with even (or zero) rotational angular momentum quantum numbers $N$, $N'$. By the anti-symmetrization postulate for fermions the nuclear spin state must be a singlet in such levels, i.e.~the total nuclear spin $I$ is  zero. As a consequence, the set of spin states (``spin structure") of each level is particularly simple: the total angular momentum arises only from rotation and electron spin and can take on only the two values $F=|N-1/2|$, $N+1/2$.  The reduced number of spin states simplifies the analysis.

The vibrational transitions we consider are electric quadrupole (E2) transitions  \cite{Korobov2018a}, a type that has recently been demonstrated on \Htwoplus{} in the team of one of the present authors \cite{Schenkel2024,Alighanbari2025}. For such transitions, the  selection rule for the connected rotational states reads $N\rightarrow N'=N,\,N\pm2$, with the case $0\rightarrow0$ being excluded. 
For the total angular momentum $F$  the selection rules are $\Delta F=0,\,\pm1\pm2$, $\Delta M_F=0,\,\pm1\pm2$, and  $F+F'\ge2$.

As reference vibrational transition we choose one from the lower level $v=0$ to the upper level $v'=2$, at a wavelength near 2.4\,$\mu$m ($f_\mathrm{vib}\simeq127\,$THz). As alternative, one may consider a transition to $v'=3$, at 1.6\,$\mu$m ($f_\mathrm{vib}\simeq185\,$THz). These are good choices, since suitable lasers are available, see e.g.~\cite{Schenkel2024a}. 
In fact, these transitions were already discussed in the context of RF traps as ``clock transitions" for fundamental physics \cite{Schiller2014,Karr2014}.

\begin{figure}[t!]
    \centering
\includegraphics[width=0.6\linewidth]{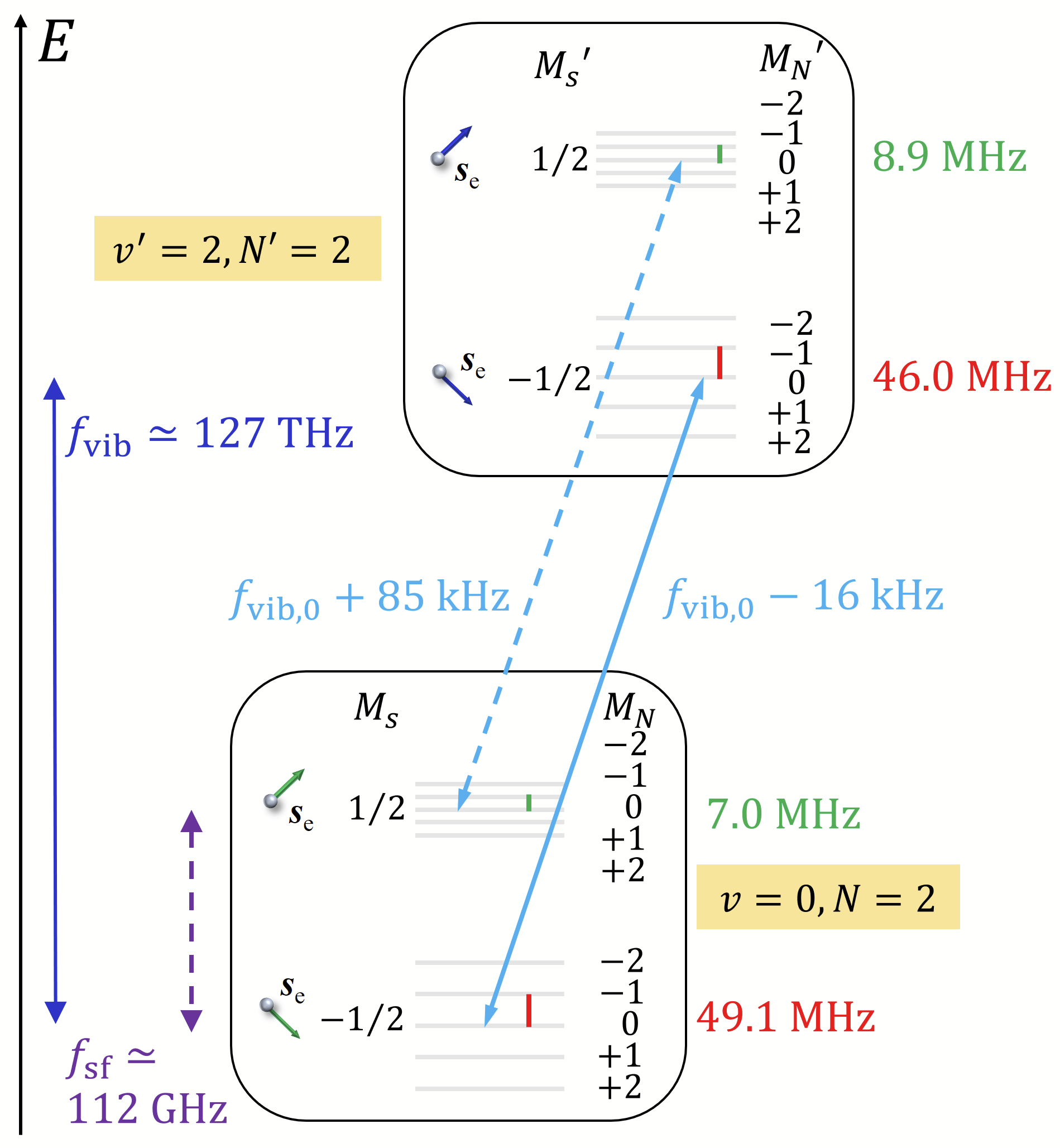}
    \caption{
    \textbf{Schematic of the spin structure of two selected rovibrational energy levels of para-\Htwoplus{} in a $\mathbf{\textit{B}_0=4}\,$Tesla magnetic field and selected transitions}. 
    $M_s$, $M_s'$ for electron-spin projection and $M_N$, $M_N'$ for rotational-angular-momentum-projection are approximate quantum numbers, used here to label the states. 
    \textcolor{cyan}{\textbf{Cyan}}: Two  transitions considered to be suitable for a CPTI test, i.e.~having small systematic shifts, are indicated. These transitions  also appear in figure\,\ref{fig:Principle of CSGE - detected vibrational transition} as $\mathrm{A}$, $\mathrm{A}'$. $f_\mathrm{vib,0}$ is the spin-averaged rovibrational transition frequency in zero $B$-field, i.e.~the hypothetical value if $H_\mathrm{tot}=0$  ($H_\mathrm{tot}$ includes the spin-rotation interaction.) 
    The indicated offsets with respect to $f_\mathrm{vib,0}$ are the total shifts $\Delta f_\mathrm{mag}$ due to the magnetic field incl.\,spin-rotation interaction (see  the case $M_N=0\rightarrow M_N'=0$ in Table\,\ref{tab:summary of the shifts}). 
    Energy splittings are indicated in \textbf{\textcolor{alizarin}{red}}, \textbf{\textcolor{ao(english)}{green}} and \textbf{\textcolor{violet}{violet}} color.
    }
    \label{fig:Energy levels}
\end{figure}

Figure~\ref{fig:Energy levels} shows schematically the spin state energies of two rovibrational levels, namely $(v=0,\,N=2)$ and $(v'=2,\,N'=2)$, in a strong magnetic field typical of a \PMT{}. The Zeeman interactions with the external field are the strongest ones (Paschen-Back regime). In each level, the electron spin can be aligned with or opposed to the magnetic field  ($M_s=\pm1/2$). The corresponding energy splitting amounts to $f_\mathrm{sf}\simeq112$\,GHz in a 4~T field. The different orientations of the rotational angular momentum with respect to the field/electron spin, $M_N=-N, \ldots,N$, lead to energy splittings of the order 10\,MHz. A particular spin state is pragmatically denoted as $(v,N,M_s,M_N)$.

As will be shown below, particularly favorable spin components of the rovibrational transitions are those between spin states for which the rotational angular momentum has (approximately) zero projection on the magnetic-field axis: $M_N=0\rightarrow M_N'=0$. These are shown as cyan arrows in the figure. 

\subsection{Perturbations of H$_2^+$}

The fine structure, hyperfine structure and magnetic structure - spin structure for short - of the rovibrational levels can be described to very good approximation by effective Hamiltonians $H_\mathrm{tot}(v,N)$ that have the same operator structure in each rovibrational level but where the different operator combinations therein are multiplied by level-dependent coefficients. These coefficients can be and have been calculated ab initio, in part to high precision.

The total Hamiltonian contains several terms: 
\begin{eqnarray}
        H_\mathrm{tot}(v,N)&=&H_\mathrm{HFS}(v,N)+H_\mathrm{Z}(v,N)+
    H_\mathrm{Z-rot}(v,N)+
    \nonumber\\ 
    &\phantom{=}&
    H_\mathrm{EQ}(v,N)+H_\mathrm{dia}(v,N)+H_\mathrm{para}(v,N)+
    \nonumber\\ 
    &\phantom{=}&
    H_\mathrm{d.c.-Stark}(v,N)+H_\mathrm{a.c.-Stark}(v,N)\ .\label{eq:H_tot1}
\end{eqnarray}
These describe, in sequence, the hyperfine structure (HFS), the electronic Zeeman (Z), the rotational Zeeman (Z-rot), the electric quadrupole (EQ), the diamagnetic, the paramagnetic, the d.c.-Stark, and the a.c.-Stark interactions. All Hamiltonians, except the first on the r.h.s., depend on external magnetic or electric fields.
In the following, we discuss the various Hamiltonians and their impact on the uncertainty achievable when performing a CPTI test.

As mentioned, the quantity of interest for a vibrational CPTI test does not need to be the unperturbed transition frequency, i.e. the one in absence of external perturbations. In fact, because the effect of the \PMT{}'s strong magnetic field on the transition frequency is substantial, there is at present no foreseeable way to obtain the unperturbed transition frequency - at the goal accuracy level of interest here ($10^{-16}$ and lower) - by applying theoretical corrections. 

In the following $f_0$ denotes any rovibrational ($f_\mathrm{vib}$) or electron-spin-flip transition frequency ($f_\mathrm{sf}$).
We will denote by $u(X)$ the fractional (normalized to $f_0$) uncertainty with which a certain shift $X$ affects the determination of the transition frequency of one species in the context of a CPTI test. $u$ may be a systematic or a statistical uncertainty.

\subsection{Spin interactions in zero field}

According to \cite{Korobov2006}, the hyperfine structure Hamiltonian in zero field is
\begin{eqnarray}
        H_\mathrm{HFS}(v,N)&=&b_\mathrm{F}(v,N)\,\mathbf{I}\cdot\mathbf{s}_\mathrm{e}+c_e(v,N)\,\mathbf{N}\cdot\mathbf{s}_\mathrm{e}+
    \nonumber\\ 
    &\phantom{=}&
c_I(v,N)\,\mathbf{I}\cdot\mathbf{N}+\ldots\ \ .
\label{eq:H_hfs}
\end{eqnarray}

\noindent Here, $\mathbf{I}$ is the (dimensionless) total nuclear spin operator. 
$\textbf{N}$ is the (dimensionless) rotational angular momentum operator and $\mathbf{s}_\mathrm{e}$ is the dimensionless electron-spin operator.
Because the two nuclei have spin 1/2 each, the total spin $I$ can take on only the value 0 or 1. The spin-statistics theorem applied to \Htwoplus{} requires that $I=0$ when $N$ is even, and $I=1$ when $N$ is odd. The nuclear-spin-singlet states are denoted as para-\Htwoplus{}, the nuclear triplet states as ortho-\Htwoplus{}.

The dominant interaction is the first term, the Fermi contact interaction, which is analogous to the interaction leading to the hyperfine structure in the 1s level of atomic hydrogen. $b_\mathrm{F}$ is of order $h\times1\,$GHz. The second term is the spin-rotation interaction, with $c_e$ of order $h\times40\,$MHz for small $v$ (see Table\,\ref{tab:bound-electron g-factor}). $c_I$ is of order $h\times0.04\,$MHz. The omitted terms are of order $h\times0.3\,$MHz.

In the detailed discussions in this work we limit ourselves to the case of para-\Htwoplus{}, since this case is identified as favorable for a CPTI test. 
The only surviving term in the effective hyperfine Hamiltonian eq.\,(\ref{eq:H_hfs}) for para-H$_2^+$ is the spin-orbit interaction $c_e(v,N)\,\mathbf{N}\cdot\mathbf{s}_\mathrm{e}$.
A preliminary analysis shows that transitions between odd-$N$ levels of ortho-\Htwoplus{} would have similar metrological performance as in para-\Htwoplus{}. However, the larger number of spin states would make experiments more complicated. 

Obviously, the above Hamiltonian does not lead to any sensitivity to external fields and thus to any systematic uncertainty. However, that Hamiltonian needs to be taken into account when it is desired to compute the systematic shifts accurately and when one wishes to consider measurement scenarios in which different spin components of a transition are measured in order to cancel (large) systematic shifts of individual components, as such cancellation may not be complete. 

The signs of the coefficients in the HFS Hamiltonian are the same for \Htwoplus{} and \antiHtwoplus{}. For example, $c_e$ arises from terms  proportional to (i) the product of the lepton's magnetic moment and the nuclear charges,  or (ii) to the product of lepton charge and baryon charge.
$b_F$, a quantity introduced in eq.\,(\ref{eq:energies of ortho-H2+}) below, is proportional to the product of the magnetic moments of lepton and baryon.
Hence these and all other coefficients are invariant under charge conjugation.

\section{Magnetic perturbations}
\label{sec:Magnetic perturbations}

\subsection{Magnetic interactions linear in the field}

\subsubsection{Spin-Zeeman interaction}

The effective Hamiltonian is approximately
\begin{equation}
\begin{array}{@{}lll}
H_{\rm Z}(v,N)  \approx
    E_{11}\,\mathbf{I}\cdot\mathbf{B}
    +E_{13}(v,N)\,\mathbf{s}_e\cdot\mathbf{B}\ .
    \label{eq:H_Z}
\end{array}
\end{equation}
$\textbf{B}$ is the magnetic field, assumed oriented in $z$-direction. This Hamiltonian has been treated by Karr \textit{et al.}~\cite{Karr2008}, focusing on the weak-field limit. The anisotropy of the electron g-factor was therefore not considered.

For $E_{11}$ we may use the expression $-\mu_\mathrm{n}\,g_p$, with $\mu_{\rm n}$ being the (positive) nuclear magneton and $g_\mathrm{p}$ the bare-nucleus proton g-factor, neglecting the shielding correction since it is tiny.   

In $E_{13}(v,N)=-g_{\mathrm{e}}(v,N)\mu_\mathrm{B}$ we instead use the level-dependent, isotropic g-factor of the bound electron, $g_\mathrm{e}(v,N)$. $\mu_\mathrm{B}$ is the (positive) Bohr magneton. $g_\mathrm{e}(v,N)$ was first computed by Hegstrom \cite{Hegstrom1979} and recently recalculated more precisely by Karr and coworkers \cite{Karr2021,Kullie2025}. A few values of  $g_\mathrm{e}(v,N)$ for levels of interest are presented in Table\,\ref{tab:bound-electron g-factor}.

The electron-Zeeman interaction also has an anisotropic contribution (because the molecule is non-spherical). It adds an approximate interaction energy \cite{Karr2021}
\begin{equation}
    -M_s\mu_\mathrm{B}B\, g_t(v,N)\frac{3 M_N^2-N(N+1)}{\sqrt{N(N+1)(2N-1)(2N+3)}}\ .
\label{eq:anisotropic g_e energy}
\end{equation}

\begin{table*}
    \caption{\justifying \textbf{Ab initio computed properties of para-\Htwoplus{} relevant to the interaction with external fields and to the spin structure.} Columns 3,\,4: the scalar and tensor bound-electron g-factor according to \cite{Karr2021}. Column 5: the rotational g-factor \cite{Karr2008}. Columns 6,\,7: the scalar and tensor d.c.~electric polarizabilities in atomic units, according to \cite{Schiller2014a}. Column 8: electron-spin-rotation coefficient in MHz{\,$\times h$} \cite{Haidar2022b}. The numbers are rounded. The free-electron g-factor is $g_\mathrm{e,free}\simeq-2.0023.$}
    \centering
\small{
\begin{tabular}{c|c||c|c||c||c|c||c}
\hline
$v$ & $N$ & $1-g_{\mathrm{e}}(v,N)/g_{\mathrm{e,free}}$ 
& $g_{\mathrm{t}}(v,N)/g_{\mathrm{e,free}}$ & $g_{r}(v,N)$& $\alpha_s(v,N)$ & $\alpha_t(v,N)$&$c_e(v,N)$ \\
\hline
0& 0&  $20.36\times10^{-6}$& 0& $0$& $3.1687$ &$0$&0 \\
0& 2&  $20.30\times10^{-6}$& $-0.446\times10^{-6}$&$0.9198$& $3.1975$&$-0.193$& 42.1635 \\
2& 0 & $19.32\times10^{-6}$& $0$ &$0$& $4.8215$ & $0$&0 \\
2& 2 & $19.27\times10^{-6}$& $-0.386\times10^{-6}$&$0.9004$& $4.8714$ &$-0.385$&37.1000 \\
3& 2&  $18.82\times10^{-6}$& $-0.358\times10^{-6}$&$0.8896$& $6.0762$ &$-0.532$&34.7303 \\
    \hline
    \end{tabular}
}
\label{tab:bound-electron g-factor}
\end{table*}

\vskip .1in
\subsubsection{Rotational Zeeman interaction}

The rotational Zeeman interaction is given by
\begin{equation}
    H_{\rm Z-rot}(v,N)=-\mu_{\rm n}g_r(v,N)\,{\bf N}\cdot{\bf B}\ .
\label{eq:H_Z-rot}
\end{equation}
Again, this is an effective Hamiltonian that holds in the subspace of the rovibrational level $(v,N)$.  For a rotation-less level, $N=0$, the interaction is zero. The numerical value of the rotational g-factor is level-specific and is of order unity.
Because the interaction is of the order of the nuclear magneton, it is tiny in small fields and has therefore not yet been observed in the MHI.

The rotational g-factors $g_{r}$ have been accurately computed by Karr \textit{et al.}~\cite{Karr2008} (see also further below). A few values relevant for the present discussion are shown in Table\,\ref{tab:bound-electron g-factor}.
Generally, for fixed $v$, the variation of $g_{r}$ with $N\ge1$ is small. The g-factor difference between two rovibrational levels mainly depends on $v'-v$ and little on $N'-N$.

It turns out that for the typical magnetic field strength in a \PMT{}, $B_0=4$\,T, the spin-rotation interaction and the rotational Zeeman interaction are similar in magnitude, e.g. $c_e/2\simeq h\times 21\,\mathrm{MHz}$ vs.~$\mu_\mathrm{n} g_{r}B_0\simeq h\times28\,\mathrm{MHz}$ for $(v=0,N=2)$.

\vskip .1in
\subsubsection{Signs}

Upon charge conjugation (C) of the particle, i.e. when transforming \Htwoplus{} into \antiHtwoplus{}, the signs of the Zeeman interactions in eqs.\,(\ref{eq:H_Z}, \ref{eq:anisotropic g_e energy}) are inverted  because the magnetic moments change sign.
We may take the g factor of free electron and the free positron to be equal (even under C); also their binding contributions (and here, both the scalar and the tensor part) are even under C, as can be seen from the treatment in \cite{Karr2021}. The sign inversion of the magnetic moment is accomplished by transforming $\mu_\mathrm{B}$ (which stands for $-q_e \hbar/2 m_e$, where $q_e$ is the electron charge) into $-\mu_\mathrm{B}$ for the antiparticle.

The spin-rotation Hamiltonian $H_{\rm Z-rot}$ is actually proportional to the particle charges and to the fine structure constant \cite{Schiller2025b}. So, if we wish to maintain the same  $g_r$ for both the molecule and the anti-molecule, then the positive $\mu_\mathrm{n}$ must must be multiplied by $-1$ upon C.

While for a given magnetic field the Zeeman interaction is odd under C, the hyperfine interaction is even (see above), and also the polarisabilities and susceptibilities of the interactions that are quadratic in the external fields (Stark effect, paramagnetic and diamagnetic effect) are even (see below). It would seem, then, that the energy levels of the molecule and anti-molecule are different in the presence of external fields. This is only an apparent contradiction to the statement that the physics of electromagnetism (the interaction relevant in this work) is invariant under C. Indeed one must consider the overall system consisting of the molecule and the apparatus that generates the external fields. Under C, the directions of both the magnetic field and the electric field will be inverted (implying that in our formulae the amplitudes must transform as $B\rightarrow-B$, $E\rightarrow-E$). This $\mathbf{B}$ direction inversion implies that the Zeeman hamiltonian --\,as also the other hamiltonians\,-- actually does not change under a ``global" C.
Under the CPT operation, $\mathbf{B}\rightarrow-\mathbf{B}$, $\mathbf{E}\rightarrow\mathbf{E}$ and again the total Hamiltonian does not change. (A possible linear Stark interaction $-\mathbf{d}\cdot\mathbf{E}$ is also even.) 

However, in future experiments, the situation may occur that one studies both \Htwoplus{} and \antiHtwoplus{} with the same \PMT{} apparatus, without changing the magnetic field direction. For this situation, the changes to be made in the present formulae are, as said, 
\begin{equation}
    \mu_\mathrm{B}\rightarrow-\mu_\mathrm{B}\ ,\mu_\mathrm{n}\rightarrow-\mu_\mathrm{n}\ , \label{eq:C transformation 1}
\end{equation}
leaving the g factors unchanged. The energy spectrum (the set of energy values of all states) will not change, but the state corresponding to a given energy value is the ``spin-opposite" one:  
\begin{equation}
    M_s\rightarrow-M_s\ ,
    M_N\rightarrow-M_N\ ,
    M_I\rightarrow-M_I\ .
    \label{eq:C transformation 2}
    \end{equation}
Indeed, it can be seen that under the transformations 
(\ref{eq:C transformation 1}, \ref{eq:C transformation 2}), the energies in Table\,\ref{tab:table expansion of energies} and App.\,\ref{app:The energies of spin states for $N=2$} remain invariant.

\subsection{Energies and basic considerations}

For fields strengths as they occur in \PMT{}s, the electron spin is almost completely decoupled from the other angular momenta because its interaction energy with the field exceeds the interaction energy with the other angular momenta. Similarly, the total nuclear angular momentum (in ortho-\Htwoplus{}) is almost decoupled from the rotational angular momentum. 

The approximation of complete decouplings already yields the main insights concerning the sensitivity of various transitions to the magnetic field. This was proposed by Myers \cite{Myers2018}. In lowest approximation the energies of \Htwoplus{} in strong field are found by replacing the scalar products in the Hamiltonian by the products of the angular momentum projection quantum numbers ($M_s$, $M_N$) or products of those and the strength of the magnetic field. Of course, $M_s$, $M_N$ are not exact quantum numbers any more, but they can still be used to label the spin states; the energies can be expressed as function of these numbers. 

For para-\Htwoplus{} one has the spin energies
\begin{eqnarray}
    E_\mathrm{HFS + Z + Z-rot}(v,N)&\approx& c_e(v,N)M_s\,M_N 
    -\mu_{\rm B}g_e(v,N)M_s B\nonumber\\
    &&-\mu_{\rm n}g_r(v,N)\,M_N\,B\ .
\end{eqnarray}
Energy eq.\,(\ref{eq:anisotropic g_e energy}) is to be added. 

At this level of approximation, we can also evaluate the shifts to follow below by replacing the operators by their respective quantum numbers. This will be done throughout.

The next-order approximation for para-\Htwoplus{} can be based on the -\,simple\,- Hamiltonian matrices. Their eigenvalues can be expanded in orders of $B^{-1}$, and the results including $B^{-1}$-terms are sufficient. The case $N=0$ is trivial; the case $N=2$ is presented in Table\,\ref{tab:table expansion of energies}.
Appendix~\ref{app:Effective Hamiltonian} summarizes the derivation. (The result in case of neglect of the anisotropic electron g-factor and of other interactions treated below was already given in \cite{Schenkel2024}.)

\begin{table*}[t]
    \centering
    \caption{ \textbf{The expansion of the energies of the $N=2$ - spin states of para-\Htwoplus{} for large magnetic field $B$}. 
     State-independent shifts are not shown.
Here, $c_e=c_e(v, N=2)$ is the spin-rotation coefficient,  $g_e'=g_e'(v, N=2)$ is the bound-electron scalar g-factor plus a contribution from the tensor part,  $g_l=g_{r}(v,N=2)\mu_\mathrm{n}/\mu_\mathrm{B}$ is the rotational
g factor, $\gamma=\gamma(v,N=2,B)$ describes - in absence of electric fields - a contribution to the diamagnetic/paramagnetic tensor correction, $Z=\zeta(v,N=2)\,B$ describes a contribution to the tensor part of the electron's Zeeman energy.  $M_F$ is the (exact) projection of the total angular momentum $F$ onto the static magnetic field direction, $M_s$ is the (approximate) electron spin projection quantum number. Also, the approximate rotation projection quantum number is $M_N=M_F-M_s$. Note that $g_e\simeq-2$ is negative. See  appendix~\ref{app:Effective Hamiltonian} for details.
}
    \label{tab:table expansion of energies}
 \small   
 \begin{tabular}{|c|c|c|}
    \hline
    \multicolumn{3}{|c|}{\vrule width0pt height16pt depth10pt Large$-B$ expansion}\\
    \hline
    \vrule width0pt height16pt depth10pt 
    $M_{F}$ & lower-energy group, $M_{s}=-1/2$ & 
    higher-energy group, $M_{s}=+1/2$
    \\
    \hline
    \vrule width0pt height16pt depth8pt 
    $-\frac{5}{2}$ & 
    $c_e+4\gamma+\frac{1}{2}\mu_{\rm B}B (g'_e+4g_l-4Z)$ &
    \\
	\hline
    \vrule width0pt height16pt depth10pt 
	$-\frac{3}{2}$ &
	$\frac{c_e}{2}+\gamma+\frac{2c^2_e}{\mu_{\rm B}B(2g'_e-2g_l-5Z)}$ &
	$-c_e+4\gamma-\frac{2c^2_e}{\mu_{\rm B}B(2g'_e-2g_l-5Z)}$
	\\
  \vrule width0pt height16pt depth10pt 
	& $+\frac{1}{2}\mu_{\rm B}B\left(g'_e+2g_l-Z\right)$ &
	$-\frac{1}{2}\mu_{\rm B}B\left(g'_e-4(g_l+Z)\right)$ 
	\\
    \hline
    \vrule width0pt height16pt depth10pt 
	$-\frac{1}{2}$ &
	$\frac{1}{2}\mu_{\rm B}B\,g'_e+\frac{3c^2_e}{\mu_{\rm B}B(2g'_e-2g_l-Z)}$ &
	$-\frac{c_e}{2}+\gamma-
		\frac{3c^2_e}{\mu_{\rm B}B(2g'_e-2g_l-Z)}$\\
    \vrule width0pt height16pt depth10pt 
	& &	$-\frac{1}{2}\mu_{\rm B}B\left(g'_e-2g_l-Z\right)$ 
	\\
	\hline
    \vrule width0pt height16pt depth10pt 
	$\frac{1}{2}$ &
	$-\frac{c_e}{2}+\gamma+
		\frac{3c^2_e}{\mu_{\rm B}B(2g'_e-2g_l-Z)}$ &
	$-\frac{1}{2}\mu_{\rm B}B\,g'_e-\frac{3c^2_e}{\mu_{\rm B}B(2g'_e-2g_l-Z)}$ 
	\\
    \vrule width0pt height16pt depth10pt 
	& 	$+\frac{1}{2}\mu_{\rm B}B\left(g'_e-2g_l-Z\right)$ &
	\\
	\hline
    \vrule width0pt height16pt depth10pt 
	$\frac{3}{2}$ &
	$-c_e+4\gamma+\frac{2c^2_e}{\mu_{\rm B}B(2g'_e-2g_l-5Z)}$ &
	$\frac{c_e}{2}+\gamma-\frac{2c^2_e}{\mu_{\rm B}B(2g'_e-2g_l-5Z)}$\\
    \vrule width0pt height16pt depth10pt 
	&	$+\frac{1}{2}\mu_{\rm B}B\left(g'_e-4(g_l+Z)\right)$ &
		$-\frac{1}{2}\mu_{\rm B}B\left(g'_e+2g_l-Z\right)$ 
	\\
    \hline
    \vrule width0pt height16pt depth8pt 
	$\frac{5}{2}$ &
	&
	$c_e+4\gamma-\frac{1}{2}\mu_{\rm B}B (g'_e+4g_l-4Z)$ 
	\\
	\hline
    \end{tabular}
\end{table*}

A glance at the table easily allows us to identify transitions that have a weak sensitivity to magnetic field.
All particle quantities appearing in the table depend on $v$ and $N$. The dependence on $N$ is small except for $c_e$ and $g_r$ between $N=0$ and $N\ne0$ since $c_e(v,N=0)=0$ and $g_r(v,N=0)=0$. Transitions $N=0\rightarrow N'=0$ are forbidden. 

Vibrational transitions that include an electron-spin flip (in the table: between a lower-energy group entry and a higher-energy group entry) are out of the question: The huge electron-spin Zeeman shift contribution ($\approx 110\,$GHz   in 4\,T, or $[10^{-4}]$ relative to $f_\mathrm{vib}$), prevents a $10^{-17}$-level CPTI test, because the magnetic field cannot be measured or kept stable at $10^{-13}$ level. Thus only transitions \emph{within} a group, $M_s=M_s'$, will be considered. We focus on the higher-energy group, $M_s=1/2$.

Transitions $M_N=0\rightarrow M_N'=0$ (the case $M_F=1/2$ in Table\,\ref{tab:table expansion of energies} if $M_s=1/2$) -- without restriction on $N$, $N'$ -- are best. They have been proposed in \cite{Schenkel2024}. There is no rotational Zeeman effect in first order. To lowest order in $B$ the magnetic energy difference is $-\mu_\mathrm{B}B\,(g_e(v',N')-g_e(v,N)) M_s$ and so there is a residual first-order sensitivity to the magnetic field, arising from (mostly) the vibrational dependence of the bound-electron g-factor. This shift is $\pm58\,$kHz for the $(0,2)\rightarrow(2,2)$ transition in $B=4\,$T (the sign depends on the sign of $M_s$). 
For comparison, the shift is one order smaller than that from the $v$-dependence of the rotational g-factor in the (not optimum) case~(\ref{eq:rot Zeeman, M_N=0->M_N'=0}) discussed in Sec.\,\ref{sec:rotational g factor}. 

There is also a contribution from the $c_e^2$ term (see Table\,\ref{tab:table expansion of energies}), that for the reference transition is approximately one order smaller.

Less favorable in terms of magnetic-field sensitivity are $N\ne0\rightarrow N'\ne0$ transitions having $M_N\ne0\rightarrow M_N'=M_N$, corresponding to the cases $M_F=M_F'=-3/2,\,-1/2,\,+3/2$ or $+5/2$ (always assuming the subspace $M_s=1/2\rightarrow M_s'=1/2$). Here, the vibrational dependence of the rotational g-factor is the dominant effect:
the magnetic energy is $M_N\,\mu_\mathrm{B}B\, (g_r(v',N')-g_r(v,N))$.
Myers \cite{Myers2018} considered a transition in this class, with $M_N=2$.

Transitions $(N,M_s,M_N)\rightarrow (N',M_s,M_N'\ne M_N)$
are even less favorable. Here, the full impact of the rotational Zeeman energy makes itself felt, the magnetic energy difference being $\mu_\mathrm{B}B\,(M_N'\,g_r(v',N')-M_N\,g_r(v,N))$, which is of order $\mu_\mathrm{B}B\,g_r$.

\subsubsection{Details of the rotational Zeeman effect}
\label{sec:rotational g factor}

In a field of 4\,T the shift of a level $N\ne0$, $M_N=1$ due to this effect is 
\begin{eqnarray}
        \Delta E_{\rm Z-rot}&\simeq&-\mu_{\rm n} g_r(v,N\ne0)\,B_0/h\nonumber\\
    &\simeq& -27\,\text{MHz}\ \  [-2\times10^{-7}]\ .
    \label{eq:estimate rotational Zeeman shift}
\end{eqnarray}
For a CPTI test aimed at the $10^{-18}$-level, the magnitude of this shift can pose a challenge to maintain it constant (at the required 0.2\,mHz level), or to follow its time drift it via repeated measurements of the magnetic field.

Therefore, it is advantageous to consider transitions $M_N\rightarrow M_N'=M_N$, that take advantage of the fact that the differential g-factor is much smaller than unity, $g_r(v'=2,N)-g_r(v=0,N)\simeq-0.02$. Specifically,
the shift is 
\begin{eqnarray}
    \Delta f_\mathrm{rot}&\simeq&-\mu_{\mathrm{n}} (g_r(v'=2,N)-g_r(v=0,N)) B_0 \,M_N/h\nonumber\\
    &\simeq&+0.60\,\mathrm{MHz}\times M_N\ \  [+5\times10^{-9} M_N]\ .
    \label{eq:rot Zeeman, M_N=0->M_N'=0}
\end{eqnarray}
Obviously, the most favorable case is $M_N=M_N'=0$, when the effect vanishes.
In the less favorable case $M_N\ne0$, eq.\,(\ref{eq:rot Zeeman, M_N=0->M_N'=0}), the measurement on each species should be accompanied by accurate,
synchronized $1\times10^{-9}$-fractional-accuracy 
determinations of $B$ as further discussed in Sec.~\ref{Summary of magnetic effects}.
The same considerations hold for the transition $(0,2)\rightarrow(3,2)$. Its higher transition frequency is offset by a larger g-factor difference.

If it is possible to measure several spin components sufficiently accurately, then the shift can be determined and canceled. For example, the mean of 
$M_N=1\rightarrow M_N'=1$ and $M_N=-1\rightarrow M_N'=-1$ cancels the rotational Zeeman shift. 
It is also canceled in the mean of $M_N=0\rightarrow M_N'=1$ and $M_N=0\rightarrow M_N'=-1$. In such scenarios one has to ensure that the variations of the magnetic field between the two measurements are sufficiently small. Clearly, this will be most difficult to achieve in the last case. Implementing such scenarios will require driving some additional transitions in order to prepare the ion for the complementary transition. A possible sequence might be ($M_s$ remains unchanged)
\begin{eqnarray}
\hskip -0cm(v,N,M_N=1)&&\xrightarrow{\mathrm{optical}}(v',N',M_N'=1)\nonumber \\
 &&\xrightarrow{\mathrm{RF}-\pi}   (v',N',M_N'=0)\xrightarrow{\mathrm{RF}-\pi}(v',N',M_N'=-1)\nonumber \\
&&\xrightarrow{\mathrm{optical}}(v,N,M_N=-1)\ .
\end{eqnarray}    
Here, the two optical transitions are driven so as to have 50\% transition probability, while the RF transitions are implemented as $\pi$-pulses.

\subsubsection{The anisotropic electron Zeeman effect}

The $M_N$-independent term in eq.\,(\ref{eq:anisotropic g_e energy}) acts as additional contribution to the scalar electron g-factor, a negligible effect.
The $M_N$-dependent term acts similarly to the rotational Zeeman effect, given that we consider vibrational transitions that leave $M_s$ unchanged. The value of the term for e.g.~$(v=2,N=2)$ in a 4\,T field is $\simeq 6\,\mathrm{kHz}\times M_N^2$. The effect of such level shifts on transition frequencies is two orders smaller than what arises from the rotational Zeeman effect, eq.\,(\ref{eq:rot Zeeman, M_N=0->M_N'=0}),  and is thus negligible in comparison.

 \subsection{Magnetic interactions quadratic in the field}

It is well-known that a diamagnetic interaction exists between charges and a magnetic field. It originates from the squared vector potential  in the interaction Hamiltonian of a charged particle and an electromagnetic field. Often, only the scalar contribution needs to be considered. However, the full treatment for \Htwoplus{} leads to an effective Hamiltonian \cite{Schiller2025b}
\begin{equation}
H_{\rm dia}(v,N) =
   -\frac{\alpha^2}{2}\Tilde{B}^2(\chi^{(d)}_s(v,N) + \chi^{(d)}_t(v,N)(N_z^2-{\textstyle\frac{1}{3}}\bold{N}^2)).
\end{equation}
This  holds for a magnetic field aligned along the quantization $(z)$ axis. Here, $\Tilde{B}$ is the magnetic field in appropriate units (see below). 
$\chi_s^{(d)}$, $\chi_t^{(d)}$ are dimensionless, order-unity, level-dependent scalar and tensor diamagnetic susceptibilities, respectively. The dependence on the fine-structure constant $\alpha$ is explicitly shown so as to clarify the order-of-magnitude of the effect. The diamagnetic susceptibility $\chi^{(d)}_{s,t}$ includes a contribution from the electron and another, much smaller one, stemming from the nuclei. 

Furthermore, there exists a paramagnetic interaction, described by an analogous effective Hamiltonian,
\begin{equation}
H_{\rm para}(v,N) =
   -\frac{\alpha^2}{2}\Tilde{B}^2(\chi^{(p)}_s(v,N) + \chi^{(p)}_t(v,N)(N_z^2-{\textstyle\frac{1}{3}}\bold{N}^2)).
\end{equation}
This Hamiltonian arises from second-order perturbation theory, in contrast to the diamagnetic one. 
The \emph{paramagnetic}  susceptibilities $\chi_s^{(p)}$, $\chi_t^{(p)}$ are smaller than the diamagnetic ones. They originate from the electron.  
Note that the susceptibilities have the same sign for \Htwoplus{} and \antiHtwoplus{}: the diamagnetic ones are independent of the electron charge, while the paramagnetic ones are quadratic in the charges \cite{Schiller2025b}.

V.\,I.\,Korobov has computed the susceptibilities in the non-relativistic approximation. High-accuracy three-body wavefunctions have been used. Whereas the diamagnetic susceptibilities of any rovibrational level are particular expectation values over the wavefunction, the paramagnetic ones arise from second-order perturbation theory and are more complicated to compute. Table\,\ref{tab:susceptibilities} displays results relevant to the present discussion. For a detailed exposition, see \cite{Schiller2025b}.
To obtain actual energies, using the susceptibility values appearing in the table, $\Tilde{B}^2$ must be replaced by
$(4\pi/\mu_0)\,a_0^3\,B^2$,
where the vacuum permeability $\mu_0$, the Bohr radius $a_0$ and $B$ are in SI units.

\newcommand{\chis}{{5*0.1}}

\begin{table*}[t!]
    \centering
\begin{tabular}{c|cccc}
\hline\hline
  & \multicolumn{2}{c}{$N=0$} & \multicolumn{2}{c}{$N=2$} \\
\hline
  &  $\chi_s$, scalar&$\chi_t$, tensor& $\chi_s$, scalar& $\chi_t$, tensor \\[0.7mm]
$v=0$  & $-0.3836419$ & 0\ \ \  & $-0.3848714$ & $-0.0081525$ \\
$v=2$  &    \ \ \        & \ \ \  & $-0.4251311$ & $-0.0102977$ \\
\hline\hline
\end{tabular}
\caption{ \textbf{Magnetic susceptibilities of a few low-lying levels of \Htwoplus{}}. Each rovibrational level has a scalar and a tensor contribution. The tensor contribution vanishes if $N=0$. 
Shown are the total susceptibilities $\chi_s=\chi_s^{(d)}+\chi_s^{(p)}$ and 
$\chi_t=\chi_t^{(d)}+\chi_t^{(p)}$, that determine the energy shifts. 
The values were computed by V.~I.~Korobov and are in atomic units. 
    \label{tab:susceptibilities}   }
\end{table*}

Let us now consider how a vibrational transition in strong field is affected. The total scalar shift of the reference transition is approximately 38\,kHz $[3\times10^{-10}]$ in 4\,Tesla. 
In a state-of-the-art \PMT{}, $B$ can be tracked with resolution better than 1 part in $10^9$. Therefore, the uncertainty due to this  shift's time variation via $B$ variation can be kept below  $1\times10^{-18}$. 

It is important to note that (at fixed magnetic field) the differential scalar shift cannot be measured directly, as it impacts all spin states of the upper and lower level equally. 
Of course, measurements at different field strengths would allow determining the effect very precisely. Varying the magnetic field in a \PMT{} via the electric current is impractical, but a \PMT{} may contain trapping areas equipped with additional permanent magnets allowing for locally different fields. 
Alternatively, measurements may be performed in different \PMT{}s operating at significantly different fields. Note that to determine the shift with $1\times10^{-17}$ fractional uncertainty contribution requires doing just that: making transition frequency determinations at that fractional level and under at least two substantially different B-field conditions. 

Also, the shift could be inferred from a comparison between experimental transition frequency in strong field and the QED prediction of the frequency for zero magnetic field. The accuracy of the determination would be limited by the uncertainty of the QED prediction, currently in the high-$10^{-12}$ range. Another option is the comparison with the experimental transition frequency in zero (or near-zero) magnetic field. 

These considerations are relevant if one is interested in determining these shifts precisely, but they are not essential for a CPTI test.

We now turn to the tensor shift of a transition. As Table\,\ref{tab:susceptibilities} shows, the differential tensor susceptibility is one order smaller, leading to a shift of $-4.1\,$kHz  in 4\,T for the $M_N=0\rightarrow M_N'=0$ component.  Thus, 
the tensor shift is not a major issue among the various magnetic sensitivities.

Moreover, a different situation exists for the tensor shift compared to the scalar shift: the tensor shift depends on the quantum numbers $M_N$, $M_N'$. 
The dependence implies that the shift can be determined experimentally from data taken at a constant $B$-field value, by measuring the transition frequencies of at least two spin components of a rovibrational transition.  
For example, the shift of the  $M_N=0\rightarrow M_N'=0$ component  has the opposite value compared the $M_N=2\rightarrow M_N'=2$ component. Thus, it can be determined from the the difference of the transition frequencies and eliminated by taking the mean. (For simplicity of the discussion we ignored the other perturbations and the diagonalization of the Hamiltonian. For more details, see appendix \ref{app:determination of tensor polarisability}.)  

\subsection{Summary of magnetic effects}
\label{Summary of magnetic effects}

\begin{table}[t!]
    \caption{ \textbf{The magnetic sensitivities $\beta$ and the magnetic shifts $\Delta f_\mathrm{mag}$} of all spin components of two first-overtone rovibrational transitions that do not involve an electron-spin flip.                 $B_0=4\,$T has been assumed.}
    \centering
    Transitions $(v=0,N=0,M_s,M_N)\rightarrow(v'=2,N'=2,M_s,M_N')$
\vskip .1in
{\small 
$
\begin{array}{cccc|cccc}
\multicolumn{4}{c}{M_s=1/2}&\multicolumn{4}{c}{M_s=-1/2}\\
\hline
M_N&M_N'&\beta\ (\mathrm{kHz/T})&\Delta f_\mathrm{mag}\mathrm{\ (kHz)}& 
M_N&M_N'&\beta\ (\mathrm{kHz/T})&\Delta f_\mathrm{mag}\mathrm{\ (kHz)}\\
\hline
 0 & -2 & \phantom{-}13.8\times 10^3 & \phantom{-}55.0\times 10^3    & 0 & -2 & \phantom{-}13.7\times 10^3 & \phantom{-}54.9\times 10^3 \\
 0 & -1 & \phantom{-}6.89\times 10^3 & \phantom{-}27.6\times 10^3    & 0 & -1 & \phantom{-}6.86\times 10^3 & \phantom{-}27.4\times 10^3 \\
 0 & 0 & 23.5 & 111.    & 0 & 0 & -3.61 & -71.2 \\
 0 & 1 & -6.83\times 10^3 & -27.3\times 10^3    & 0 & 1 & -6.86\times 10^3 & -27.5\times 10^3 \\
 0 & 2 & -13.7\times 10^3 & -54.8\times 10^3    & 0 & 2 & -13.7\times 10^3 & -54.9\times 10^3 \\
\end{array}
$
}
\vskip 0.3in
        Transitions $(v=0,N=2,M_s,M_N)\rightarrow(v'=2,N'=2,M_s,M_N')$
\vskip .1in
{\small 
$
\begin{array}{cccc|cccc}
\multicolumn{4}{c}{M_s=1/2}&\multicolumn{4}{c}{M_s=-1/2}\\
\hline
M_N&M_N'&\beta\ (\mathrm{kHz/T})&\Delta f_\mathrm{mag}\mathrm{\ (kHz)}&
M_N&M_N'&\beta\ (\mathrm{kHz/T})&\Delta f_\mathrm{mag}\mathrm{\ (kHz)}\\
\hline
 -2 & -2 & -259. & -1.08\times 10^3    & -2 & -2 & -289. & -1.20\times 10^3 \\
 -1 & -1 & -114. & -504.    & -1 & -1 & -145. & -608. \\
 0 & 0 & \phantom{-}\phantom{-}32.5 & \phantom{-}\phantom{-}84.8    & 0 & 0 & \phantom{-}\phantom{-}\phantom{-}1.81 & \phantom{-}-16.3 \\
 1 & 1 & \phantom{-}181. & \phantom{-}681.    & 1 & 1 & \phantom{-}150. & \phantom{-}576. \\
 2 & 2 & \phantom{-}333. & \phantom{-}1.28\times 10^3    & 2 & 2 & \phantom{-}301. & \phantom{-}1.17\times 10^3 \\
 \hline
 -2 & -1 & -7.13\times 10^3 & -28.5\times 10^3    & -2 & -1 & -7.17\times 10^3 & -28.7\times 10^3 \\
 -1 & -2 & \phantom{-}6.76\times 10^3 & \phantom{-}27.0\times 10^3    & -1 & -2 & \phantom{-}6.73\times 10^3 & \phantom{-}26.9\times 10^3 \\
 -1 & 0 & -6.98\times 10^3 & -28.0\times 10^3    & -1 & 0 & -7.01\times 10^3 & -28.1\times 10^3 \\
 0 & -1 & \phantom{-}6.90\times 10^3 & \phantom{-}27.5\times 10^3    & 0 & -1 & \phantom{-}6.87\times 10^3 & \phantom{-}27.5\times 10^3 \\
 0 & 1 & -6.83\times 10^3 & -27.4\times 10^3    & 0 & 1 & -6.86\times 10^3 & -27.5\times 10^3 \\
 1 & 0 & \phantom{-}7.04\times 10^3 & \phantom{-}28.1\times 10^3    & 1 & 0 & \phantom{-}7.01\times 10^3 & \phantom{-}28.0\times 10^3 \\
 1 & 2 & -6.70\times 10^3 & -26.8\times 10^3    & 1 & 2 & -6.70\times 10^3 & -26.8\times 10^3 \\
 2 & 1 & \phantom{-}7.19\times 10^3 & \phantom{-}28.7\times 10^3    & 2 & 1 & \phantom{-}7.15\times 10^3 & \phantom{-}28.6\times 10^3 \\
\hline
-2 & 0 & -14.0\times 10^3 & -56.0\times 10^3    & -2 & 0 & -14.0\times 10^3 & -56.2\times 10^3 \\
 -1 & 1 & -13.8\times 10^3 & -55.4\times 10^3    & -1 & 1 & -13.9\times 10^3 & -55.5\times 10^3 \\
 0 & -2 & \phantom{-}13.8\times 10^3 & \phantom{-}55.0\times 10^3    & 0 & -2 & \phantom{-}13.7\times 10^3 & \phantom{-}55.0\times 10^3 \\
 0 & 2 & -13.7\times 10^3 & -54.8\times 10^3    & 0 & 2 & -13.7\times 10^3 & -54.9\times 10^3 \\
 1 & -1 & \phantom{-}13.9\times 10^3 & \phantom{-}55.6\times 10^3    & 1 & -1 & \phantom{-}13.9\times 10^3 & \phantom{-}55.5\times 10^3 \\
 2 & 0 & \phantom{-}14.0\times 10^3 & \phantom{-}56.2\times 10^3    & 2 & 0 & \phantom{-}14.0\times 10^3 & \phantom{-}56.0\times 10^3 \\
\end{array}
$
}    \label{tab:sensitvities beta}
\end{table}

\begin{figure}[t]
    \centering
    \includegraphics[width=1\linewidth]{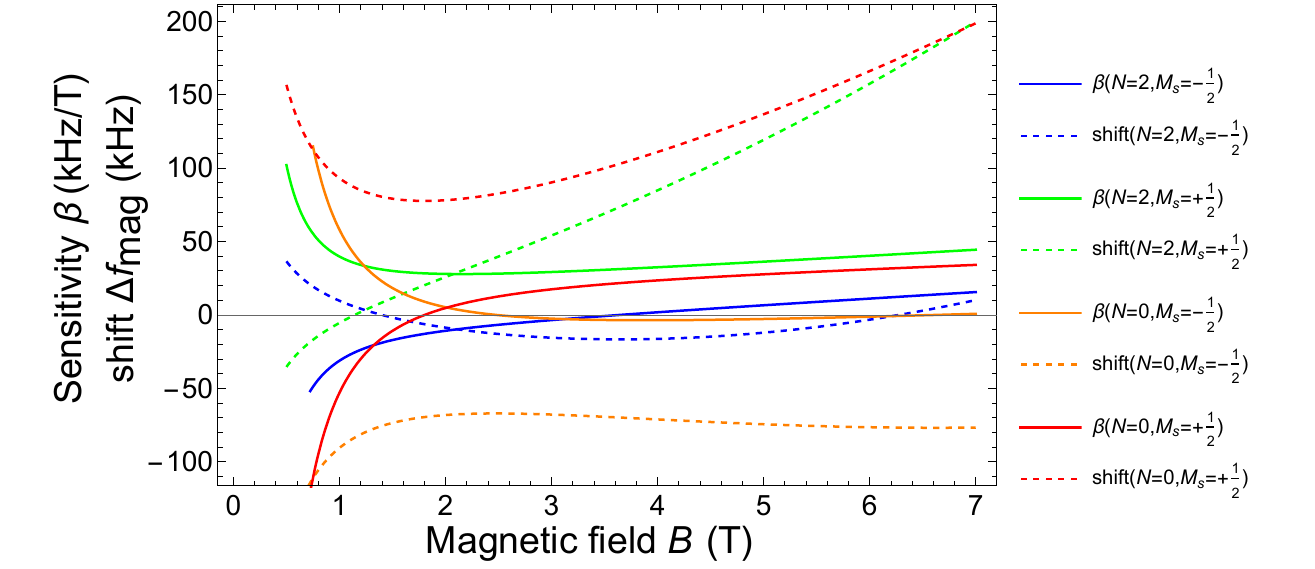}
    \caption{ \textbf{Sensitivities $\beta$ (full lines) and magnetic-field shifts $\Delta f_\mathrm{mag}$ (dashed lines)} of the $(v=0,N,M_s,M_N=0)\rightarrow(v'=2,N'=2,M_s'=M_s,M_N'=0)$ transitions as a function of magnetic field $B$. The cases of lower rotational quantum number $N=0$ and $N=2$ are shown.   }
    \label{fig:Plot of sensitivities and shifts versus B-field}
\end{figure}

Table\,\ref{tab:sensitvities beta} summarizes for a number of spin components the total magnetic shift $\Delta f_\mathrm{mag}$ and the sensitivity $\beta$ to the magnetic field, at $B_0=4\,$Tesla. The definition is $\beta=d \Delta f_\mathrm{mag}(B_0)/d B$. 

$\Delta f_\mathrm{mag}$ includes all interactions in eq.\,(\ref{eq:H_tot1}) except the Stark effect. It includes the spin-rotation interaction, which becomes $B$-dependent in large $B$, see the $c_e^2$ contribution in Table\,\ref{tab:table expansion of energies}. This justifies the naming ``magnetic shift". 

In addition, we show in figure~\ref{fig:Plot of sensitivities and shifts versus B-field} the shifts and sensitivities $\beta$ for the particularly favorable $M_N=0\rightarrow M_N'=0$ spin components, for a range of magnetic fields as they may be implemented in a \PMT{}. 

The compilation of sensitivities allows a first selection of suitable transitions. A CPTI test is implemented by comparing the frequencies of \Htwoplus{} and \antiHtwoplus{}. There are two scenarios: (1) The measurements are performed in different \PMT{}s. Then one requires magnetic field metrology with absolute uncertainty.
(2) If the comparison is made in the same \PMT{}, for example by shuttling the ion of each species back and forth to the interrogation section and performing alternating measurements on each species, the requirements are further relaxed: it is then only the magnetic field's \emph{relative} uncertainty (or instability on the time scale of the cycle time) that must be considered. 
(A third scenario, two precision traps located close-by in the same trap stack, for simultaneous interrogation of the dissimilar ions, is briefly considered in the final section.)

For the purpose of the following discussion, we assume that in a \PMT{} the absolute accuracy with which the magnetic field can be tracked, with a time resolution of 90\,s, is at a level of $\Delta B/B\approx5\times10^{-10}$. This is e.g.~the typical value achieved in \cite{borchert202216} using cyclotron peak detection. 

We set the goal of a magnetic-field related uncertainty of $u_\mathrm{sys}(\Delta f_\mathrm{mag})\le3\times10^{-18}$, where we understand this to refer to repeatability: the uncertainty related to the ability of setting the same magnetic field value at the ion (observed via  the cyclotron frequency or an ESR frequency).
Taking advantage of the above \PMT{}-magnetic-field determination capabilities, we find that transitions with sensitivities $\beta$ up to $\pm 200\,$kHz/T at 4\,T are acceptable. 

Figure\,\ref{fig:Plot of sensitivities and shifts versus B-field} shows that the $\Delta v=2$ transitions considered so far offer $M_N=0\rightarrow M_N'=0$ components, for both electron-spin orientations, that satisfy this condition. 
Note that there are three special magnetic fields values at which one of the considered transitions is completely insensitive to the magnetic field. 

For concreteness, we shall consider a specific value of magnetic field, e.g. $B_0=4\,$T, and only the $N=2\rightarrow N'=2$ transition. According to the table, there are six spin components that satisfy the condition $|\beta|<200\,$kHz/T.
The smallest sensitivity, $\beta_\mathrm{min}=1.8\,$kHz/T, occurs for $M_N=0\rightarrow M_N'=0$ and spin projection $M_s=-1/2$. If that transition is used, the requirement to precisely measure or maintain constant the  magnetic field is particularly uncritical.
This small value (and the one for $M_s=+1/2$, still usefully small) occurs because of partial cancellation of the various contributions to the total magnetic field shift. 

For the implementation of the QDS cancellation concept (see Sec.~\ref{sec:QDS cancellation}), transitions with $\beta>\beta_\mathrm{min}$ may be required.

\subsection{Theoretical uncertainties}
\label{Theoretical uncertainties}

The above results are affected by the uncertainties of the theoretical coefficients.
The magnetic susceptibilities $\chi$ have so far been calculated only in the non-relativistic approximation and are therefore expected to have a fractional uncertainty of $10^{-4}$ (order $\alpha^2$). The diamagnetic plus paramagnetic energy shift of a level is on the order of 400\,kHz at 4\,T. Assuming conservatively no correlation between the uncertainties of the upper and lower level, the uncertainty of the transition frequency is of order 40\,Hz. Similarly the uncertainty of the sensitivity is conservatively of order 20\,Hz/T.  

Less relevant are the uncertainties of $g_r$ and $g_e$. For the rotational g-factor the relativistic corrections have been computed \cite{Schiller2025b}, bringing the fractional uncertainties of $g_r$ to the $10^{-6}$ level. However, for the purpose of the present analysis, it is sufficient to use the older values \cite{Karr2008}.

The ab initio theory of the (scalar) bound-electron $g$ factor is currently the most advanced; here also QED corrections have recently been computed and $g_e(v,N)$ is available with $5\times10^{-11}$ fractional uncertainty \cite{Karr2021,Kullie2025}. The theory has been confirmed experimentally for the case of HD$^+$ \cite{Koenig2025b},  albeit at a higher level of uncertainty.

The uncertainty of $\chi$ also implies that the determination of the total magnetic shift from theoretical coefficients and experimentally measured magnetic field -- and thus of the zero-field transition frequency -- would have an uncertainty of the order of 40\,Hz, or $3\times10^{-13}$ relative to $f_0$. 
This is not a limiting factor to a CPTI test at the $10^{-18}$ level. 

   A comparison of the transition frequency in a \PMT{} (e.g. on \Htwoplus{}) and an RF-trap based determination would allow a strong consistency check of the understanding of all systematics in both trapping environments if there also is an independent (experimental or theoretical) determination of the diamagnetic and paramagnetic shift effect. Thus, in the medium-term future, assuming strong experimental progress in \PMT{}s, an improved diamagnetism/paramagnetism theory, including relativistic effects, might be desirable.

Therefore, it seems worthwhile to develop a theory of the relativistic corrections to the susceptibilities. Once available, the theoretical uncertainty of the corresponding shift will be reduced by approximately two orders. 

\section{Electric perturbations}
\label{sec:Electric perturbations}

\subsection{D.c.~Stark shift}
\label{sec: d.c. Stark shift}
The effective Hamiltonian for the d.c. Stark shift has been discussed in \cite{Schiller2014a}. It is given by
\begin{eqnarray}
    &&\hskip -2 cm H_{\mathrm{dc-Stark}}(v,N)=\label{eq:H_dc stark}\\ 
    &&-{\textstyle\frac{1}{2}}
    \alpha_s(v,N)(E_x^2+E_y^2+E_z^2)\nonumber\\
    &&-{\textstyle\frac{1}{2}}\alpha_t(v,N)(E_z^2-{\textstyle\frac{1}{2}}(E_x^2+E_y^2))(2 N_z^2-{\textstyle\frac{2}{3}} \bold{N}^2)\ .\nonumber
\end{eqnarray}
The scalar polarizability $\alpha_s$ and the tensor polarizability $\alpha_t$ have been computed in the same reference. The values in atomic units are listed in Table\,\ref{tab:bound-electron g-factor}.
Note that the polarizabilities are quadratic in the electric dipole moment of the molecule. Hence, they have the same sign for \Htwoplus{} and \antiHtwoplus{}.

We recall that the electric field generated by the quadrupolar electrostatic potential of a \PMT{} can be expressed in terms of the axial oscillation frequency $\omega_z$, as
\begin{equation}
{\bf E}(x,y,z)= \frac{m}{2\,q} \omega_z^2 
\begin{bmatrix}
x \\
y \\
-2 z
\end{bmatrix}\ .
\label{eq:electric field in a PMT}
\end{equation}
$m$ and $q$ are the mass and charge of the particle, respectively. Since the axial temperature is defined as $k_\mathrm{B}T_z/2=m\,\omega_z^2 \langle z^2\rangle/2$ (the angled brackets denote statistical expectation values), the mean squared axial electric field can be expressed in terms of $T_\mathrm{z}$,
\begin{equation}
    \langle E_z^2\rangle=k_\mathrm{B}\,T_\mathrm{z}\,\frac{m}{q^2}\,(2\pi\nu_z)^2\ .
    \label{eq:mean squared axial electric field}
\end{equation}
The corresponding Stark shift according to eq.\,(\ref{eq:H_dc stark}) is $<1\times10^{-6}\,$Hz for $T_z=0.2\,$K and $\nu_z=1\,$MHz.
It is negligible. 

The radial electric field $E_r=(E_x^2+E_y^2)^{1/2}$ has two contributions. First, one generated by the quadrupolar electrostatic potential, 
\begin{equation}
    |\bold{E}_r|=\frac{m}{2 |q|}\omega_z^2 \,r\ ,
    \label{eq:E radial due to static quadrupole potential}
\end{equation}
where $r$ is the distance from the $z$ axis. It is negligible compared to the second, the motionally induced electric field, $\bold{E}_{r,\mathrm{motion}}=
-\bold{v}\times\bold{B}$ (in non-relativistic approximation). (This field is directed radially inward for a positively charged ion. In case of magnetron motion it provides the radial trapping force that compensates the outward force, eq.\,(\ref{eq:E radial due to static quadrupole potential})). A detailed treatment of the action of this time-dependent field on the energy levels of a molecule is presented in \cite{Kathrein2024}. Here a simplified treatment suffices. The squared field is
\begin{equation}
    |\bold{E}_{r,\mathrm{motion}}|^2=(v_x^2+v_y^2)B^2\ .
    \label{eq:E_r,motion as a fct of vx, vy}
\end{equation}
If cyclotron motion provides the dominant velocity contribution, then the induced field may be expressed in terms of the cyclotron motion orbital radius:
\begin{equation}
    |\bold{E}_{r,\mathrm{motion}}|=\frac{|q|}{m}B^2 \,r_{\mathrm{orbital}}\ .
    \label{eq:E_r,motion}
\end{equation}

Thus, considering only eq.\,(\ref{eq:E_r,motion}) and eq.\,(\ref{eq:H_dc stark}), a transition $(v=0,N=2)\rightarrow(v'=2,N'=2)$ without change in magnetic quantum number, $M_N=0\rightarrow M_N'=0$, has a shift (in 4\,T)
\begin{equation}
    \Delta f_{\rm dc-Stark}/f_0\simeq-3\times10^{-16}\times (r_{\mathrm{orbital}}/1\,\mu\mathrm{m})^2\ .
\label{eq:Delta f_dc-Stark}
\end{equation}
In terms of the thermal energy of transverse motion, $k_B T_r/2=m\langle v_x^2+v_y^2\rangle/2$,  
\begin{equation}
    \Delta f_{\rm dc-Stark}/f_0\simeq-4\times10^{-17}\times (T_r/1\,\mathrm{K})\ .
\label{eq:Delta f_dc-Stark thermal}
\end{equation}

The question arises whether the field $|\mathbf{E}_{r,\mathrm{motion}}|$ or at least its variations in time can  principle be determined via spectroscopy. Their  coupling to $\alpha_s$ affects the ``spin-averaged" vibrational transition frequency (and thus all spin components equally). But this frequency is also the actual quantity of interest in the CPTI test. Therefore an independent spectroscopic signature would be of interest. This is discussed in Appendix~\ref{app:determination of tensor polarisability}.
It appears that RF transitions (change in $M_N$, no change in $M_s$) might be suitable. 
Such transitions could also be detected by the CSGE effect, by driving an electron-spin flip transition tuned to the spin state excited after the RF pulse. Myers has considered RF transitions and their detection, for a different purpose \cite{Myers2018}.

We may also consider ways to determine in an independent way the experimental parameter that determines the transverse Stark shift: the orbital radius $r_\mathrm{orbital}$ or the transverse temperature. 

A measurement could occur by measuring the strength of cyclotron motional sidebands on the optical transition, when the spectroscopy wave is oriented in the radial direction. For a single atomic ion, these have been observed \cite{Mavadia2014}.
It is reasonable to assume that using such a strategy in a CSGE-\PMT{} one can achieve a fractional accuracy (or just reproducibility) for $r_\mathrm{orbital}$ at the 5\% level, giving an uncertainty 
$\simeq3\times 10^{-17}$ for a 1-$\mu$m radius. 

Alternatively, we consider cooling of the cyclotron mode, a technique that routinely allows reaching  radii below $50\,$nm for protons and antiprotons (see Sec.\,\ref{sec:The cyclotron mode}). 
We shall assume that also a \Htwoplus{}/\antiHtwoplus{} ion can be prepared at 50\,mK cyclotron temperature.
The corresponding d.c.~Stark shift is $u_\mathrm{sys}(\Delta f^\mathrm{CSGE}_{\mathrm{dc-Stark}})=2\times10^{-18}$. 

The Stark shift from magnetron motion is negligible because 
the net electric field experienced, $\mathbf{E}_r+\mathbf{E}_{r,\mathrm{motion}}$, is nearly zero. Moreover this mode can also be cooled, to mK-temperatures (Sec.\,\ref{sec:The cyclotron mode}).

Possibly, one could also perform measurements of the rovibrational transition spin component as a function of $r_\mathrm{orbital}$ and extrapolate to zero radius. This would then allow determining the actual shift rather than merely keeping it constant.

In case of the QLS-\PMT{}, the molecular ion will be sympathetically cooled by a laser-cooled beryllium ion. The laser cooling will be implemented on the axial mode and the two radial modes (magnetron, cyclotron). It is expected that cooling leads to the quantum-mechanical ground state of the three modes, see Sec.~\ref{sec:The sympathetic cooling process}.
 Nevertheless, conservatively, we shall assume that the uncertainty of the motional Stark shift arises from an uncertainty in the quantum numbers associated with the radial motion. We re-express eq.\,(\ref{eq:E_r,motion as a fct of vx, vy}) using the quantum-mechanical virial theorem ($E_\mathrm{kin}=E_\mathrm{pot}=E_\mathrm{tot}/2$) as
 \begin{equation}
    |\bold{E}_{r,\mathrm{motion}}|^2=2\,E_{\mathrm{kin},r}B^2/m= E_{\mathrm{tot},r}B^2/m\ .
    \label{eq:E_r,motion as a fct of Etot}
\end{equation}
Let the uncertainty of the total radial energy $E_{\mathrm{tot},r}$ be one quantum for each radial mode, $u(E_{\mathrm{tot},r})=[(\hbar \omega_+)^2+(\hbar\omega_-)^2]^{1/2}\approx\hbar\omega_+$. Here $\omega_-$, $\omega_+$ are the angular frequencies of the magnetron and cyclotron modes, respectively. With $\omega_+\approx |q|B/m$ we find the uncertainty of the shift to be negligible, $u_\mathrm{sys}(\Delta f^\mathrm{QLS}_{\mathrm{dc-Stark}})<1\times10^{-18}$.

\subsection{The a.c. Stark shift (light shift)}
\label{sec:A.C. Stark shift (light shift)}
Here we consider the light shift that occurs during the excitation of the ion. The spectroscopy radiation shifts both the lower and the upper level.

We describe the interaction with radiation by the effective Hamiltonian analogous to the d.c. Stark shift,
\begin{eqnarray}
       &&\hskip -2 cm H_{\mathrm{ac-Stark}}(v,N,\omega)=-\frac{F^2}{2}\times\nonumber\\
        &&(\alpha_s(v,N,\omega)+\alpha_t(v,N,\omega)(2\,N_z^2-{\textstyle \frac{2}{3}}\,\bold{N}^2))  \ .
    \label{eq:H_ac-Stark}
\end{eqnarray}
Here, $F$ is the electric field amplitude of the laser wave, assumed to be polarized along the magnetic field direction. 

The a.c.~polarizabilities $\alpha_s$, $\alpha_t$ for a few low-lying levels $v=0,1,3$ and $N=0,1,3$ are given in \cite{Schiller2014}. 
(Also the a.c.~polarisabilities have the same sign for \Htwoplus{} and \antiHtwoplus{}). At the spectroscopy frequency $\omega=\omega_{if}$ corresponding to 2.4\,$\mu$m wavelength the polarizabilities were found to differ by about 1\% from their d.c.~values ($\omega=0$), which are given in Table\,\ref{tab:bound-electron g-factor}.
V.I.\,Korobov extended these computations to include a few more levels and also obtained the vector polarizability. It was found to be of the order $1\times10^{-3}$ atomic units, and can therefore safely be neglected.

In order to estimate the shift according to eq.\,(\ref{eq:H_ac-Stark}) we consider the situation where the Rabi frequency of the excitation is chosen to be $\Omega_{if}\simeq0.2\,\mathrm{rad/s}$. Such a value is compatible with a spectroscopic  linewidth of 0.1\,Hz, a value assumed also further below. 
We find the corresponding intensity of the spectroscopy wave from eq.\,(\ref{eq:Rabi frequency 45 degree incidence}) below.
For the $(0,2,M_s,M_F=0)\rightarrow(2,2,M_s'=M_s,M_F'=0)$ transition,
the corresponding intensity is 
0.1\,W/m$^2$.
The light shift is then of order 
$10^{-6}\,$Hz,
negligible. We may also consider the second overtone, $(0,2)\rightarrow(3,2)$ at $1.62\,\mu$m.
The intensity is then 
2.5\,W/m$^2$,
and the light shift is 
$-1\times10^{-4}\,$Hz,
still negligible.

A light shift could also occur from the 313\,nm beryllium cooling laser wave. However, that wave can be turned off during interrogation of the molecular ion.

\subsection{Electric quadrupole shift (EQS)}

The molecule \Htwoplus{}/\antiHtwoplus{} has an electric quadrupole moment due to its nonspherical charge distribution. The nonzero moment leads to an electric quadrupole interaction with electric field gradients. Since the moment depends on the rovibrational level,  transition frequencies are shifted \cite{Bakalov2014}.

The EQS depends on the second derivatives of the electric potential.
In a \PMT{}, contributions to the potential are of two origins: from the electrostatic potential eq.\,(\ref{eq:electric field in a PMT}) and from the motionally induced electric field $\mathbf{E}_{r,\mathrm{motion}}$. The second-order derivatives from the first potential depend on the very stable applied electric voltage and not on the motional state of the ion. 
This is approximately also the case for the motionally induced field (see approximation eq.\,(\ref{eq:E_r,motion})), although the derivatives  are $10^3$ times larger in this case.

Ref.~\cite{Bakalov2014} treated the case of an RF~trap and operation in near-zero magnetic field. In its Sec.\,4.4 it was discussed that if $z$ is the quantization axis, the transverse components of the field gradient ($V_{xx}$, $V_{yy}$) contribute in second-order perturbation only. This remains the case also in strong magnetic field and we shall therefore neglect them.

In that approximation we may introduce the effective Hamiltonian
\begin{equation}
        H_{\rm EQ}(v,N)=(3/2)^{3/2}E_{14}(v,N)\,V_{zz}(N_z^2-{\textstyle \frac{1}{3}}\bold{N}^2)\,,
\end{equation}
where $V_{zz}$ denotes the relevant second derivative of the electric potential at the location of the ion. 
The quadrupole coefficients $E_{14}$ for \Htwoplus\, have been computed in \cite{Bakalov2014}. For example, 
\begin{eqnarray}
        E_{14}(v=0,N=2)&=&0.434\times10^{-4}\,\mathrm{MHz\,m^2/GV}\ ,\nonumber\\
E_{14}(v'=2,N'=2)&=&0.562\times10^{-4}\,\mathrm{MHz\,m^2/GV}\ .\nonumber
\end{eqnarray}
 A typical magnitude of the axial gradient $V_{zz}=m\,\omega_z^2/q$ is  $8\times10^{-4}\,$GV/m$^2$, for $\omega_z\simeq2\pi\times1\,\mathrm{MHz}$. A transition $(v=0,N=2,M_N=0)\rightarrow(v'=2,N'=2,M_N'=0)$ then has a shift of $\simeq-0.04\,$Hz [$-3\times10^{-16}$]. Since the axial frequency can be precisely measured, the uncertainty of the shift is negligible, $u_\mathrm{sys}(\Delta f_{\rm EQ})\simeq0$
\footnote{Myers \cite{Myers2018} states a value $1\,$Hz/T$^2$ for the EQS shift of a $(v=0,N=2,M_N=2)\rightarrow(v'=1,N'=2,M_N'=2)$ transition. Apparently, he used the radial gradient ($\simeq  0.05\,\mathrm{GV}/\mathrm{m}^2\times (B/\mathrm{T})^2$) in the evaluation. In our analysis, the EQS has essentially no dependence on the radial gradient. In any case, the uncertainty of the EQS is negligible.}. 

For \antiHtwoplus{}, the sign of $E_{14}$ is opposite. Under the transformation C the second derivative $V_{zz}$ also changes sign, so that $H_\mathrm{EQ}$ is even under C.

\section{Electric quadrupole transitions}
\label{sec:Electric quadrupole transitions}

\subsection{Derivation of the Rabi frequencies}

The spectrum of rovibrational transitions in \Htwoplus{} is dominated by electric quadrupole (E2) transitions \cite{Bates1953a,Korobov2018a}; the electric-dipole (E1) transitions -- also referred to as ``forbidden transitions'' -- are strongly suppressed \cite{Korobov2023}, and the intensity of magnetic-dipole (M1) transitions is nearly five orders of magnitude lower \cite{Aznabayev2023}.
In this subsection we evaluate the Rabi frequencies of E2 transitions to serve as a guide for experimentalists to provide suitable spectroscopy sources and also to permit evaluation of the light shift (see Sec.\,\ref{sec:A.C. Stark shift (light shift)}).

We take the basis set for perturbative calculations of the spectrum of  para-\Htwoplus{} 
in strong magnetic field in the form $|NM_N\rangle |s_eM_s\rangle$, where, depending on the context, $|vNM_N\rangle$ denotes either the space-variable-dependent rovibrational wave function of \Htwoplus{} or -- when used with the effective Hamiltonian eq.\,(\ref{eq:H_tot1}) -- the constant $(2N+1)$-dimensional eigenvector of $\mathbf{N}^2$ and $N_z$ (the index $v$ being dropped). We label the eigenstates 
$|(NM_NM_s)M_F\rangle$ 
of {the full Hamiltonian} eq.\,(\ref{eq:H_tot1}) 
with the exact quantum number $M_F=M_s+M_N$ and the approximate 
$N$,  $M_N$, and $M_s$. 
{For given $N$, the set of eigenstates $\{|(NM_NM_s)M_F\rangle\}$ encompasses those with values $M_N=-N,\ldots,N$, $M_s=-1/2,+1/2.$}
As long as the transitions amplitudes are evaluated in the Born-Oppenheimer approximation, the mixing of various $N$ states can be neglected. We have
\begin{align*}
&|(NM_NM_s)M_F\rangle=\\
&\quad \sum\limits_{M'_s=-1/2}^{1/2} X_{M'_s}^{(NM_NM_s)M_F} 
|N,M_F\!-\!M'_s\rangle |s_eM'_s\rangle,\\ 
&\quad M_N+M_s=M_F.
\end{align*}
The mixing amplitudes, normalized by 
$\sum_{M'_s}(X_{M'_s}^{(NM_NM_s)M_F})^2=1$, are obtained by diagonalizing the matrix of the total effective Hamiltonian, 
{but in practice only the sum of eqs.\,(\ref{eq:H_hfs}), (\ref{eq:H_Z}), (\ref{eq:anisotropic g_e energy}), and (\ref{eq:H_Z-rot})} 
{is relevant.}

In strong magnetic field the hyperfine splitting is much smaller than the Zeeman shift, and to a good approximation $X_{M'_s}^{(NM_NM_s)M_F}=\delta_{M'_sM_s}$, 
so that the basis states are a good approximation to the exact eigenstates.

Following \cite{Korobov2018a}, we put the amplitude of the laser-stimulated 
E2-transition  in para-\Htwoplus{},
$\langle (v'N'M'_{N}M'_{s})M'_F|\leftarrow|(vNM_{N}M_{s})M_F\rangle$, in the form 
\begin{align*}
{\cal A}_{i\to f}&=\langle (v'N'M'_{N}M'_{s})M'_F|H^{\rm (E2)}|(vNM_{N}M_{s})M_F\rangle\\
&=i\frac{\omega_{if}|\mathbf{E}_0|}{2c}
\sum_{q=-2}^2
\left(\widehat{T}^{(2)q}e^{-i\omega t}+\widehat{T}^{(2)q*}e^{i\omega t}\right)\\
&\times
\sum_{\bar{M}',\bar{M}''}
X^{(v'N'M'_{N}M'_{s})M'_F}_{\bar{M}'}
X^{(vNM_{N}M_{s})M_F}_{\bar{M}''} \\
&\times
\langle s_e\bar{M}'|\langle v'N',M'_F\!-\!\bar{M}'| Q^{(2)}_q|vN,M_F\!-\!\bar{M}''\rangle
|s_e\bar{M}''\rangle,
\end{align*}
where $H^{(\rm E2)}$ is the operator of the quadrupole interaction of \Htwoplus{} with   oscillating electric field of frequency $\omega$ and (complex) amplitude $\mathbf{E}_0$ 
and wave vector 
$\mathbf{k}$, $\omega_{if}$ is the resonance transition frequency, $Q^{(2)}_q,q=-2,...,2$ are the cyclic components of the tensor of the electric quadrupole moment of \Htwoplus{}, and 
$\widehat{T}^{(2)}$ is a rank-2 traceless tensor, composed from the unit vectors 
$\hat{\boldsymbol{\epsilon}}$ and $\hat{\mathbf{k}}$ along $\mathbf{E}_0$ and $\mathbf{k}$, respectively: 
$\widehat{T}^{(2)}_{ij}=(\hat{k}_i\hat{\epsilon}_j+\hat{k}_j\hat{\epsilon}_i)/2$.

In strong magnetic field electron-spin-flip transitions induced by a laser are strongly suppressed, because the spin-rotation coupling is much smaller than the electron-Zeeman interaction. Therefore, in what follows we consider only electron-spin-conserving transitions and neglect the non-diagonal elements of $X_{ M'_s}^{(NM_NM_s)M_F}$, which leads to
\begin{align*}
{\cal A}_{i\to f}=&
i\frac{\omega_{if}|\mathbf{E}_0|}{2c}
\sum_{q=-2}^2
\left(\widehat{T}^{(2)q}e^{-i\omega t}+\widehat{T}^{(2)q*}e^{i\omega t}\right)\\
&\times
\langle v'N'M'_{N}| Q^{(2)}_q|vNM_{N}\rangle.
\end{align*}
Thus, for the Rabi frequency $\Omega_{if}$ of the E2 transition 
$\langle (v'N'M'_{N}M'_{s})M'_F|\leftarrow|(vNM_{N}M_{s})M_F\rangle$ we obtain 
\begin{align}
\Omega_{if}&=
\frac{\omega_{if}}{2\hbar c}\frac{1}{\sqrt{2N'+1}}
\left|\langle v'N'||Q^{(2)}||vN\rangle\right|\nonumber\\
&\times\left| C_{NM_{N},2q}^{N'M'_{N}}\right|\,
\left|\widehat{T}^{(2)q}\right|\,|\mathbf{E}_0|, \quad q=M'_N-M_N.
\label{eq:rabi_if}
\end{align}
The dependence on the quantum numbers $M_{N},M'_{N}$ is described by the Clebsch-Gordan coefficients $C_{NM_{N},2q}^{N'M'_{N}}$. 
The dependence on polarization and irradiation geometry is described by the factor $\widehat{T}^{(2)q}$.
The expression in the first line of eq.~(\ref{eq:rabi_if})
(denoted by ${\cal F}_{if}$, see eq.~(\ref{eq:ref_to_tab})) is common to all spin components of a given rovibrational transition. The values of this factor for 
$v=0,N\le 2$ and $v'\le6,N'\le2$
are given in Table\,\ref{tab:redq}. The reduced matrix elements of $Q^{(2)}$ were evaluated in the Born-Oppenheimer approximation. To assess their accuracy, we juxtapose them with the more accurate values, calculated in \cite{Korobov2018a} in the variational approach for some of the transitions of interest.

The electric field amplitude may be expressed in terms of the laser 
intensity $I_{\rm las}$. 
In case the latter is given in units W/m$^{2}$, the value of $|\mathbf{E}_0|$ in units 
V/m is related to $I_{\rm las}$ by $|\mathbf{E}_0|=27.45\sqrt{I_{\rm las}}$. 
To help avoid mismatch of units when using the table, we also put eq.~(\ref{eq:rabi_if}) in the form 
\begin{align}
&\Omega_{if}=
27.45\,{\cal F}_{if} \, 
\sqrt{I_{\rm las}}\,\left| C_{NM_{N},2q}^{N'M'_{N}}\right|\,
\left|\widehat{T}^{(2)q}\right|, \ q=M'_{N}\!-\!M_{N}\ ,\nonumber\\
&{\cal F}_{if}=\frac{\omega_{if}}{2\hbar c}\frac{1}{\sqrt{2N'+1}}
\left|\langle v'N'||Q^{(2)}||vN\rangle\right|\ .
\label{eq:ref_to_tab}
\end{align}
The values of ${\cal F}_{if}$ are given in column 3 of the table
and the Rabi frequency is obtained in units rad/s. 

\subsection{Polarization dependence}

The dimensionless factor $\left|\widehat{T}^{(2)q}\right|$ describes the geometry and laser light polarization effect, which vary with the difference $q=M'_N-M_N$. To evaluate 
$\left|\widehat{T}^{(2)q}\right|$ we
introduce the laboratory reference frames $K$ with $z$ axis along
the external magnetic field $\mathbf{B}$, and $K'$ with
$z$ axis along $\hat{\mathbf{k}}$, and take the Cartesian coordinates $(\epsilon'_x,
\epsilon'_y,\epsilon'_z)$ of $\hat{\boldsymbol{\epsilon}}=\mathbf{E}_0/|\mathbf{E}_0|$ in $K'$ to be 
$(\cos\theta,\sin\theta\,e^{i\varphi},0)$.
Linear polarization of the
incident light is described by $\varphi=0$, circular polarization by
$\varphi=\pm\pi/2,\theta=\pi/4$, and all other combinations correspond
to a general elliptic polarization. 

Let $(\alpha,\xi,\gamma)$ be the Euler
angles of the rotation that transforms $K$ into $K'$, and denote
by $M(\alpha,\xi,\gamma)$ the matrix relating the Cartesian coordinates
$(a_x,a_y,a_z)$ and $(a'_x,a'_y,a'_z)$ of an arbitrary vector $\mathbf{a}$ in $K$ and $K'$,
respectively: $a_i = \sum_j M_{ij}(\alpha,\xi,\gamma)\, a'_j$. (To avoid mismatch of
$M$ with $M^{-1}$, note that, e.g., $M_{xz}=-\sin\xi\,\cos\gamma$.) 
The absolute values of the components of $\widehat{T}^{(2)q}$ in the laboratory
frame $K$, appearing in eq.\,(\ref{eq:rabi_if}), are expressed in closed form
in terms of the angles $\xi,\gamma,\theta$, and $\varphi$ (the dependence
on $\alpha$ is canceled),
where $\xi$ is the angle between $\mathbf{B}$ and 
$\hat{\mathbf{k}}$.
The angles $\gamma,\theta$, and $\varphi$ are not independent, but the relation between them can be put in a simple form in a few particular cases only. 
The explicit expressions of $\left|\widehat{T}^{(2)q}\right|$ in the most general case read:
\begin{eqnarray*}
\left|\widehat{T}^{(2)0}\right|^2&=&\frac{1}{8} \sin^2\! 2\xi\,
\big(\cos^2(\gamma-\theta)+\cos^2(\gamma+\theta)\nonumber\\
&-&\sin 2\gamma\,\sin 2\theta\,\cos\varphi\big)\nonumber\\
\left|\widehat{T}^{(2)\pm1}\right|^2&=&
\frac{1}{12}\cos^2\!2\xi\,(\cos^2(\gamma+\theta)+\cos^2(\gamma-\theta))\\
&+&\frac{1}{12}\cos^2\!\xi\,(\sin^2(\gamma+\theta)+\sin^2(\gamma-\theta))\\
&+&\frac{1}{12}(1+2\cos2\xi)\sin^2\!\xi\,\sin2\gamma\,\sin2\theta\,\cos\varphi\\
&\mp&\frac{1}{6}\cos\xi\,\cos2\xi\,\sin2\theta\,\sin\varphi\\
%
\left|\widehat{T}^{(2)\pm2}\right|^2&=&
\frac{1}{6}\sin^2\!\xi
-\frac{1}{12}\sin^4\!\xi(\cos^2(\gamma\!+\!\theta)\!+\!\cos^2(\gamma\!-\!\theta))\\
&+&\frac{1}{12}\sin^4\!\xi\,\sin2\gamma\,\sin2\theta\cos\varphi\\
&\mp&\frac{1}{12}\sin\xi\,\sin2\xi\sin2\theta\sin\varphi
\end{eqnarray*}

\subsection{Application to a particular beam geometry}

In the concrete \PMT{}s discussed in Sec.\,\ref{sec:Penning traps}, the opto-mechanical design permits both an axial path for the spectroscopy wave  (for $\Delta M_F=\pm1$ transitions), as well as a path at 45~degree to the magnetic field. The latter is supposed to permit $\Delta M_F=0$ transitions. Since $\Delta M_s=0$ for the stronger transitions, one can thus address both $\Delta M_N=\pm1$ and $\Delta M_N=0$ transitions.

For the 45-degree case $\mathbf{k}$ is oriented at 45~degree with respect to the  magnetic field $\mathbf{B}$. The linearly polarized  electric field $\mathbf{E}_0$ lies in the plane defined by $\mathbf{k}$ and $\mathbf{B}$. 
We set $\varphi=0$, $\theta=\gamma=0$, $\xi=\pi/4$, and obtain
\begin{align*}
&\left|\widehat{T}^{(2)0}\right|^2=\frac{1}{4},\ 
\left|\widehat{T}^{(2)\pm1}\right|^2=0,\ 
\left|\widehat{T}^{(2)\pm2}\right|^2=\frac{1}{24}.
\end{align*}
For the transitions $(v=0,N=2)\to(v'=2,N'=2)$ with $q=0$, we obtain for the Rabi frequency 
\begin{equation}
    \Omega_{if}\approx
    \begin{cases}
0.589 \sqrt{I_{\rm las}}\, \mathrm{rad/s}\,, &M'_{N}=M_{N}=0,\pm2\\
0.295 \sqrt{I_{\rm las}}\, \mathrm{rad/s}\,, &M'_{N}=M_{N}=\pm1\,.
    \end{cases}
    \label{eq:Rabi frequency 45 degree incidence}
\end{equation}
\begin{table}[h!]
\caption{ \textbf{Numerical values pertinent to E2 transitions of \Htwoplus{}.} Listed are the reduced electric quadrupole transition matrix elements (in units $ea^2_0$) for selected pairs of rovibrational states, calculated in the Born-Oppenheimer approximation (present work), and in the variational approach (Ref.~\cite{Korobov2018a}), and of the dimensional factor ${\cal F}_{if}$ (see eq.~\ref{eq:ref_to_tab}), adjusted to values of the laser power flux $I_{\rm las}$ in units W/m$^{2}$. The explicit units of ${\cal F}$ are omitted for clarity. 
}
\label{tab:redq}
\begin{center}
\begin{tabular}{c@{\hspace{6mm}}l@{\hspace{4mm}}l@{\hspace{4mm}}l@{\hspace{4mm}}}
\\
\hline
\vrule width0pt height20pt depth3pt Transition & 
\multicolumn{2}{c}
{$\frac{\left|\langle v'N'||Q^{(2)}||vN\rangle\right|}{\sqrt{2N'+1}}$ 
($e\,a_0^2$)}
& \multicolumn{1}{c}{${\cal F}_{if}$}\\
$\langle vN|\rightarrow|v'N'\rangle$ & 
Present work
&
Ref.~\cite{Korobov2018a}
& 
 {see eq.~(\ref{eq:ref_to_tab})} \\
\hline
$(0,0) \to (0,2)$ &    0.7350	& 0.735648 &	0.1713	\\
$(0,0) \to (1,2)$ &    0.1402	& 0.140356 &	0.4417	\\
$(0,0) \to (2,2)$ &    0.1293[$-$1]	& &	0.7626[$-$1]	\\
$(0,0) \to (3,2)$ &    0.2081[$-$2]	& &	0.1765[$-$1]	\\
$(0,0) \to (4,2)$ &    0.4690[$-$3]	& 0.468812[$-$3] &	0.5115[$-$2]	\\
$(0,0) \to (5,2)$ &    0.1335[$-$3]	& &	0.1759[$-$2]	\\
$(0,0) \to (6,2)$ &    0.4494[$-$4]	& 0.449061[$-$4] &	0.6871[$-$3]	\\
$(0,2) \to (0,0)$ &    1.644	& 1.644998 &	0.3831	\\
$(0,2) \to (1,0)$ &    0.3733	& &	1.0067	\\
$(0,2) \to (2,0)$ &    0.2312[$-$1]	& &	0.1262	\\
$(0,2) \to (3,0)$ &    0.2662[$-$2]	& &	0.2143[$-$1]	\\
$(0,2) \to (4,0)$ &    0.3980[$-$3]	& &	0.4174[$-$2]	\\
$(0,2) \to (5,0)$ &    0.5781[$-$4]	& &	0.7380[$-$3]	\\
$(0,2) \to (6,0)$ &    0.1730[$-$6]	& &	0.2576[$-$5]	\\
$(0,2) \to (1,2)$ &    0.1840	& 0.183195 &	0.5368	\\
$(0,2) \to (2,2)$ &    0.1417[$-$1]	& 0.141706[$-$1] &	0.8028[$-$1]	\\
$(0,2) \to (3,2)$ &    0.1990[$-$2]	& &	0.1641[$-$1] 	\\
$(0,2) \to (4,2)$ &    0.3910[$-$3]	& 0.390765[$-$3] &	0.4173[$-$2]	\\
$(0,2) \to (5,2)$ &    0.9518[$-$4]	& &	0.1232[$-$2]	\\
$(0,2) \to (6,2)$ &    0.2630[$-$4]	& &	0.3960[$-$4]
\end{tabular}
\end{center}
\end{table}

\section{Quadratic Doppler shift}
\label{sec:Quadratic Doppler shift}

Myers has pointed out that the quadratic (relativistic) Doppler shift (QDS) - arising from  the finite temperature of the stored ion - could be the dominant shift in \Htwoplus{} spectroscopy in a \PMT{}, even if the temperature is sub-Kelvin. In fact,
the QDS due to thermal motion in \PMT{}s has been of interest in the early days of frequency standards development, when the ions were not cooled to a sufficiently low temperature \cite{Bollinger1985}. The QDS  remained an issue for an early microwave clock, albeit for an ensemble of clock ions in a \PMT{}, embedded in a Coulomb crystal \cite{Bollinger1991}. Later, this issue faded, as the interest in clocks moved to RF traps. 

A related effect is the relativistic shift of the modified cyclotron frequency $\Delta\nu_+/\nu_+=-E_+/(m\,c^2)$, where $E_+$ is the mode energy in the cyclotron oscillator. This is one of the dominant systematic shifts in Penning trap precision experiments \cite{gabrielse1999precision, borchert202216}. Its uncertainty will enter the error budget of magnetic field determination at the 10$\,$ppt level. 

The QDS is independent of the angle between the radiation wave vector and the particle velocity.
This precludes a determination of the shift by performing spectroscopy with different laser wave propagation directions.
Thus, it is necessary to determine the mean motional energy by other means. A CPTI measurement campaign will need to budget an additional measurement time for motional energy determination. Conservatively, the necessary time will be similar to the one required for determining the line center precisely.

The statistical uncertainties of the determination of the perturbed or unperturbed line center arising from the spectroscopy itself are discussed in Appendix \ref{app:Determination of line center in presence of QDS}.

We consider here two trap types and concomitant ion temperature regimes.

\subsection{The QDS in a CSGE-\PMT{}}
\label{sec:Estimate of the QDS shift}

The first type is a CSGE-equipped \PMT{} with a 4$\,$K cryogenic environment for the ion. 
No sympathetic cooling is implemented. 
The ion is in the classical regime.

Whereas the instantaneous QDS is
\begin{equation}
 \delta f_\mathrm{QDS}(t)/f_0=-\frac{\bold{v}(t)^2}{2\,c^2}\ ,
    \label{eq:instantaneous QDS - fundamental equation}
\end{equation}
the mean shift of a transition line will be given by the time average of the QDS,  
\begin{equation}
 \Delta f_\mathrm{QDS}/f_0=-\frac{\langle \bold{v}(t)^2\rangle}{2\,c^2}\ .
    \label{eq:mean QDS - fundamental equation}
    \end{equation}
For a classical harmonic oscillator, this is $-{E_\mathrm{kin}}/{m\,c^2}$. However, it is known that 
in experiments in which resistive cooling is applied \cite{wineland1975principles}, for each mode $i$ (axial, magnetron, cyclotron), the particle energy $E_{\mathrm{kin},i}$ follows a Boltzmann distribution with temperature $T_i$, 
with the statistical average $\langle E_{\mathrm{kin},i}\rangle=\frac{1}{2}k_\mathrm{B} T_i$. Therefore,
\begin{equation}
    \Delta f_\mathrm{QDS}/f_0=-\frac{k_\mathrm{B} \sum_i{T_i}}{2\,m\,c^2}\ .
    \label{eq:mean QDS}
\end{equation}
 By determining the temperatures $T_{i}$ and their uncertainties by appropriate techniques (which may or may not involve the spectroscopy of the vibrational transition of interest), the  mean QDS and its uncertainty can be computed from this expression.
Since the kinetic energies - in particular the axial one in case of the CSGE-\PMT{} - exhibit Boltzmann distributions, a distribution of QDSs will occur, with impact on the spectral line shape (see App.\,\ref{app:Determination of line center in presence of QDS}).
\subsubsection{Axial mode}
In the classical regime, the ion oscillates along the trap's magnetic field with a classical axial motion whose amplitude changes stochastically on a time scale $\tau_{\rm th}$ due to coupling to the thermal bath of the axial detector. This coupling \cite{Nagahama2016HighlyAntiprotons} is required to implement the CSGE.

Consequently, the QDS  also varies stochastically. (See App.\,\ref{app:Stark shift and QDS correlation} for a connection with the Stark shift.) 
More precisely, the axial mode is  resistively cooled and thus in equilibrium with the thermal bath of the detection resistor $R_p$. The axial energy $E_{\mathrm{kin},z}$ fluctuates with the particle-to-detector correlation time constant  $\tau_\text{th}=(m D^2)/(R_p\, q^2)$, following Boltzmann statistics. Here $D$ is a trap-specific length. This axial energy fluctuation is convolved into the measured line-profiles \cite{brown1986geonium}. 

 We may assume a baseline temperature $T_z=4.2\,$K for the axial degree of freedom, and a typical correlation time constant is $\tau_\text{th}=30$\,ms.  
 
\subsubsection{The radial modes}
\label{sec:The cyclotron mode}
Unlike the axial mode, the cyclotron ($E_+$) and magnetron ($E_-$) mode energies can be cooled via so-called "sub-thermal-cooling" \cite{latacz2024orders}. For practical reasons, for cyclotron-mode-cooling a cyclotron detector is used, while for the magnetron mode sideband-cooling is applied \cite{Cornell1990}.  \\
In this stochastic cooling method the respective mode is coupled for a certain interaction time $t_\text{int}$  to a resonant thermal resistive reservoir  at temperature $T_R$. 
Afterwards, the radial mode energy is analyzed using the CSGE.  Repeating this process multiple times gives a Boltzmann-type reference histogram with an expectation value $T_R$, and a lower temperature that defines absolute zero radial energies,  with a resolution-limit that depends on the statistics of the histogram sampling, magnitude of the magnetic bottle $B_2$ in the CSGE trap, and voltage stability of the trap power supply \cite{nagahama2017sixfold}. 
For example, the analysis trap of the BASE experiment exhibits a particularly large $B_2=268\,$kT/m$^2$, while $B=1.21\,$T. With this trap and its {high-Q} axial detector the axial frequency of a single antiproton was resolved to within 0.25$\,$Hz within 6$\,$s measurement time. 
This translates to a radial energy resolution of $\sigma(E_+)/k_\mathrm{B}=3.5\,$mK for the cyclotron temperature \cite{Smorra2015}. \\
In dedicated small-diameter cooling traps equipped with high-Q cyclotron detection circuits at $T_R=4.8\,$K, thermalization constants of  $\tau_\text{th}\approx5\,$s have been achieved \cite{latacz2024orders}. Using a double-trap cooling scheme a particle at an absolute cyclotron energy of $E_+/k_\text{B}=T_+<200\,$mK was prepared in a time of about 8$\,$min. \\
With further optimized detectors, magnetic-bottle single traps, and the application of adaptive cooling schemes, we expect that the cooling performance of these devices could be improved by another factor of 10.

The magnetron mode can be cooled in exactly the same way, however, using axial-to-magnetron sideband drives that couple the mode to the axial detector, where a temperature $T_-=(\nu_-/\nu_z) T_z$ is achieved \cite{Cornell1990}.  Using another technique, feedback cooling (cooling the detection resonator with its own phase-shifted signal \cite{DUrso2003}), the preparation of a particle with $E_-/k_\text{B}=T_-<5\,$mK typically takes two minutes \cite{smorra2017observation}. 

\subsubsection{Summary}

We may safely assume that magnetron and cyclotron temperatures are negligible and small, respectively, compared to the axial temperature.
With only one relevant mode, the mean QDS for a 4.2$\,$K trap is 
\begin{equation}
    \Delta f_{\mathrm{QDS}}^{\mathrm{CSGE}}/f_0=-1.0\times10^{-13}\quad\text{at }T_z=4.2\,\mathrm{K}.
\label{eq:Delta fQDS for 4.2 K}
\end{equation}
With feedback-cooled high-Q axial resonators at high signal-to-noise ratio, axial temperatures $T_z$ of order 0.4$\,$K are expected to be possible, reducing the shift by a factor of 10, 
\begin{equation}
    \Delta f_{\mathrm{QDS}}^{\mathrm{CSGE}}/f_0\simeq-1\times10^{-14}\quad\text{at }T_z=0.4\,\mathrm{K}.
\label{eq:Delta fQDS for 0.4 K}
\end{equation}
Reduction factors of 100 - 200 might be reached in the future, if ultra-cryogenic temperature, e.g.~as provided by dilution refrigerators, is combined with feedback cooling. 
Then, a mean QDS in the low $10^{-15}$ range appears possible. However, we shall not consider this option further in this work. \\ 

\subsubsection{Determination of the QDS}
\label{sec:Determination of the QDS in CSGE traps}
While the QDS would be crucially important for the aim of absolute frequency determination, it might be suppressed if the experimental approach is the  comparison of \Htwoplus{} and \antiHtwoplus{} frequencies in the same trap. Then, the uncertainty of the temperature difference of the two particles, $\Delta T$(\Htwoplus{}-\antiHtwoplus{}), provides the figure of merit that will limit the resolution of the CPTI test. 
This information is implicit in the resonator noise spectra recorded in each axial trap frequency $(\omega_z)$ measurement carried out in the precision trap.

The thermal noise power $S$ of the axial detector is $S\propto {4 k_B T_z \mathrm{Re}(Z(\omega))}$, where $T_z$ is the detector's electronic temperature and $\mathrm{Re}(Z(\omega))$ is the real part of its impedance. Thus, $S$ contains information about the detector temperature, which equals the particle temperature. 
If a spectrum is recorded first for \Htwoplus{} and subsequently for \antiHtwoplus{}, assuming detector performance is similar to that described in \cite{Nagahama2016HighlyAntiprotons}, the axial temperature difference between the two measurements can be determined to an uncertainty $\approx60\,$mK$\times\sqrt{T_z}$ \cite{borchert202216}.

A typical measurement campaign would include some $10^3$ ion pair measurements. Then, the mean axial temperature difference $\Delta T_z$(\Htwoplus{}-\antiHtwoplus{}) would be determined with sub-10$\,$mK resolution \cite{borchert202216}.
This resolution leads to an uncertainty of the mean QDS approximately a factor $10^3$ smaller than eq.\,(\ref{eq:Delta fQDS for 4.2 K}). (If the calibration of the noise is not performed, the uncertainty should be understood as statistical, i.e.~an instability.)

In summary, based on our experience in  systematic studies 
of the axial temperature, and given that sub-thermal-cooling of the magnetron and the cyclotron modes provides tracking of the radial temperatures (which will be in the few-mK range), we believe that for same-trap \Htwoplus{}/\antiHtwoplus{} comparisons, an \emph{instability} of the differential QDS at the level 
$u^{(1)}=u_\mathrm{stat,T}(\Delta f^\mathrm{CSGE}_\mathrm{QDS,diff})=2\times10^{-16}$ will become possible after an integration time of a few weeks.

The second approach for tracking the temperatures is to measure the strength of the motional sidebands (caused by the first-order Doppler effect) by spectroscopy of the ion, as proposed by Wineland \textit{et al.}~\cite{Wineland1987} for RF traps and shown for a \PMT{} by Mavadia \textit{et al.}~\cite{Mavadia2014} and more recently by \cite{cornejoOpticalStimulatedRamanSideband2023}. (A cyclotron sideband measurement was suggested above in Sec.\,\ref{sec: d.c. Stark shift} concerning the transverse-motion-induced Stark shift.) We may assume to be able to measure the strength of axial, cyclotron and magnetron sidebands relative to the carrier with 5\% accuracy, implying approximately 5\% fractional reproducibility of the QDS.  In case of a future axial temperature of 0.4\,K as mentioned above, this leads to the estimate
$u_\mathrm{stat,T}(\Delta f^{\mathrm{CSGE}}_\mathrm{QDS})\simeq5\times10^{-16}$, since the axial QDS dominates. 
It is higher than the projected  $u^{(1)}$ above.

\subsection{The QDS in a QLS-\PMT{}}
\label{The QDS in a QLS-PT}

This type of trap will provide a much lower ion temperature thanks to  sympathetic cooling by a laser-cooled atomic ion. We follow a similar consideration as in Sec.~\ref{sec: d.c. Stark shift}, after eq.\,(\ref{eq:E_r,motion as a fct of Etot}). 
Assume that the cyclotron mode is not perfectly cooled to the ground state, leaving a residual excitation $\langle n_+\rangle=1$.  The corresponding QDS is $\Delta f_\mathrm{QDS}^{\mathrm{QLS}}/f_0=-3\hbar\,\omega_+/(4\,m\,c^2)$ $\simeq -5\times10^{-17}$.
\label{sec:Determination of QDS shift via temperature measurement: Case QLS}
Also in a QLS-\PMT{} the motional sideband strengths can be measured. This should allow determining the temperature (especially the deviation from the ground state of cyclotron motion) to 5\% uncertainty, bringing the systematic uncertainty to
$u_\mathrm{sys}(\Delta f^{\mathrm{QLS}}_\mathrm{QDS})\simeq3\times10^{-18}$.
In a recent experiment, the axial temperature has been measured with 6\% uncertainty \cite{Boehn2025}.

\subsection{Summary}
In an advanced CSGE-\PMT{} the QDS caused by axial motion may be dominant. One may consider canceling the effect by implementing a magnetic bottle. This idea is discussed in Appendix\,\ref{sec:QDS cancellation}. However, we conclude that the idea is still uncertain in its viability. 
We are thus left with the total QDS uncertainty $u^{(1)}$, provided that motional sideband characterization or long-term direct temperature measurements are implemented so as to resolve the  {cyclotron} and magnetron mean temperatures at the 1\,mK level. 

In the QLS-\PMT{}, the use of a magnetic bottle might be useful in case that the total QDS is  dominated by the contribution from the radial motion.

\section{Penning traps: Introduction}
\label{sec:Penning traps}

\subsection{Overview}

A wide range of studies has been performed on particles trapped in \PMT{}s. These include ESR on single electrons/positrons and on single hydrogen-like atomic ions as well as NMR on single protons/antiprotons. Earlier work on RF spectroscopy of atomic ion clouds was aimed at developing a frequency standard. Laser spectroscopy of a single atomic ion trapped in a \PMT{} and cooled thermally and by feedback to 1\,K was reported in \cite{Egl2019}.

Laser cooling of a single atomic ion in a classic \PMT{} has been pioneered at Imperial College London. 
Axial, cyclotron and magnetron sidebands were laser-excited and observed via fluorescence \cite{Mavadia2014,Goodwin2016}. The work indicated the usefulness of providing a non-axial laser beam irradiation, which is also important in the present context. 

A new-generation \PMT{} spectrometer is the project BASE-QLEDS, where sympathetic cooling of a single proton/antiproton by a single laser-cooled beryllium ion is pursued \cite{Cornejo2021}. Laser cooling of a Be ion to the ground state of the axial mode has been achieved \cite{cornejoResolvedsidebandCoolingSingle2024}.
Nowadays, ions in \PMT{} arrays are intensively studied as a platform for quantum computing~\cite{Jain2020}.

Molecular ions - even MHI - have been trapped and studied in \PMT{}s in previous studies. Their goal has usually been to measure the masses of the light nuclei contained within \cite{Rau2020,Fink2020,Fink2021}. 
In early work, the ESR of \Htwoplus{} was measured on a cloud of (warm) ions stored in a \PMT{} \cite{Loch1988}.
The most advanced work in the context of the present discussion are experiments at the ALPHATRAP facility, mentioned in Sec.\,\ref{sec:CPTI tests using PTs}. 

Appendix~\ref{app:experimental studies in PMTs with narrow linewidth} presents some details about selected high-resolution RF, microwave and optical spectroscopy work performed in \PMT{}s. 

There are two approaches for non-destructive detection of rovibrational transitions. In the approach based on CSGE the particle necessitates a  sufficiently large magnetic moment - thus preferably an electron-magnetic moment - and whose  value is level-dependent. A tiny dependence suffices. 

\Htwoplus{} and \antiHtwoplus{} do provide such a  magnetic moment: the  magnetic moment of the bound electron or positron. 
\footnote{We take the opportunity to correct a minor mistake in the work of Myers \cite{Myers2018}: the identification of the vibrational quantum number is also possible for $N=0$ states, since $g_e(v,N=0)$ is $v$-dependent, see Table\,\ref{tab:bound-electron g-factor}.}

In contrast, in the approach based on QLS the existence of a magnetic moment is not a necessity. Its presence, however, is not a hindrance towards metrological performance, as the discussion in the sections above has shown.

The above two types of \PMT{} will be described in detail in Sec.\,\ref{sec:CSGE-PT} and \ref{sec:QLS-PT}.

\subsection{Multi-Penning Traps}
The invention of multi-\PMT{} experiments emerged from ideas developed in the group of G.\,Werth at the University of Mainz, for measurements of the magnetic moment of the electron bound in hydrogen-like spinless nuclei, to test bound-state quantum electrodynamics. 
In particular, the double~\PMT{}, where state analysis and precision frequency measurements  are separated to two traps was first described by Häffner \textit{et al.}
\,\cite{Haeffner_PhysRevLett.85.5308}. Out of this work more complex \PMT{}s have been developed, for example the LIONTRAP experiment \cite{Heisse2017}, the ALPHATRAP experiment \cite{Sturm2019} or the $\mu$-TEX experiment \cite{dickopf2024precision}. Also \PMT{}s that include much stronger magnetic bottles were implemented, that reach nuclear-magnetic-moment resolution. Some of these trap designs also include more traps. For example, the BASE 
\PMT{} apparatus at CERN contains a total of four traps \cite{Smorra2015} including reservoir, precision, cooling, and analysis traps. The BASE-QLEDS \PMT{} contains four traps \cite{Cornejo2021}, see below. The mass spectrometer PENTATRAP \cite{schussler2020detection} includes five traps, while the BASE-Mainz experiment has even six trapping regions that are engineered for different purposes \cite{bohman2021sympathetic}. 
\subsection{Production of \Htwoplus{}}
The production of \Htwoplus{} can proceed in two ways. One is axial injection from an ion source.  In the experiment \cite{Koenig2025b} the related molecule HD$^+$ was injected into a cryogenic \PMT{}. This approach is also chosen in the proposed apparatus shown in Fig.\,\ref{fig:CSGETRAP}.

\begin{figure}
    \centering
    \includegraphics[width=0.75\linewidth]{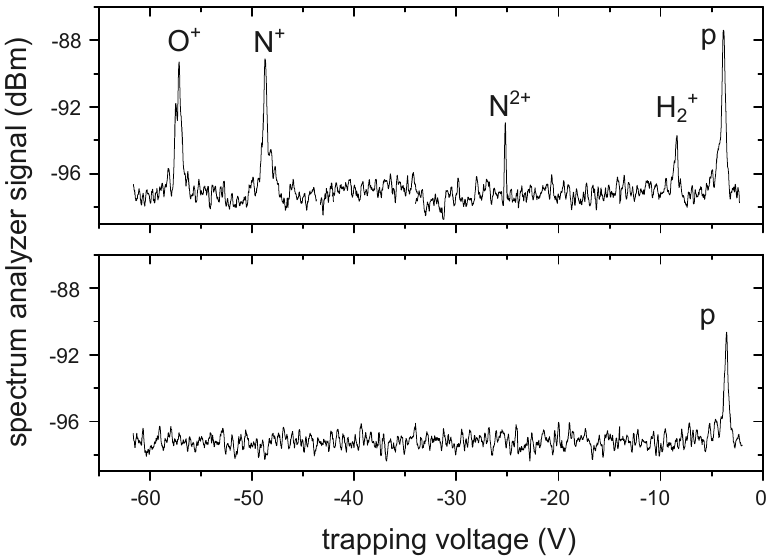}
    \caption{ 
    \textbf{Example of ion preparation in a \PMT{}.} The trap is loaded with particles; afterwards the resonant center of the axial detector spectrum is observed with a certain bandwidth, while a weak axial resonant drive is applied and the trap voltage is being ramped to bring different ion species to resonance with the detector. In between the upper and the lower spectrum resonant axial cleaning drives are applied, to remove particles from the trap. The lower spectrum is recorded after cleaning all particles out, except for a cloud of protons. Selecting the resonant cleaning drive frequencies differently, any particle species can be prepared. }
    \label{fig:mass spectrum of ions}
\end{figure}

The other possibility is in-situ production within the trap system. This has been realized for example in the BASE-CERN and the BASE-Mainz experiments, which are closed systems. Fig.\,\ref{fig:mass spectrum of ions} shows an example of \Htwoplus{} production in such a case. Here an electron beam bombards a cryogenic target made out of black polyethylene. This evaporates hydrogen molecules frozen out on the target surface which are then ionized by the electron current.  Shown on the plot are measurements where a spectrum analyzer is tuned to the axial detector, while a weak axial drive is applied, and the trap voltage is swept, which brings different particle species to resonance with the detector. Applying resonant axial drives, unwanted particles can be removed. This is shown in the lower figure, where all particles have been cleaned out of the trap, except protons. Selecting the drives appropriately, any species that comes out of the target can be  prepared in the trap.  
\vskip .03in
There are practical issues worth mentioning. 
Firstly, after loading into the trap, a given \Htwoplus{} ion may be in the ortho configuration - which is not the preferred one. Even if in para configuration, the ion's rovibrational level may be neither the lower nor the upper spectroscopy level. If the actual level is an excited level $v>2$, it may take an average wait time of several weeks \cite{Fink2021,Karr2021} for spontaneous emission to populate one of the two desired levels. 
In particular cases, the time may even be longer. 
Secondly, during a spectroscopy campaign on the above transition, spontaneous emission from $v=2$ to $v=1$ may occur, interrupting the measurements. 

These features suggest to employ a widely tunable optical parametric oscillator capable of addressing a larger set of vibrational transitions. 
In combination with non-destructive internal state identification it should be possible to transfer the ion (back) into a desired level within a reasonable time.  

\subsection{Injection of \antiHtwoplus{}}

The synthesis of \antiHtwoplus{} will only be possible at the AD/ELENA facility of CERN, using synthesis techniques that will likely rely on antihydrogen trapping, as successfully demonstrated by the ALPHA collaboration \cite{Akbari2025},  as a staring point. After production, the particles can be ejected into the transportable Penning-trap system BASE-STEP \cite{leonhardt2025proton}, where they will be kept under the excellent vacuum conditions available in precision Penning trap experiments \cite{sellner2017improved}. Using this system, the particles can be transported to dedicated receiver experiments as described below. The injection of \antiHtwoplus{} into these experiments will be managed via standard beam optics such as quadrupole triplets and bending structures. The optimization of the beamline before \antiHtwoplus{} injection will be carried out with an offline ion source capable of delivering protons and negatively charged hydrogen ions. To keep the vacuum at levels $<10^{-18}\,$mbar, the beamline will be NEG coated, appropriate absorber material will be inserted into the trap chambers and cryogenic valves will be implemented similar to the ones described in \cite{sturm2019alphatrap} and \cite{Smorra2023}. 

\section{A \PMT{} employing the CSGE}
\label{sec:CSGE-PT}

\begin{figure*}[t]
\centering
\includegraphics[width=0.95\linewidth]{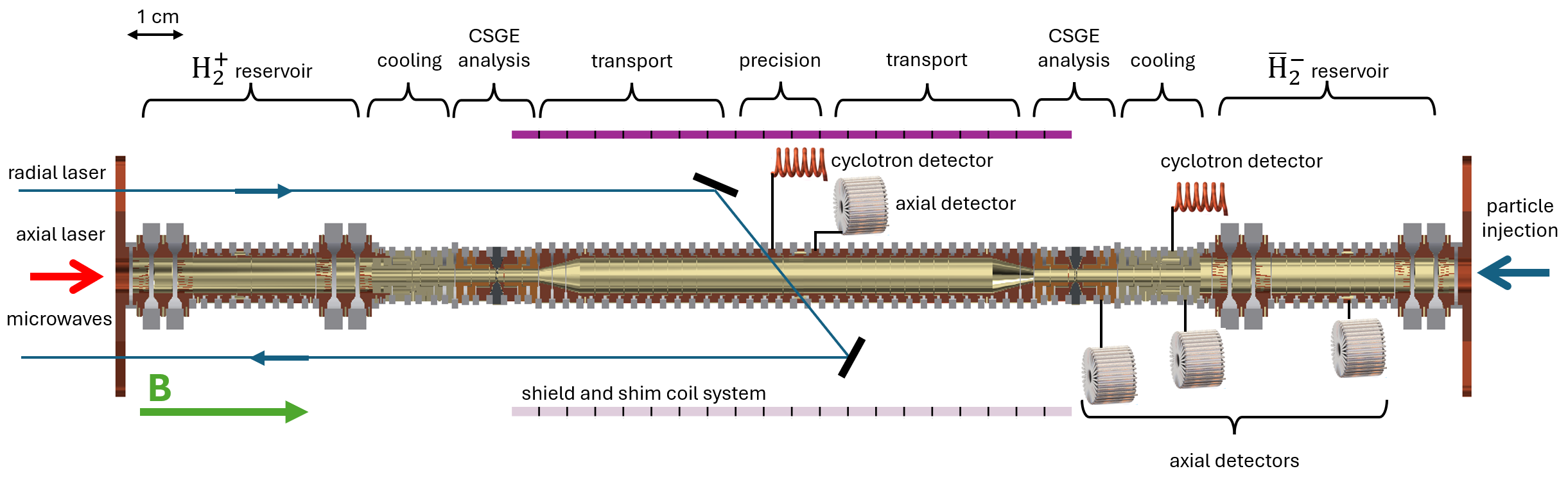}
\caption{\textbf{Proposed multi-Penning-trap stack for precision spectroscopy on \Htwoplus{} and \antiHtwoplus{}}. The stack consists of seven traps. In the center is a highly shielded and homogenized precision trap with radial and axial laser access. Moving outwards, analysis traps for CSGE detection and cooling traps for sub-thermal cooling are located. The outer traps are reservoir traps. All traps are equipped with resonant detection systems for pick-up of axial and cyclotron image currents and resistive cooling (not shown for the left traps in order to keep the figure organized).  
}
\label{fig:CSGETRAP}
\end{figure*}
\subsection{Apparatus}
In an experiment purpose-built to compare the properties of \Htwoplus{} and \antiHtwoplus{} a multi-\PMT{} system may be used similar to the one  shown in Fig$.\,$\ref{fig:CSGETRAP}. This system would be engineered such that in the interest of high data rate synchronization of quantum state analysis of one particle  while doing spectroscopy on the other particle will be implemented. The trap system would consist of two reservoir trap regions \cite{smorra2015reservoir}, left and right to the trap system, one for \Htwoplus{} and the other for \antiHtwoplus{}. In the center of the trap system, a highly homogenized precision measurement trap will be placed, shielded with a multi-layer self-shielding coil system \cite{devlin2019superconducting}, and local persistent magnetic gradient coils to homogenize the trap's magnetic field \cite{Erlewein2024Magneticthesis}. 
Spectroscopy laser wave access in axial and non-axial direction will be implemented, similar to experiments like BASE-Mainz \cite{bohman2021sympathetic} and BASE-QLEDS \cite{cornejoOpticalStimulatedRamanSideband2023}. The laser windows will also allow microwave injection to perform bound-electron and bound-positron spin resonance experiments. On each side, between the reservoir traps and the precision trap two additional traps will be placed. The first is an analysis trap \cite{ulmer2011observation} with a superimposed magnetic bottle for radial angular momentum and spin state detection, the second will be a cooling trap \cite{latacz2024orders} for efficient cyclotron mode cooling.  

For efficient systematic studies and co-operation of \Htwoplus{} and \antiHtwoplus{} experiments the entire trap stack will be operated on bi-polar high-precision power supplies, as available in BASE for proton/antiproton experiments. 
Experimental procedures to be demonstrated  are particle injection through cryogenic valves, as established in ALPHATRAP \cite{Sturm2019},
ARTEMIS at GSI \cite{vogel2019electron}, and recently also with antiprotons in the BASE-STEP setup \cite{smorra2023base}. 

Another important ingredient of the experiment would be to establish seamless reservoir operation, including reservoir maintenance, efficient particle extraction and re-merging. These are meanwhile routine techniques developed by some of the present authors \cite{smorra2015reservoir}. 
Based on those, continuous antiproton trap-experiment operation has been demonstrated for years, at typical particle consumption rates of one particle in a month, in measurement mode even as low as one particle in four months. 

The design of the analysis traps will be based on the layouts available in BASE \cite{ulmer2011observation}, where magnetic bottles of 42$\,$kT/m$^2$ are used for cooling traps and 268$\,$kT/m$^2$ for analysis traps. As described above, for mode temperature analysis higher magnetic bottles provide higher resolution, and with the available 268$\,$kT/m$^2$-layout temperature analysis with $<5\,$mK resolution for characterization times of a few minutes have been achieved. The magnetic bottles of the sub-thermal cooling traps can be adjusted to values optimal for QDS compensation. 

To establish more efficient sub-thermal cooling than demonstrated in \cite{latacz2024orders}, improved cyclotron cooling detectors need to be developed \cite{Ulmer2013AEnergy}. While in sub-thermal cooling the frequency readout, particle shuttling and trap geometry schemes have been fully optimized, feedback-cooled high-Q detectors have headroom to further improve the cooling performance by up to a factor of 10 for the cyclotron mode.

Another interesting approach to more efficient cooling is currently being developed in the ELCOTRAP setup at MPIK, being based on cyclotron cooling of electrons and resonant coupling of the axial mode of ground-state cooled electrons to the cyclotron mode of single trapped ions, a method that could also be considered for future radial mode cooling.
\\
\subsection{Experimental spectroscopy protocol}
The experiment protocol that would be implemented in that case would be similar to protocols that are applied in high-precision multi-trap magnetic moment measurement experiments such as \cite{Haeffner_PhysRevLett.85.5308, mooser2013demonstration, Sturm2014, smorra2017parts, schneider2017double, dickopf2024precision}. 
\begin{figure}
    \centering
        \includegraphics[width=.9\linewidth]{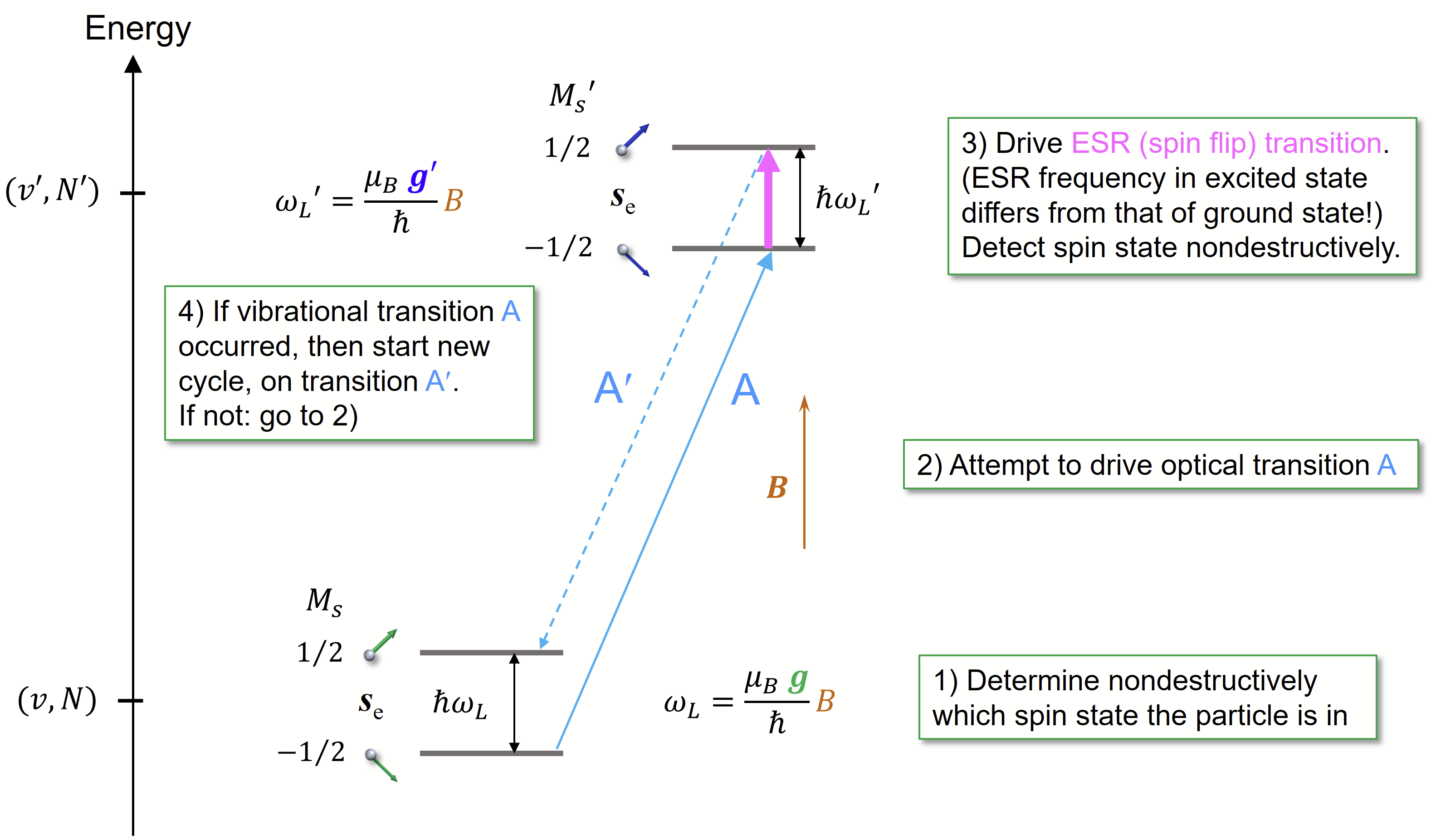}
    \caption{ 
    \textbf{The CSGE procedure for nondestructive detection and spectroscopy of a rovibrational transition.} The concept is from Myers \cite{Myers2018}. For clarity, only the minimum number of states required is shown.
    The figure shows the case where in step 1
    it was found that the molecule is in the lower spin state of $(v,N)$. Then step 2 (vibrational excitation A) is taken, followed by step 3 (detection). If no spin flip in $(v',N')$ is detected (by CSGE), step 2 is repeated. This continues until a spin flip is detected. Then, the ``downward" excitation $\mathrm{A}'$, step 4, is attempted. The following steps are analogous to those connected with the upward excitation. Note that $f_{\mathrm{A}}\ne f_{\mathrm{A'}}$;  example values are indicated in Fig.\,\ref{fig:Energy levels}. $g$, $g'$ denote the bound-electron g factors in the lower and upper rovibrational level, respectively. The fact that they differ substantially is key to the procedure.
}
    \label{fig:Principle of CSGE - detected vibrational transition}
\end{figure}
As illustrated in Fig.$\,$\ref{fig:Principle of CSGE - detected vibrational transition} the basic idea of the experimental technique is to read out the electron-spin state of the trapped \antiHtwoplus{}/\Htwoplus{}, and to upward and downward excite the trapped particle's internal state as a function of the laser frequency. The fact that the electron g-factor depends on the vibrational state, allows to probe whether the laser wave has induced a transition, be it upward or downward. 
For example, starting in the electron spin-down state of the lower rovibrational level,  the transition $(v, N, M_s=-1/2) \rightarrow (v', N', M_s'=-1/2)$ (arrow \textcolor{lightskyblue}{\textbf{A}} in the figure) is addressed by the appropriately tuned laser.  Irradiating subsequently  a microwave tuned exactly to the  ESR transition $(v', N', M_s'=-1/2) \rightarrow (v', N', M_s''=+1/2)$ and having a field strength corresponding to a $\pi$ pulse, will lead to a spin flip  if the antecedent laser excitation attempt 
\textcolor{lightskyblue}{\textbf{A}} was successful. Whether a spin-flip occurred or not is read out by electron-spin projection analysis in the CSGE trap. 
If no spin flip is detected, the procedure is repeated, possibly with a different laser detuning. If the spin flip is detected, the procedure is analogously applied to transition \textcolor{teal}{$\mathrm{A}'$} with spin flip in the lower level $(v,N)$. 

This concept is a standard technique applied in many different trap experiments to access different fundamental constants. To the best of our knowledge, it was first proposed by W.\,Quint in the context of the ARTEMIS experiment \cite{vogel2019electron}, and later further developed by Mooser \cite{dickopf2024precision} for schemes to measure the magnetic moment of the $^3$He$^+$ ion. Myers applied the concept in his proposal of  \antiHtwoplus{} spectroscopy \cite{Myers2018}. Its application to read out laser-induced transitions was first applied in Egl \textit{et al.} \cite{Egl2019}. 
In that case, the upper spectroscopy level spontaneously decayed to a third level with concomitant electron-spin flip. The accompanying change in the ion's axial frequency in the bottle trap (CSGE) indicated that the laser excitation had been successful. In the present case of \Htwoplus{}/\antiHtwoplus{} there is no spontaneous decay and it is also not desirable to drive optical transitions that flip the electron spin. \\
In a planned experiment scheme, a single particle, for example \antiHtwoplus{}, is first extracted from the reservoir and its radial modes are cooled using the CT/AT-cycle, similar to the protocols described in \cite{latacz2024orders}. Depending on CT performance and defined cooling temperature threshold, this would require typically about 1 to 10 minutes. 
Afterwards the particle's radial state would be analyzed by measuring the axial frequency in this trap. Note that these measurements track radial temperature information, and can be used to deconvolve the measured optical resonance line.  

Synchronously, another particle (in that case \Htwoplus{}) would continuously sample the magnetic field in the \precisiontrap{}, using standard cyclotron frequency measurement methods. 
After electron spin-state identification of \antiHtwoplus{} in its dedicated analysis trap, the particle would be moved to the \precisiontrap{}, where the laser/microwave interrogation would take place and the cyclotron frequency of the particle will be measured. 
Conversely, while preparing a cold \Htwoplus{} in the respective CT/AT cycle and analyzing the electron-spin state of that particle, the \antiHtwoplus{} would be used as a magnetic-field probe to continue monitoring the field of the \precisiontrap{}.

Concerning the systematic studies of such comparison measurements we note that the continuous magnetic field measurements in the central measurement trap will allow interleaving of magnetic field data, which would ensure magnetic field reconstruction on the 10$\,$pT level. 
 
We emphasize that the three ion temperatures (axial, cyclotron, magnetron) can be tracked. In practice it was found that the radial temperatures are long-term-stable on time scales of days \cite{Smorra2017}. The axial temperature is implicitly tracked by measuring its noise spectrum each 30\,s. Its peak value reflects the particle temperature (see Sec.\,\ref{sec:Determination of the QDS in CSGE traps}). While obtaining an absolute value of the temperature requires a separate, one-time calibration, variations in temperature are automatically tracked by sampling the detector's thermal noise spectra. 

\section{A \PMT{} employing QLS}
\label{sec:QLS-PT}
QLS has emerged as a powerful technique to enhance both precision and sampling rates in high-precision spectroscopic measurements by enabling full motional control and high-fidelity detection of single trapped ions. This method~\cite{Schmidt2005}, already implemented on singly-charged atomic ions in atomic clocks~\cite{Rosenband2007, Chou2010}, on molecular ions~\cite{Wolf2016a,chouPreparationCoherentManipulation2017} and on highly charged ions~\cite{mickeCoherentLaserSpectroscopy2020} in radio-frequency traps, was originally proposed for antiprotons in \PMT{}s~\cite{heinzenQuantumlimitedCoolingDetection1990,winelandExperimentalIssuesCoherent1998}. 
A comprehensive proposal for implementing QLS in proton/antiproton comparison measurements using \PMT{}s was discussed in Ref.~\cite{Cornejo2021}. Its implementation for \Htwoplus{}/\antiHtwoplus{} requires further considerations, which are discussed below.

\subsection{Apparatus}
\begin{figure*}[t]
\centering
\includegraphics[width=0.8\linewidth]{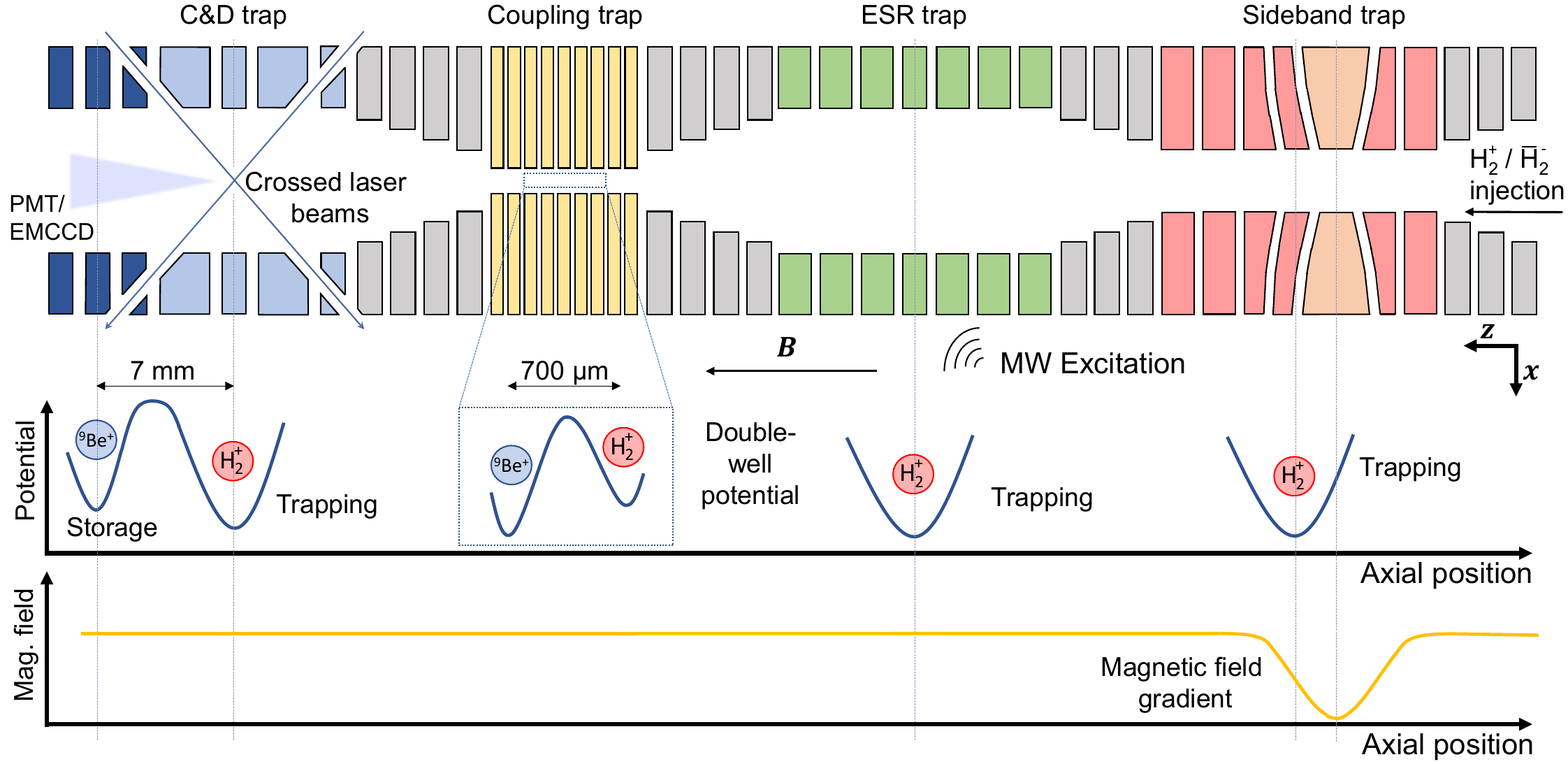}
\caption{\textbf{ Sketch of the \PMT{} setup for implementing QLS in \Htwoplus{} and \antiHtwoplus{}.}
\textbf{Top panel}: Longitudinal cross section of a cylindrical multi–PM-trap system with several dedicated zones. The \textbf{cooling \& detection (C\&D) trap} hosts beryllium ions and \Htwoplus{} and \antiHtwoplus{} for individual manipulation with multiple laser beams. 
A \textbf{coupling trap} enables energy exchange between ions via a double-well potential. The \textbf{ESR trap} is used to probe the electron-spin transition of the molecule, while the \textbf{sideband trap} couples the spin and motional states of \Htwoplus{} or \antiHtwoplus{} using a magnetic-field gradient. Gray electrodes enable adiabatic ion transport between the traps. \textbf{Middle panel}: Axial electric trapping potential in the different zones for \Htwoplus{} and beryllium ions. The potential plot shows the situation where the beryllium ion is stored while the \Htwoplus{} is individually manipulated with lasers in the \textbf{C\&D} trap. In the case of individual manipulation of beryllium ions at the center of the \textbf{C\&D} trap for laser manipulation and detection (not shown), \Htwoplus{} would be shuttled to another part of the trap stack, e.g. \textbf{ESR} trap. For the case of \antiHtwoplus{} (not shown), the trapping potential would be inverted due to the negative charge.  
\textbf{Bottom panel}: Magnetic-field strength across the different trapping zones. PMT: Photomultiplier Tube. EMCCD: Electron Multiplying Charge-Coupled Device. MW: Microwave.} 
\label{fig:trap}
\end{figure*}

Figure~\ref{fig:trap} shows a possible multi-PM-trap setup for implementing QLS with \Htwoplus{}/\antiHtwoplus{}. 
Both the experimental setup and the associated protocol follow the design principles of the BASE QLEDS experiments developed for the implementation of QLS with (anti-)protons~\cite{niemannCryogenic9BePenning2019, Cornejo2021}, but are adapted here for molecular ions. 
As in the BASE QLEDS experiment, the complete trap stack is enclosed in a trap can positioned at the center of a superconducting magnet that provides a magnetic field of several Tesla~\cite{cornejoOpticalStimulatedRamanSideband2023}. The enclosure is thermally anchored to a two-stage, low-vibration cryocooler~\cite{dubielzig2021}, which maintains cryogenic temperatures of about 4\,K in the case of BASE QLEDS, thereby ensuring ultra-high vacuum conditions suitable for storing antimatter ions over several weeks without annihilation, while simultaneously minimizing motional heating rates~\cite{borchert2019measurement}.
As in proton/antiproton QLS experiments, beryllium ions are the preferred choice for both sympathetic cooling of and logic operations on \Htwoplus{} and \antiHtwoplus{}, as they are the lightest ions that can be addressed with lasers~\cite{Cornejo2021}. These atomic ions can be produced by ablation of a beryllium target followed by photo-ionization in the trap, and then cooled to their motional ground state, first by Doppler cooling and subsequently by sideband cooling~\cite{niemannCryogenic9BePenning2019,cornejoResolvedsidebandCoolingSingle2024}.

The qubit in $^9$Be$^+$ is encoded in the two Zeeman-split ground-state electronic levels $\ket{\downarrow}_\mathrm{Be} \equiv \ket{^2S_{1/2}, m_j=-1/2}$ and $\ket{\uparrow}_\mathrm{Be} \equiv \ket{^2S_{1/2}, m_j=+1/2}$, which can be manipulated and detected via two-photon stimulated Raman transitions~\cite{cornejoOpticalStimulatedRamanSideband2023}. The analysis is restricted to the \mbox{$m_I = +3/2$} nuclear spin state, ensuring a closed cooling cycle~\cite{Nakamura2002} and assuming the nuclear spin remains unchanged after loading and state preparation.  
These operations can be carried out in the ``C\&D trap" shown in Fig.~\ref{fig:trap} by introducing two laser beams for Doppler cooling and repumping, along with two Raman laser beams arranged in a 90$^\circ$ crossing configuration. 
For the two-photon stimulated Raman transition, this geometry ensures a wavevector difference along the axial trap vibrational mode, which couples to the atomic qubit state.

Turning now to the far-right end of the trap stack, both \Htwoplus{} and \antiHtwoplus{} can be injected into the trap by appropriate means, while in the case of \Htwoplus{}, the ions can also be produced in situ within a dedicated production region of the setup (not shown in Fig.~\ref{fig:trap}). These ions can be transported adiabatically~\cite{meiners2024fast,Boehn2025} through the trap stack to  the ``coupling trap", where sympathetic cooling via free-space Coulomb interaction with a laser-cooled $^9$Be$^+$ ion will enable a single \Htwoplus{} or \antiHtwoplus{} ion to reach submillikelvin temperatures, or even their motional ground state, thereby enabling the implementation of logic operations to probe rovibrational and electron-spin transitions (see Fig.~\ref{fig:Energy levels}). To access these transitions non-destructively, both the internal and motional states of \Htwoplus{} and \antiHtwoplus{} must be manipulated, and their motional states coupled to the logic ion via free-space Coulomb interaction, enabling high-fidelity state detection. Therefore, the coupling between a single laser-cooled $^9$Be$^+$ ion and a single \Htwoplus{}/\antiHtwoplus{} ion via Coulomb interaction is a key feature for both sympathetic cooling and state detection. This is the purpose of the ``coupling trap" in Fig.~\ref{fig:trap}, where a double-well potential is generated to exchange the energy of the common axial mode between two particles. 

Returning to the ``C\&D trap", this trap provides laser access for probing the rovibrational carrier transition and applying sideband pulses to the molecular ions. 
These sideband pulses correspond to driven transitions that couple the internal and motional modes of the ion, enabling coherent control over both internal and motional quantum states. 

Adjacent to the ``coupling trap" lies the ``ESR trap", where microwave excitation enables probing the electron-spin transition on \Htwoplus{}/\antiHtwoplus{}.
Although a dedicated trap for microwave spectroscopy is not strictly required, it provides a practical solution, as combining laser access and microwave excitation within the same trap presents significant experimental challenges. 

Next to the ``ESR trap" is the ``sideband trap", where a magnetic field gradient is generated by a ferromagnetic ring electrode (shown in light orange in Fig.~\ref{fig:trap}) used as correction electrode. 
In the microwave regime, the variation of the electromagnetic field over the spatial extent of the ion's ground-state  wavefunction of the axial mode is negligible, preventing direct coupling between the axial vibrational mode and the electron-spin transition. 
Introducing a magnetic-field gradient in the ``sideband trap" overcomes this limitation by enabling an effective coupling between the ion’s axial motion and its spin degree of freedom~\cite{ mintert2001}.  
Magnetic-field gradients on the order of a few hundred Tesla per meter would be required to achieve effective Rabi rates on the sideband transition under realistic experimental conditions, such us feasible pulse intensities and durations~\cite{nitzschkeElementaryLaserLessQuantum2020}. Since such gradients are more easily achieved in traps with smaller inner diameters, the sideband trap in Fig.~\ref{fig:trap} is designed accordingly.

\subsection{The sympathetic cooling process: harmonic regime}
\label{sec:The sympathetic cooling process}

\begin{figure}[t]
    \centering
    \includegraphics[width=\linewidth]{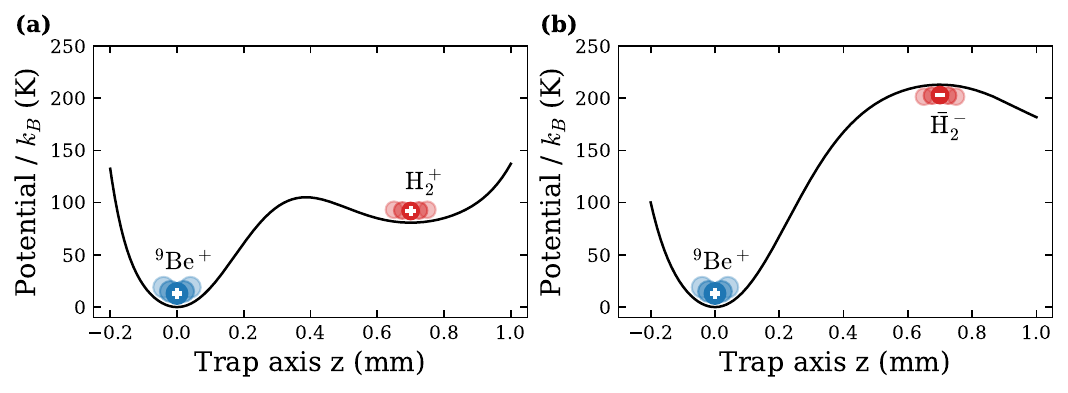}
    \caption{ \textbf{Double-well potentials for sympathetic cooling of \Htwoplus{}/\antiHtwoplus{}  
    using a $^9$Be$^+$ ion.
    }
    \textbf{(a)} Example of a double-well potential for \Htwoplus{} using parameters $s_0 = 0.7$ mm and $\omega_z = 2\pi\times300$ kHz. 
    \textbf{(b)} Example of a double-well potential for \antiHtwoplus{} using the same parameters. The \antiHtwoplus{} potential well is inverted due to its negative charge.}
    \label{fig:double_well}
\end{figure}

Examples of the double-well potentials used in the sympathetic cooling simulations of $^9$Be$^+$/\Htwoplus{} and $^9$Be$^+$/\antiHtwoplus{} pairs are shown in Figures~\ref{fig:double_well}a and \ref{fig:double_well}b, respectively. Assuming that both trapped particles, labeled $a$ and $b$, remain within the harmonic regions of their respective potentials, complete axial energy exchange between them occurs after a  duration $\tau_{ex}$~\cite{brownCoupledQuantizedMechanical2011},
\begin{equation}
    \tau_{ex} = \frac{2\pi^2 \epsilon_0 s_0^3 \sqrt{m_a m_b} \sqrt{\omega_{z,a} \omega_{z,b}}}{q_a q_b}\ ,
\label{eqn:transfer_time}
\end{equation}
where $m_a$ and $m_b$ are particle masses, $q_a$ and $q_b$ are their charges, $\omega_{z,a}$ and $\omega_{z,a}$ are their axial angular oscillation frequencies, $\epsilon_0$ is the vacuum permittivity, and $s_0$ is the particle separation. In our case, we use the resonance condition $\omega_z=\omega_{z,a}\approx\omega_{z,b}$. 

Moreover, the coupling can be precisely timed by bringing the particles into and out of resonance using tailored voltage waveforms, such that the interaction duration matches eq.~\ref{eqn:transfer_time}. 
Since shorter exchange times are more robust against trapping frequency fluctuations (e.g. those arising from electrode-voltage noise), it follows from eq.~\ref{eqn:transfer_time}  that a small particle separation is essential to achieve efficient energy exchange under realistic experimental conditions.
This requirement can only be fulfilled by using small electrodes fabricated using micro-fabrication techniques~\cite{cornejoOptimizedGeometryMicro2016}. 

The coupling \PMT{} used to generate the double-well potential, shown in Fig.~\ref{fig:double_well}, consists of nine hollow cylindrical electrodes, each 200\,$\mu$m thick with an inner diameter of 800\,$\mu$m, allowing particle separations of only a few hundred micrometers. 
It is important to note that the radial component of the coupling can be neglected, as the radial motion of the particles can be reduced to a few micrometers even in a thermal state~\cite{brown1986geonium}. 
Furthermore, after sympathetic cooling of the axial mode, a $\pi$-pulse between the radial 
and axial modes~\cite{Cornell1990} can be used to cool the radial motions of both ions to the ground state. 
Since particle separation during coupling is on the order of a few hundred micrometers, the coupling interaction can be treated as effectively one-dimensional along the trap axis.
Taking this into account, full motional exchange is expected when the particles are close to their motional ground state~\cite{brownCoupledQuantizedMechanical2011}, which is the case during state detection. 

However, during sympathetic cooling for particle initialization, we assume that \Htwoplus{}/\antiHtwoplus{} is initially in thermal equilibrium with the trap environment at \mbox{$T_z \approx$ 4\,K}, and thus the axial energies $E_z$ follow the Boltzmann distribution,
\begin{equation}
    g(E_z) = \frac{1}{k_B T_z} \exp{\left[-E_z/(k_B T_z)\right]}\ ,
\label{eqn:boltzmann}
\end{equation}
where $k_\mathrm{B}$ is the Boltzmann constant. 
In this regime, there is a substantial probability that ions have an energy as high as \mbox{$10\,\mathrm{K}\times k_\mathrm{B}$} and therefore they may sample non-harmonic regions of the coupling potential, requiring a more detailed analysis. 
For this, we simulate the classical motion of two particles subject to Coulomb interaction and electrostatic trapping forces, computed via Newton's laws of motion, and integrated using the Verlet algorithm with a 50\,ns timestep.
The system is treated as one-dimensional along the trap axis, as discussed previously. 
For each simulation, the $^9$Be$^+$ ion is initialized in the motional ground state. Note that during coupling to the molecular ion, the $^9$Be$^+$ ion is not laser cooled.
The total energy of each particle is evaluated as the sum of its kinetic and electrostatic potential energy along the trap axis $z$. 

\begin{figure}[t]
    \centering
    \includegraphics[width=\linewidth]{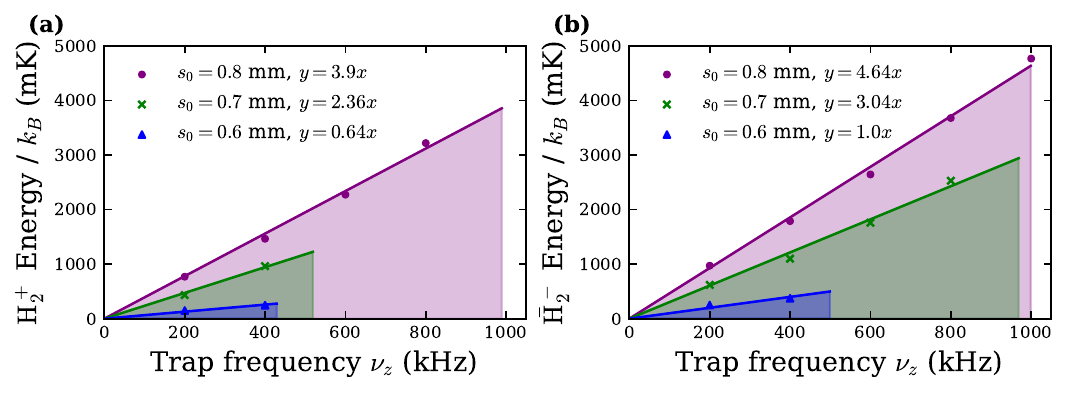}
    \caption{ \textbf{Cooling of \Htwoplus{}/\antiHtwoplus{} using the harmonic region of the potential as a function of the axial trap frequency $\nu_z$.} \textbf{(a)} \Htwoplus{}. \textbf{(b)} \antiHtwoplus{}. Molecules with initial energies within the shaded regions can be cooled to below \mbox{$1\,\mathrm{mK}\times k_\mathrm{B}$}. The top boundary of these regions is described by a linear fit (see legend), with coefficients given in mK\,kHz$^{-1}$. Higher trapping frequencies ($\nu_z$) and larger interparticle separations ($s_0$) allow cooling from higher initial energies.}
    \label{fig:cooling}
\end{figure}

Fig.~\ref{fig:cooling} shows the energy range of \Htwoplus{}/\antiHtwoplus{} that can be cooled to below \mbox{$1\,\mathrm{mK}\times k_\mathrm{B}$} within the harmonic region of the double-well potential illustrated in Figure~\ref{fig:double_well}. The parameters considered correspond to electrode voltages of up to 10\,V. Due to anharmonicities and frequency-stability constraints, efficient cooling is limited at higher initial particle energies for realistic oscillation frequencies and interparticle separations.

\begin{figure}[t]
    \centering
    \includegraphics[width=\linewidth]{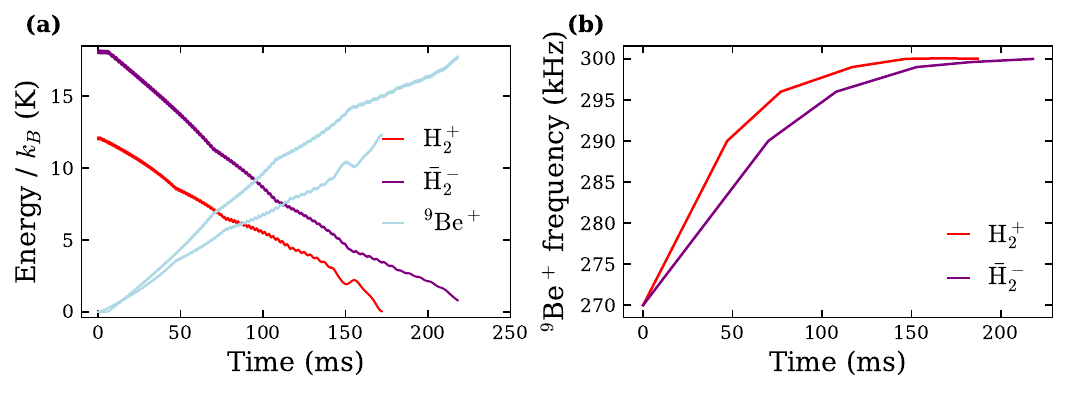}
    \caption{
    \textbf{Cooling of \Htwoplus{}/\antiHtwoplus{} using frequency sweeping of the $^9$Be$^+$ potential.} \textbf{(a)} Examples of the energy evolution during the $\nu_z$ sweep. The top light blue line illustrates the energy evolution of the $^9\mathrm{Be}^+$ ion coupled to \antiHtwoplus{}, and the bottom light blue line is for the $^9\mathrm{Be}^+$ ion coupled to \Htwoplus{}. \textbf{(b)} Frequency sweep from 270\,kHz to 300\,kHz, implemented via a series of voltage ramps. 
    }
    \label{fig:sweep}
\end{figure}

\subsection{The sympathetic cooling process: anharmonic regime}
\label{sec:The sympathetic cooling process - anharmonic regime}

To mitigate the effect of anharmonicity, we introduce a time-dependent potential in which the oscillation frequency of the $^9$Be$^+$ ion is swept to maintain resonance with \Htwoplus{}/\antiHtwoplus{} throughout the coupling process, as illustrated in Fig.~\ref{fig:sweep}a. Such a frequency sweep can be realized experimentally by applying a sequence of voltage ramps to the trap electrodes. 
As shown in Fig.~\ref{fig:sweep}b, \Htwoplus{} can be cooled to below \mbox{$1\,\mathrm{K}\times k_\mathrm{B}$} from initial energies up to \mbox{$12\,\mathrm{K}\times k_\mathrm{B}$}, and \antiHtwoplus{} can be cooled from energies up to \mbox{$17\,\mathrm{K}\times k_\mathrm{B}$}. Importantly, any energy below the illustrated initial energies can be cooled using the same sweep parameters without requiring prior knowledge of the initial energy of \Htwoplus{}/\antiHtwoplus{}. 
As a result, more than 95\% of a 4\,K Boltzmann distribution can be efficiently cooled. 

Therefore, the frequency sweep can serve as an initial pre-cooling stage, after which, as shown in Fig.~\ref{fig:cooling}, time-independent harmonic coupling can be applied to further reduce the energy of \Htwoplus{}/\antiHtwoplus{} below \mbox{$1\,\mathrm{mK}\times k_\mathrm{B}$} and ultimately reach the motional ground state by repeating the same coupling. 
If the available power supply does not provide sufficient voltage precision, the efficiency of the energy transfer is reduced. This limitation can be mitigated by repeating the frequency-sweep and time-independent coupling steps. Finally, to verify that \Htwoplus{}/\antiHtwoplus{} has reached the ground state, the motional state of the co-trapped $^9$Be$^+$ ion can be assessed via a blue-sideband interrogation. A detailed treatment of sympathetic cooling using the double-well potential technique, together with a robustness analysis, will be presented in ref.~\cite{Poljakov2026} for the case of the \mbox{(anti-)proton}.

\subsection{Rovibrational spectroscopy of $\mathrm{H}_2^+$/\antiHtwoplus{}}

\begin{figure*}[t!]
\centering
\includegraphics[width=.85\linewidth]{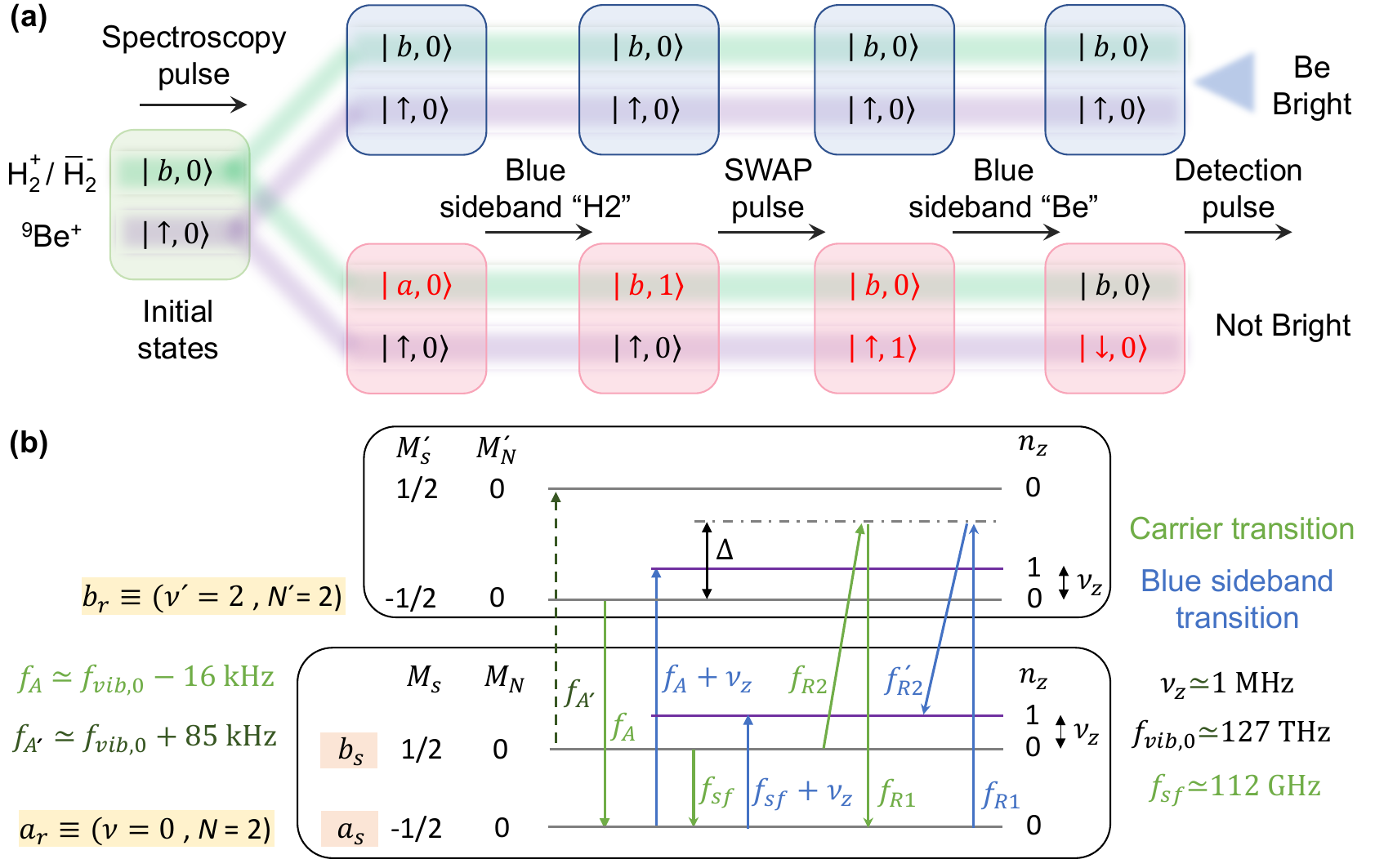}
\caption{
Quantum logic spectroscopy. \textbf{(a)} 
The motional ($\ket{0}$, $\ket{1}$) states and internal states, ($\ket{\uparrow}$, $\ket{\downarrow}$) for $^9$Be$^+$ and ($\ket{a}$, $\ket{b}$) for \Htwoplus{}/\antiHtwoplus{}, are shown throughout the QLS sequence. States highlighted in red indicate changes with respect to the previous step. The initial states are indicated by the filled green rounded square. The cases corresponding to the $\ket{b}$ and $\ket{a}$ states of the molecular ion after the spectroscopy pulse are indicated by filled blue and red rounded squares, respectively. The green and purple bands serve as a visual guide through the sequence for \Htwoplus{}/\antiHtwoplus{} and $^9$Be$^+$, respectively. The ``Spectroscopy'' pulse is used to probe the carrier transition, while  blue sideband labels (``H2'' and ``Be'') denote the blue-sideband transition for the molecular and beryllium ions, respectively. \textbf{(b)} Simplified energy-level diagram showing selected states for QLS. For  rovibrational spectroscopy these are $|a_r,n_z=0\rangle$, $|b_r,0\rangle$, $|b_r,1\rangle$, where $a_r\equiv(v=0,N=2,M_s=-1/2,M_N=0)$ and $b_r\equiv(2,2,-1/2,0)$. For ESR these are  $|a_s,0\rangle$, $|b_s,0\rangle$, $|b_s,1\rangle$, where $a_s\equiv(0,2,-1/2,0)$,  $b_s\equiv(0,2,+1/2,0)$. 
The motional quantum number is $n_z$. Energy levels are not to scale. $f_{R1} = f_\mathrm{A} + \Delta$, $f_{R2} = f_\mathrm{A} + \Delta - f_{sf}$ and $f^{\prime}_{R2} = f_\mathrm{A} + \Delta - (f_{sf} + \nu_z$) are the transition frequencies required to perform ESR via stimulated Raman transitions.}
\label{fig:QLS}
\end{figure*}

Once each particle has been stored in an individual trapping potential along the trap stack and prepared in its motional ground state $\ket{n_+ = 0, n_z = 0, n_- = 0}$, where $n_+$, $n_z$ and $n_-$ are the phonon number of the modified cyclotron, axial and magnetron modes, respectively, QLS of the molecular ion can be performed. 
Figure~\ref{fig:QLS} shows the sequence for QLS of \Htwoplus{}/\antiHtwoplus{}. A single beryllium ion is initially prepared in the $\ket{\uparrow_{\mathrm{Be}}, n_z = 0}$ state and stored at the side of the ``C\&D trap" (see Fig.~\ref{fig:trap}). 

A single \Htwoplus{}/\antiHtwoplus{} ion is initially prepared in the upper $b_r \equiv (v^{\prime}=2, N^{\prime}=2)$ rovibrational state and confined at the center of the ``C\&D trap" (see Fig.~\ref{fig:trap}). To probe the downward $b_r \rightarrow a_r \equiv (v=0, N=2)$ rovibrational carrier transition (see Fig.~\ref{fig:Energy levels}), a single ``spectroscopy" laser pulse at a frequency of approximately $f_0\simeq$ 127\,THz is applied to the \Htwoplus{}/\antiHtwoplus{} ion in the ``C\&D trap" (see Fig.~\ref{fig:trap}). A subsequent ``mapping" laser pulse, tuned to the motional blue sideband at frequency $f_0 + \nu_z$, where $\nu_z$ is the axial oscillation frequency of the molecular ion in the trap, coherently couples the rovibrational and motional states, adding one quantum of motion and transferring the rovibrational state back to $b_r$ only if the molecular ion was transferred to state $a_r$ by the previous spectroscopy pulse. We will thus have a motional excitation conditioned on the
rovibrational state established after the spectroscopy pulse. This procedure can work for any upper and lower molecule spin state $(M_s,M_N)$, $(M_s',M_N')$; therefore these quantum numbers were omitted above. Since the rovibrational transition frequencies $f_0$ differ for different spin state pairs, the actual transition to be driven is selected by the choice of laser frequency.

Subsequently, the molecular and beryllium ions are transported to the coupling trap (see Fig.~\ref{fig:trap})  by means of adiabatic transport techniques, in which the harmonic trapping potential is smoothly moved along the trap stack. This is a common method in ion-based quantum computing experiments~\cite{blakestad2011, blakestad2009, furst2014}, and it has recently been demonstrated in a cryogenic multi-\PMT-trap stack~\cite{meiners2024fast,Boehn2025}. When both particles are sufficiently close in the coupling trap and their axial frequencies are matched, the axial modes of the ions become coupled. This process, known as a SWAP pulse, transfers the quantum of motion of the molecular ion, if previously added, to the $^9$Be$^+$ ion.

Afterwards, the beryllium ion is transferred back to the ``C\&D trap", where another ``mapping" Raman laser pulse tuned to the motional blue sideband of the atomic ion, couples its spin and motional states, flipping the spin state and removing one quantum of motion, if and only if it was previously added after the SWAP pulse. 
Finally, a detection pulse is applied to the $^9$Be$^+$ ion to read out its spin state. If the ion scatters photons from the detection laser, it was in the $\ket{\uparrow}_{\mathrm{Be}}$ state, and in the $\ket{\downarrow}_{\mathrm{Be}}$ state otherwise.  

After an interrogation cycle, re-initialization of the molecular ion is not required, since after the SWAP pulse the molecular ion is already in the axial mode ground state and in the upper $b_r$ rovibrational state. The initial preparation of the molecular ion after sympathetic cooling can be achieved by applying a mapping pulse followed by a SWAP pulse with the atomic ion. The $^9$Be$^+$ ion must be reinitialized in the $\ket{\uparrow_{\mathrm{Be}}, n_z = 0}$ state in the ``C\&D trap'' and then parked in the storage side to start a new interrogation cycle with the molecular ion. The latter is already in the initial spectroscopy state.

By sweeping the frequency of the ``spectroscopy" pulse, the resonance line can be measured. Although the ``spectroscopy'' pulse will be weak (e.g.~$\pi$/2),  the ``mapping'' pulse  should be strong, power-broadening the transition, so as to ensure robustness of the scheme. 

\subsection{Electron-spin resonance of $\mathrm{H}_2^+$/\antiHtwoplus{}}

The case is similar to rovibrational spectroscopy except for the fact that the transition used to probe the state can be an electron-spin resonance or a vibrational transition. In this case, to probe the electron-spin transition $a_s \equiv (M_s = -1/2) \leftrightarrow b_s \equiv (M_s = 1/2)$ in the rovibrational state $(v=0, N=2)$ (see Fig.~\ref{fig:Energy levels}), the molecular ion must be prepared in the $b_s$ upper state.
As in rovibrational spectroscopy case, a single beryllium ion must also be prepared in the $\ket{\uparrow_{\mathrm{Be}}, n_z = 0}$ state and stored in the ``C\&D trap'' (see Fig.~\ref{fig:trap}).  
\vskip .05 in
\noindent\emph{A. Detection via ESR}

The molecular ion must be in the ``ESR trap'' (see Fig.~\ref{fig:trap}), where a ``spectroscopy" microwave carrier pulse excitation at a frequency of approximately $f_\mathrm{sf} \simeq 112$\,GHz attempts to drive the molecule into the $a_s$ down state. Subsequently, the molecular ion is transported to the ``sideband trap'', where a ``mapping" microwave pulse tuned to the motional blue sideband $f_\mathrm{sf} + \nu_z$ of the molecular ion in the trap couples the electron-spin state and the axial motional mode in the presence of the magnetic-field gradient, adding one quantum of motion and transferring the electron-spin state back to $b_s$ only if the molecular ion was transferred to $a_s$ state by the previous ``spectroscopy" pulse. The rest of the sequence will be identical as in the case of the rovibrational spectroscopy, moving both ions to the coupling trap, where a SWAP pulse transfers this quantum of motion to the beryllium ion, if it was previously added. Finally, the beryllium ion is transferred to the ``C\&D trap'', where a blue-sideband Raman pulse and subsequent fluorescence detection reveal the spin state of the beryllium ion, and thus the electron-spin state of the molecular ion after carrier probing. The interrogation cycle then restarts with the single beryllium ion prepared in the $\ket{\uparrow_{\mathrm{Be}}, n_z = 0}$ state. The advantage of this detection technique is that no laser resonant with a molecular ion vibrational transition is necessary. 

\vskip .05 in
\noindent\emph{B. Detection via a vibrational transition}

In this case, after the ``spectroscopy" microwave pulse in the ``ESR trap'', the molecular ion must be transported to the ``C\&D trap'', where a resonant laser pulse will attempt to excite the molecule from the $a_r$ state to $b_r$ state. Since this transition frequency differs depending on whether the molecular ion is in the $a_s$ or $b_s$ state, mainly due to spin-rotational coupling, rovibrational-state detection can be used to infer the electron-spin state after the spectroscopy pulse. The process then follows the same sequence as for the rovibrational spectroscopy: a ``mapping'' pulse is applied, followed by the transport of the ions for the SWAP pulse, and finally the detection of the beryllium ion. A big advantage of this scheme is that no magnetic bottle is required in the trap stack to implement the mapping pulse, which reduces key systematic effects.

\vskip .05 in
\noindent\emph{C. ESR Spectroscopy via a Raman transition}

 Another alternative to probe the electron-spin transition $a_s \leftrightarrow b_s$ in the lower rovibrational state $(0,2)$ would be to use a stimulated-Raman transition via a virtual level detuned by, for example, $\Delta \simeq$ 10\,MHz from another state, such as the rovibrational level $(2,2)$. (A small detuning is possible because the levels are metastable.) Two laser beams near $f_0\simeq$ 127\,THz would be required, with a frequency difference of approximately $f_\mathrm{sf} \simeq$ 112\,GHz. These two laser beams would be introduced into the ``C\&D trap'' in a 90° crossing configuration, as shown in Fig.~\ref{fig:trap}, to produce a wavevector difference along the axial mode, thereby enabling coupling between the electron-spin and axial motional modes. In this scenario, the molecular ion must be in the ``C\&D trap'', where a "spectroscopy" carrier Raman laser pulse attempts to drive the $b_s \rightarrow a_s$ transition in the molecule ion. For this carrier excitation, the  lasers' frequency difference must be tuned close to the carrier transition frequency $f_\mathrm{sf}$. Subsequently, a ``mapping'' pulse is applied on the molecular ion by detuning the lasers' frequency difference to the motional blue sideband, $f_\mathrm{sf} + \nu_z$, of the molecular ion in the trap. This pulse adds one quantum of motion and transfers the state back to $b_s$ only if the molecular ion was transferred to $a_s$ state by the previous spectroscopy pulse. The remainder of the sequence follows the same procedure as in the previous cases.

Since these are electric quadrupole transitions, the transition probability will be low; therefore, higher laser power intensities will be required. 
Further calculations are necessary to determine the appropriate parameters. Importantly, this approach would allow the implementation of QLS directly in the ``C\&D trap'', simplifying the experimental setup, and removing the need for the ``ESR trap'' and ``sideband traps''. One issue to consider is whether the required laser intensities would cause a relevant light shift on the ESR transition.

\subsection{State preparation and time consumption}
\label{sec:QLS-PT: State preparation and time consumption}

The above schemes rely on having the molecular ion in the initial state $b_r$ or $b_s$ for the intended spectroscopy transition (Fig.\,\ref{fig:QLS}). In order to achieve this, one can use QLS for state preparation. It is necessary to probe sequentially all states in which the ion presumably could be, until a signal appears that indicates that the states has been found. To speed up the search, instead of using a weak spectroscopy pulse one will use a $\pi$ pulse. After finding the state, a series of rovibrational transitions will have to be applied to reach the rovibrational level $(v,N)$ corresponding to the initial state $b$. 

The ion may well be in a state with quantum number $M_N$ that differs from the preferred value ``0" for spectroscopy. In this case one will have to apply one or more laser (or ESR or Raman) excitations that change $M_N$. These transitions have relatively large magnetic shifts $\Delta f_\mathrm{mag}$ (Table\,\ref{tab:sensitvities beta}). However, they can be predicted with sufficient accuracy to allow finding the exact transition frequency experimentally, by systematic searches over small frequency intervals. 

Regarding time consumption, the state preparation of the molecular and beryllium ions will not be taken into account in the time budget, as it is performed only once at the beginning of the spectroscopy measurement. However, a re-cooling process may be applied at any time to ensure proper ion states. For the complete spectroscopy measurement, multiple interrogation cycles would be performed, in which a spectroscopy pulse at a frequency close to the target transition is applied. For this spectroscopy pulse, a Ramsey sequence with two $\pi$/2 pulses separated by a free-precession time $t_s$ can be used. 
This approach reduces systematic shifts arising from laser intensity fluctuations during the excitation pulses. The interval $t_s$ constitutes the effective measurement time and should be maximized, since any remaining steps contribute only as dead time. Sideband pulses on the atomic ion can be performed within a few tens of microseconds~\cite{cornejoResolvedsidebandCoolingSingle2024}, and state detection can be completed in a few hundred microseconds~\cite{meiners2024fast}. 
Transport of particles between traps typically requires a few milliseconds~\cite{Boehn2025}, although this time can be reduced to below 1\,ms with optimized transport waveforms. SWAP coupling between the molecular and atomic ion for motional-state readout can be achieved within a few milliseconds (see Section~\ref{sec:The sympathetic cooling process}). 
Sideband operations on the molecular ion can likewise be performed on the millisecond timescale. 

Under these conditions, and considering the excellent ion-heating performance typically provided by \PMT{} systems~\cite{borchert2019measurement}, a complete interrogation cycle would take on the order of 10\,s per frequency point (depending on the choice of $\delta_\mathrm{int}$), with only a few tens of milliseconds devoted to the detection sequence. This leads to a dead-time fraction below 1\%.

\section{Summary and Conclusion}
\label{sec:Discussion}
\begin{table*}
    \caption{ \textbf{Summary of the projected uncertainties occurring in vibrational spectroscopy.} The transition $(v=0,N=2,M_s=-1/2)\rightarrow(v'=2,N'=2,M_s'=-1/2)$ at $B_0=4\,$T is assumed. Uncertainty values are fractional and refer to the transition frequency of one molecule - \Htwoplus{} or \antiHtwoplus{}. Some shifts can be determined with different uncertainties, depending on the type of \PMT{} (CSGE-\PMT{} or QLS-\PMT{}); in those cases the table presents two.  The CSGE-\PMT{} is assumed to be equipped with advanced cryo-electronics-based cooling techniques. BBR: black-body radiation. $(^{*})$ outside the magnetic bottle. $(\dagger)$ based on current state-of-the-art (few-week-long integration time).
    }
    \centering
 {\small
    \begin{tabular}{c|m{3cm}|m{7.5cm}}
        \textbf{effect} & \quad\textbf{uncertainty} & \quad\quad\quad\textbf{main requirements} \\
        \hline
total magnetic &\phantom{$<$}\,$3\times10^{-18}$ & $M_N=1\rightarrow M_N'=1$, mean B-field measured at $5\times10^{-10}$ level$^*$ \\
& \phantom{$<$}\,$3\times 10^{-18}$ &  $M_N=0\rightarrow M_N'=0$, mean B-field measured at $5\times10^{-8}$ level   \\
        EQS & $<1\times10^{-18}$  & none in particular  \\
        d.c. Stark shift & \phantom{$<\,$}$2\times10^{-18}$ &CSGE-\PMT{} 
        \\
  &   $<1\times10^{-18}$
  & QLS-\PMT{} 
\\
light shift &$<1\times10^{-18}$  &Rabi angular frequency $\Omega_{if}<2\,\mathrm{rad/s}^{-1}$   
\\
QDS&{\phantom{$<$}\,$2\times10^{-16}$} ($u^{(1)}$)&CSGE-\PMT{}, high-resolution temperature measurement$^\dagger$ 
\\
&{\phantom{$<$}\,\text{$3\times10^{-18}$~\phantom{(syst.)}}\ \ \ \ \ \ \ \ \ \ \ \ \ }&QLS-\PMT{}, measurement of sidebands or \Htwoplus{}, \antiHtwoplus{} in equal motional condition\\
BBR&$<1\times10^{-19}$&cryogenic environment \\
\hline\hline
line resolution&\phantom{$<$}\,$3\times10^{-17}$ (stat.) &CSGE-\PMT{}, 0.05\,Hz linewidth laser, integration time: {several }weeks \\
&\phantom{$<$}\,$1\times10^{-17}$ (stat.)&QLS-\PMT{}, 0.05\,Hz linewidth laser, integration time: {few}\,days\\
    \end{tabular}
    }
      \label{tab:summary of the shifts}
\end{table*}

\subsection{CPTI test of the vibrational transition frequency} Table\,\ref{tab:summary of the shifts} summarizes the relevant shifts discussed above.
Also included is the BBR shift, discussed before for RF traps \cite{Schiller2014,Karr2014,Karr2016}. At 4\,K environmental temperature it is less than $1\times10^{-19}$ fractionally for the vibrational transition. Additionally, it would be common-mode and therefore suppressed in the frequency difference $\delta f_\mathrm{vib}$, if the two molecular species are 
evaluated in the same \PMT{}.
The collision shift is expected to be negligible because of  the cryogenic vacuum.

The (equal) uncertainty of the vibrational frequency of each type of ion - \Htwoplus{} and \antiHtwoplus{}\,- may be estimated as the r.m.s.\,combination of the contributions in the table, since we may assume that they are uncorrelated. From that uncertainty, 
the accuracy of a CPTI test,
i.e.\,the uncertainty of the difference of the transition frequencies 
$\delta f_\mathrm{vib}=$ $f_\mathrm{vib}(\mathrm{H}_2^+)-f_\mathrm{vib}($\antiHtwoplus{}), is found by multiplication with $\sqrt{2}$. 

We find a test accuracy of approximately 
$2\times10^{-16}$ in the case CSGE-\PMT{}, employing an integration time of a few weeks.
For the QLS-\PMT{} we estimate approximately $1\times10^{-17}$ , if we --\,reasonably\,-- assume integration times longer than 2 days.

\subsection{CPTI test of the electron/positron g factor} 
\label{sec:CPTI test of the electron/positron g factor}
As proposed by Myers, this test is on the bound electron/positron, rather than on the free particles, as was the case in the classic experiment in ref.~\cite{VanDyck1987}. 
{That experiment achieved an uncertainty of $4\times10^{-12}$. This level or below should be the target for any new experiment.} 

The test amounts to measuring very precisely the frequency $f_\mathrm{sf}$ of an ESR transition $\Delta M_s=\pm1$, within any rovibrational level of \Htwoplus{}/\antiHtwoplus{}. If a test were performed with the techniques of sympathetic cooling and QLS, it would be vastly more complex than the classic approach, but the higher speed and lower particle temperature could give it an edge.

We have shown here that suitably chosen ESR transitions have no relevant nonmagnetic perturbation, a fact that is relevant for achieving a high accuracy. Appendix~\ref{app:determination of tensor polarisability} indicates that ESR transitions with $\Delta M_N=0$ have no Stark shift contribution. 
While there is a correction from spin-rotation interaction, eq.\,(\ref{eq:Appendix spin-rotation contribution}), it is $B$-field dependent. For small variations of $B$ the correction will change linearly, i.e.~with the same algebraic dependence as the dominant electron-Zeeman splitting. 

Because of the long wavelength of the microwave necessary for ESR transitions, they occur deep in the Dicke regime, without substantial motional sidebands. 

The QDS is present, but its level - even without a $B_2$-cancellation technique - is negligible compared to the main systematic uncertainty. The latter arises from the fact that the ESR frequency is of course directly proportional to the magnetic field. 

The accuracy of the CPTI test thus - apparently - depends on the ability to precisely measure and track the magnetic field via cyclotron frequency measurements, which is a separate topic, e.g. described in detail in \cite{borchert202216, schussler2020detection}. 

A first  implementation option would be to measure alternately \Htwoplus{} and \antiHtwoplus{} in the same trap, taking advantage of common-mode rejection of slow field drifts. 
Suppose that in the QLS-\PMT{} the cycle time for one ESR interrogation of \Htwoplus{} and subsequently one on \antiHtwoplus{} will be on the order of one minute. 
{Assume that  $\delta_\mathrm{int}\simeq 0.1\,\mathrm{Hz}/f_\mathrm{sf}\simeq1\times10^{-12}$}. (The microwave must exhibit a sufficiently narrow linewidth - fractionally at the $10^{-13}$ level; the wave could be  derived from an ultra-stable laser, via a frequency comb or from a cryogenic microwave oscillator.)
Assume further  10 days integration time, so that the fluctuations of the magnetic field on the time-scale of the alternation, $u(\delta B_a)$ could in the optimum case average down by two orders. If the experiment aims for  a total uncertainty of $1\times10^{-13}$ and hence requires a uncertainty in the mean differential magnetic field at that level, this translates into the requirement $u(\delta B_a)\simeq1\times10^{-11}$. The best fluctuation levels achieved in state-of-the-art experiments are about two to five times worse \cite{dickopf2024precision, schussler2020detection}; therefore this scenario is uncertain.

To overcome these limitations one could  consider a more complex trap stack architecture, containing two precision traps next to each other, in which one \Htwoplus{} ion and one \antiHtwoplus{} ion are simultaneously undergoing ESR spectroscopy.
Then, temporal fluctuations of the B-field potentially cancel to a certain extent in the differential ESR frequency. 
This concept is already implemented in the precision mass spectrometer PENTATRAP at MPIK for synchronized cyclotron frequency measurements. This apparatus does not have CSGE detectors, only homogeneous precision traps. The trap stack of Fig.\,\ref{fig:CSGETRAP} would then be extended to a total of 8 traps. The implementation of similar ideas in a QLS-\PMT{} would be technically complex.

The price to pay in the dual-precision-trap concept is that an axial gradient of the ideally homogeneous axial B-field will now affect the differential measurement. This effect would need to be characterized by periodically swapping the ion/anti-ion occupation in the two traps during the measurement campaign, or by performing an additional measurement campaign that compares two ions of the same species in each trap.

In summary, a CPTI test of the bound electron g-factor in such an advanced multi-trap CSGE-\PMT{} appears challenging but possible at the low-$10^{-13}$ level, with {weeks-long integration times. If the magnetic field is characterized very well, a further reduction of that level might be feasible.
Note that the dual-precision-trap scheme could also be advantageous for the vibrational CPTI test in a CSGE-\PMT{}.

\subsection{CPTI test of the nuclear g factor} 

The rovibrational transition frequencies depend on the nuclear charge radius of proton/antiproton at the $10^{-10}$ fractional level. Thus, a CPTI test performed at the $1\times10^{-17}$ level
would test the equality of this baryonic property at the sub-ppm level.

An additional option has been put forward by Myers \cite{Myers2018}: compare the hyperfine structure of \Htwoplus{} and \antiHtwoplus{}. This requires working with the ortho-configurations, having total nuclear spin $I=1$. He stated that fractional uncertainties of below $10^{-13}$ should be possible by measuring particular transitions that are insensitive to the magnetic field in first order. In \cite{Myers2018a} the level of $10^{-11}$ or below is mentioned.

In order to fill in the details, we have calculated the hyperfine energies as function of $B$ and find that such transitions indeed occur, with values of approximately 1\,GHz, at magnetic field values of approximately 0.5\,T. These would be unusually small for a \PMT{}. 
Because the viability of such a \PMT{} is uncertain (cf.\,App.\ref{sec:QDS cancellation}) we here propose  an alternative approach, suitable also for typical fields of several Tesla.

Ortho-\Htwoplus{} has the approximate spin energies in strong field \cite{Korobov2006}
\begin{eqnarray}
    E_\mathrm{s}(v,N,M_s,M_I,M_N)&\approx& c_e(v,N)M_s\,M_N + b_F(v,N)M_s\,M_I
    \label{eq:energies of ortho-H2+}
    \\
    &&+\frac{d_1(v,N)}{(2N-1)(2N+3)}\left(\frac{2}{3}N(N+1)M_N-2 M_N^2\,M_I\right)M_s\nonumber\\
    &&-\mu_{\rm B}g_e(v,N)M_s B
    -\mu_{\rm n}g_p M_I B\nonumber\\
    &&-\mu_{\rm n}g_r(v,N)\,M_N B\ .\nonumber
\end{eqnarray}
Here, the subscript ``s" on the l.h.s.~stands for ``HFS + Z + Z-rot", $g_p$ is the bare-proton g factor, $M_I$ is the approximate quantum number of the nuclear spin projection. The hyperfine coupling constants have the values $b_F(v=0,N=1)\simeq h\times0.93\,$GHz and $d_1(v=0,N=1)\simeq h\times0.13\,$GHz \cite{Korobov2006}. Importantly, $b_F$ is proportional to the product of magnetic moments of the baryon and the lepton.

RF transitions within a given electron spin state (fixed $M_s$) are magnetic-field dependent. However, consider the sequential measurement of two nuclear-spin-flip transitions with opposite electron spin orientation,
\begin{eqnarray}    
    h\,f_{\uparrow}&=E_\mathrm{s}(v,N,M_s=\phantom{-}1/2,M_I=1,M_N)-E_\mathrm{s}(v,N,M_s=\phantom{-}1/2,M_I=0,M_N)\\
    h\,f_{\downarrow}&=E_\mathrm{s}(v,N,M_s=-1/2,M_I=1,M_N)-E_\mathrm{s}(v,N,M_s=-1/2,M_I=0,M_N)\ .\nonumber
\end{eqnarray}
The difference in transition frequencies is
\begin{equation}
    h\,\delta f_{\uparrow \downarrow}=h(f_{\uparrow} - f_{\downarrow})=b_F(v,N) -\frac{2 d_1(v,N)}{4 N(N+1)-3}M_N^2\ ,
\end{equation}
and is insensitive to magnetic field. Of course, an exact treatment will find a small sensitivity due to the finiteness of $B$, from terms similar to those in App.\,\ref{app:The energies of spin states for $N=2$}.
The sensitivity is presumably minimized by choosing states with $M_N=0$.

We propose that $\delta f_{\uparrow\downarrow}$ is determined by repeated measurement of the two transitions in alternation, on the same species, so as to suppress the magnetic-field-drift effect. In more detail, assume the first RF transition, ``$\uparrow$", is to be measured. After irradiation of the RF wave, one must test whether the transition has occurred. Similar to a suggestion by Myers, in a CSGE-\PMT{} this can conveniently be done by a $\pi$-pulse electron-spin flip excitation having frequency adapted to the final state of the  ``$\uparrow$" transition. If an electron-spin reversal is detected by CSGE, then the molecule is already in the initial state for the subsequent ``$\uparrow$" RF transition and the procedure can continue. If no electron-spin flip reversal is detected, then the molecule has to be prepared in the initial state for the ``$\downarrow$" transition, and that is done by an electron-spin-flipping $\pi$-pulse tuned to the initial state of the ``$\downarrow$" transition. Then the procedure continues.

Suppose interrogation of the RF transitions achieves a spectroscopic linewidth $\delta_\mathrm{int}=h\times0.1\,\mathrm{Hz}/b_F\simeq1\times10^{-10}$. A one-hundredfold smaller statistical uncertainty of line center determination should be achievable in both a CSGE-\PMT{} and in a QLS-\PMT{}, within  realistic integration times. Thus, a CPTI test of a nuclear spin-related property -\,the g factor\,- at level $1\times10^{-12}$ should be feasible. At present, this test has been done on bare protons/anti-protons at $1.5\times10^{-9}$-level in the BASE collaboration \cite{Smorra2017}. Recent progress on coherent antiproton spin spectroscopy \cite{Latacz2025} could eventually lead to the $10^{-11}$-level.

Achieving a sufficient common-mode rejection of the magnetic sensitivity of the individual transitions by performing alternating measurements relies, as for the CPTI test of the lepton g factor, on a sufficiently small statistical variation $\delta B_a$ of the field on the time scale of the alternation. The numerical requirement is $\delta B_a/B<\delta_\mathrm{int}$, which is feasible (Sec.\,\ref{sec:CPTI test of the electron/positron g factor}).  An option that alleviates this requirement would be to use a dual-precision-trap apparatus and measure the first transition on one molecule and simultaneously the second transition on the other molecule (of the same species). With such an apparatus, one could also do simultaneous measurements on \Htwoplus{} and \antiHtwoplus{}, as proposed for the lepton g factor measurements. 

A CPTI test based on measurements of $\delta f_{\uparrow \downarrow}$ has several advantages compared to the currently performed experiments on bare protons/antiprotons. (1) Due to the near-cancellation of sensitivity to magnetic field, no precise measurement of the field is necessary (although this is standard). (2) For the same reason, measurements on \Htwoplus{} and on \antiHtwoplus{} can be performed in different campaigns, and even in different set-ups. This is particularly attractive for the QLS-\PMT{}, avoiding the complexity of handling simultaneously two species in an already complex apparatus and allowing to play out the advantage of much shorter dead time between spectroscopy cycles compared to a CSGE-\PMT{}. (3) The observable is $b_F/h$, whose value for $(v=0,N=1)$ is approximately 5.5\,times as large as the NMR frequency of the proton in a typical field $B_0=4$\,T. This gives an enhancement in resolution compared to the latter type of measurements. (4) Other systematics discussed for the vibrational transition, in particular the QDS, play no role at the test accuracy level forecast above. (5) There appears to be no need for advanced cooling to be implemented.

\subsection{Conclusions}

We have shown that the uncertainty levels of \Htwoplus{}/\antiHtwoplus{} vibrational spectroscopy achievable with advanced \PMT{}s are highly attractive. They would be more than competitive with the goal uncertainty level of future hydrogen/antihydrogen comparisons. Furthermore, the vibrational spectroscopy is complementary in scope to the electronic spectroscopy of the atoms.

The key requirements for pushing the systematic uncertainties of a vibrational-CTPI test to the $1\times10^{-16}$ level and lower are firstly, a proper choice of transition and secondly a sufficiently precise tracking of the magnetic field. These are straightforward to implement, given the current state of the field. Thirdly, a {highly sensitive determination of} axial ion temperature 
is required. This has not yet been achieved for a molecular ion in a \PMT{}.

A CSGE-\PMT{}, a type of trap that is currently implemented in several units for 4\,K operation, appears to be a viable approach. 
Key aspects will be the use of advanced classical cooling techniques and detectors to enable mean differential axial temperature determination at the sub-10-mK level and the optional inclusion of an appropriately designed magnetic bottle. The aim of these approaches is to minimize the axial-QDS uncertainty. 
Cyclotron and magnetron modes must be cooled as well and their temperatures should ideally be determined to the sub-mK level. 
Some of these techniques are routinely in use 
on protons and antiprotons \cite{latacz2024orders} and are available. We may even envisage a further performance enhancement: a CSGE-\PMT{} operated in a lower-temperature cryostat, e.g.~at 0.3\,K, as in the LSYM project \cite{Raab2026}, leading to further reduction of the QDS-related uncertainty.

A potentially more powerful approach, developed within the framework of the BASE collaboration \cite{Cornejo2021}, is the QLS-\PMT{}, because it is designed to  cool light ions sympathetically to the ground state of motion. This will minimize a number of shifts. Such cooling has not yet been achieved, either. Nevertheless, important progress has recently been made towards this goal. Once the technique is available, it seems that the spectroscopy of \Htwoplus{}/\antiHtwoplus{} can be implemented with a performance level similar to that of single-ion optical atomic clocks utilizing RF traps. The short cycle time possible in a QLS-\PMT{} is crucial, here. The performance would be to a large extent limited by the linewidth of the clock laser, which determines the achievable statistical uncertainty. 

{CPTI tests of the bound-lepton and bound-proton g factors are also attractive. Due to the substantially lower transition frequencies, the fractional resolutions of such tests will be correspondingly lower compared to the vibrational-transition test. However, a large improvement compared to the present state of the art appears possible.} 

The key conclusion from this study  is therefore that a research programme aiming at studying \Htwoplus{} in advanced \PMT{}s should begin as soon as possible. There is no radically new technique or technology that needs to be developed. {We suggest that, even though the QLS-PT approach may turn out to be superior in the long term, at this time both the CSGE-PT and the QLS-PT approach should be pursued in parallel to ensure rapid progress even in case of delays in one of the approaches. } The potential outstanding accuracy of the vibrational spectroscopy makes this work worthwhile even if the production of \antiHtwoplus{} is still far off. 

\section*{Acknowledgments}
S.\,S.~thanks V.I.~Korobov for providing the magnetic susceptibility and polarisability values prior to publication, and J.-Ph.~Karr for a discussion. 
The work of S.\,S.~performed under a grant of Deutsche Forschungsgemeinschaft includes  a collaboration with S.\,Sturm and K.\,Blaum and their team 
on spectroscopy of HD$^+$ in ALPHATRAP. 
Joint discussions about spectroscopy in ALPHATRAP have been helpful.

\section*{Funding}
The work of S.\,S.~was  performed under  grant Schi\,431/29-1 of Deutsche Forschungsgemeinschaft. The work of S.\,S.\, and D.\,B.\, was also supported by the European Research Council (ERC) under the European Union’s Horizon 2020 research and innovation program (Grant Agreement No.~786306, “PREMOL”). J.\,M.\,C.\, acknowledges the grant “RYC2023-042535-I” funded by MICIU/AEI/10.13039/501100011033 and by “ESF+”.

\bibliographystyle{tfo} 
\bibliography{Bibliography_all}
\vfill
\eject
\appendix
\section{Effective Hamiltonian matrices }
\label{app:Effective Hamiltonian}

The Hamiltonian $H_\mathrm{tot}(v,N)$ of eq.~(\ref{eq:H_tot1}) can be solved exactly. 
We discuss only the case $N=2$, so $\textbf{I}=0$.
Since in strong magnetic field the angular momenta are nearly decoupled, we conveniently choose a basis that is the direct product of the electron spin $s_e$ basis set and the molecule rotational momentum $N$ basis set, $|N M_N\rangle |s_e M_s\rangle$, abbreviated in this appendix as $|M_N;M_s\rangle$. They are enumerated with $M_s$ and $M_N$, the respective projections of
the momenta on the $z$-axis along the magnetic field $B$. The projection $M_F$ of the total angular momentum $\mathbf{F}=\mathbf{s_e}+\mathbf{N}$ is a good quantum number of the total Hamiltonian. To an eigenstate of given $M_F$, those basis states $|M_N;M_s\rangle$ contribute for which $M_F=M_s+M_N$. The basis states  are
\vskip .1in
    \begin{tabular}{c|c|c|c|c|c|}
$M_F$&$-{\textstyle\frac{5}{2}}$&\multicolumn{2}{c|}{$-{\textstyle\frac{3}{2}}$}&\multicolumn{2}{c|}{$-{\textstyle\frac{1}{2}}$}  \\
\hline
number&1&2&3&4&5\\ 
state & $|-2;-{\textstyle\frac{1}{2}}\rangle$ & $|-1;-{\textstyle\frac{1}{2}}\rangle $& 
$|-2;{\textstyle\frac{1}{2}}\rangle$&$|0;-{\textstyle\frac{1}{2}}\rangle$ & 
      $|-1;{\textstyle\frac{1}{2}}\rangle$ \\
    \end{tabular}
\vskip .05in
\begin{equation}
        \begin{tabular}{c|c|c|c|c|c|}
$M_F$&\multicolumn{2}{c|}{$+{\textstyle\frac{1}{2}}$}&\multicolumn{2}{c|}{$+{\textstyle\frac{3}{2}}$}&$+{\textstyle\frac{5}{2}}$  \\
\hline
number&6&7&8&9&10\\ 
state &       $|1;-{\textstyle\frac{1}{2}}\rangle$ &  $|0;{\textstyle\frac{1}{2}}\rangle$ & $|2;-{\textstyle\frac{1}{2}}\rangle$ & $|1;{\textstyle\frac{1}{2}}\rangle$ &$|2;{\textstyle\frac{1}{2}}\rangle$\label{eq:list of states}\\
    \end{tabular}
\end{equation}

\noindent These basis states are also the approximate eigenstates in strong magnetic field. 

The spin-rotation Hamiltonian $H_\mathrm{spin-rot}=c_e(\textbf{s}_e\cdot\textbf{N})$ is the only non-diagonal one. Expressing the operator as $s_{e,z}N_z+(s_e^+N^- + s_e^-N^+)/2$, and using the general rules
\begin{eqnarray*}
    J^+|J,M_J\rangle&=&\sqrt{J(J+1)-M_J(M_J+1)}|J,M_J+1\rangle,\\ 
    J^-|J,M_J\rangle&=&\sqrt{J(J+1)-M_J(M_J-1)}|J,M_J-1\rangle,
\end{eqnarray*} 
 we obtain
\begin{eqnarray}
&H_\mathrm{spin-rot}(v,N=2)=c_e(v,N=2)\times\qquad\qquad\qquad\nonumber\\    
&\quad\left(
\begin{array}{cccccccccc}
 1 & 0 & 0 & 0 & 0 & 0 & 0 & 0 & 0 & 0 \\
 0 & \frac{1}{2} & 1 & 0 & 0 & 0 & 0 & 0 & 0 & 0 \\
 0 & 1 & -1 & 0 & 0 & 0 & 0 & 0 & 0 & 0 \\
 0 & 0 & 0 & 0 & \sqrt{\frac{3}{2}} & 0 & 0 & 0 & 0 & 0 \\
 0 & 0 & 0 & \sqrt{\frac{3}{2}} & -\frac{1}{2} & 0 & 0 & 0 & 0 & 0 \\
 0 & 0 & 0 & 0 & 0 & -\frac{1}{2} & \sqrt{\frac{3}{2}} & 0 & 0 & 0 \\
 0 & 0 & 0 & 0 & 0 & \sqrt{\frac{3}{2}} & 0 & 0 & 0 & 0 \\
 0 & 0 & 0 & 0 & 0 & 0 & 0 & -1 & 1 & 0 \\
 0 & 0 & 0 & 0 & 0 & 0 & 0 & 1 & \frac{1}{2} & 0 \\
 0 & 0 & 0 & 0 & 0 & 0 & 0 & 0 & 0 & 1 \\
\end{array}
\right)&\ .
\label{eq:H_eff_matrix}
\end{eqnarray}
Columns from leftmost to rightmost and rows from top to bottom correspond to the sequence of states in Table\,(\ref{eq:list of states}).

All other Hamiltonians together can be brought into the diagonal form
\begin{eqnarray}
    H_\mathrm{diag}&=&-g_e'(v,N)\mu_\mathrm{B} B s_{e,z} -  g_r(v,N)\mu_\mathrm{n} B N_z +\nonumber \\
    &&\gamma(v,N,\mathbf{E},B,V_{zz}) N_z^2 + \zeta(v,N) B s_{e,z} N_z^2+\nonumber \\
    &&\xi(v,N,\mathbf{E},B,V_{zz})\ ,
    \label{eq:diagonal Hamiltonian}
\end{eqnarray}
where $N=2$, $g_e'(v,N)=g_e(v,N)-g_t(v,N)N(N+1)/{\sqrt{N(N+1)(2N-1)(2N+3)}}$, and $\gamma$, $\zeta$, $\xi$ are appropriate field-dependent interaction strengths describing anisotropic electron-Zeeman energy, diamagnetism, paramagnetism, electric quadrupole interaction, Stark shift, and light shift.

Thus, the total Hamiltonian $H_\mathrm{tot}=H_\mathrm{spin-rot}+H_\mathrm{diag}$ is block-diagonal, with the largest blocks being $2\times2$. The reason is that basis states with different values of $M_F$ do not mix, since the angular momentum projection operator $F_z$ commutes with $H_\mathrm{tot}$.
It can be trivially diagonalized. However, because the electron-spin-Zeeman interaction is by far the dominant one, it is reasonable to consider approximations. 

In the simplest approximation, we may assume that only $g_e'$, $g_r$ and $c_e$ - they determine the dominant interactions - are nonzero. When the eigenvalues are Taylor-expanded to first order in $B^{-1}$ one obtains the results of Extended Data Table\,2~(lower) in ref.\,\cite{Schenkel2024}. The largest neglected terms are of order $c_e^3/(g_e\mu_\mathrm{B}B)^2\approx5\,$Hz at 4\,T. 

To include the neglected further (diagonal) interactions -- the last three terms in eq.\,(\ref{eq:diagonal Hamiltonian}) -- we add three energy contributions obtained by replacing the operators 
by the respective (approximate) projection quantum numbers.  A partial result is shown in the following appendix\,\ref{app:The energies of spin states for $N=2$}.

If we consider all coefficients of $H_\mathrm{spin-rot}+H_\mathrm{diag}$ on an equal footing and Taylor-expand the eigenvalues up to first order in $B^{-1}$ we obtain the results of Table\,\ref{tab:table expansion of energies}. In practice, they are numerically equivalent to the previous approximation, since $Z$ is tiny. It can therefore be neglected in the denominators appearing in the table.

\section{The energies of spin states for $N=2$}
\label{app:The energies of spin states for $N=2$}
The contributions to the energy of a spin state $(v,N=2,M_N,M_s)$ are, approximately
\begin{eqnarray}
&-&g_\mathrm{e}(v,N)\mu_\mathrm{B}B\,M_s \label{eq:Appendix ge contribution}\\ 
&-&g_{r}(v,N)\mu_\mathrm{n}B\,M_N +c_e(v,N)M_s\,M_N
\\
&-&{\textstyle \left(
\frac{\chi_t(v,N)\alpha^2\Tilde{B}^2}{2}
+\frac{3\,g_t(v,N)\,\mu_\mathrm{B}B\,M_s}
{\sqrt{N(N+1)(2N-1)(2N+3)}}
\right)}M_N^2 \label{eq:diamagnetic contributon}\\  
&+&{\textstyle \frac{c_e(v,N)^2}{2(\mu_\mathrm{B}g_\mathrm{e}(v,N)
-\mu_\mathrm{n}g_{r}(v,N))B}}
\left((M_s+M_N)^2-{\textstyle\frac{25}{4}}\right)M_s
\qquad\label{eq:Appendix spin-rotation contribution}
\\
&+&{\textstyle\frac{N(N+1)}{\sqrt{N(N+1)(2N-1)(2N+3)}}}\,g_t(v,N)\mu_\mathrm{B}B\,M_s \label{eq:electron-spin tensor contribution}\\ 
&-&\alpha_t(v,N)\Tilde{E}^2\,M_N^2 \label{eq:Stark tensor contribution}\\  
&+&\text{terms independent of $M_s$, $M_N$},
\label{eq:terms independent of proj qu numbers}
\end{eqnarray}
with $\Tilde{E}^2=E_z^2-\frac{1}{2}(E_x^2+E_y^2)$ and $
\chi_t=\chi_t^{(\mathrm{d})}+\chi_t^{(\mathrm{p})}$. This holds for $N=2$.
The terms(\ref{eq:diamagnetic contributon}-\ref{eq:electron-spin tensor contribution}) are of the same order. Contribution eq.\,(\ref{eq:Stark tensor contribution}) is by far the smallest.
To obtain the above result, the eigenvalues were expanded to first order in $B^{-1}$ and only the dominant term of that order, \ref{eq:Appendix spin-rotation contribution}, was kept. In its denominator, a contribution containing $\zeta$ (or $Z$) was neglected.

\vskip .1in
\section{Combining frequencies of spin components}
\label{app:determination of tensor polarisability}
Here we discuss what possibilities arise if several spin components are measured accurately.

We recognize that the terms in eqs.~(\ref{eq:Appendix ge contribution}-\ref{eq:Stark tensor contribution})
 depend on different combinations of the projection quantum numbers: $M_s$, $M_N$, $M_s M_N$, $M_N^2$, $M_s^3$, $M_s^2 M_N$, $M_s M_N^2$.
Their prefactors can in principle be extracted by combining suitably the optical frequencies of different spin components  $(v,N,M_s,\,M_N)\rightarrow (v',N',M_s',\,M_N')$. 
RF spectroscopy (transitions without change in $v,N,M_s$, only in $M_N$) could also furnish information. Such an extraction was mentioned also by Myers \cite{Myers2018}.

 Other combinations of frequencies will allow to determine the terms in eq.\,(\ref{eq:terms independent of proj qu numbers}) independent of $M_s$, $M_N$, i.e a combination of $\chi_s \Tilde{B}^2$, $\chi_t \Tilde{B}^2$, $\alpha_s \mathbf{E}^2$, $\alpha_t\Tilde{E}^2$,  and the spin-averaged vibrational transition frequency $f_\mathrm{vib,0}$.

One characteristic feature of the above energy expression is that the $\chi_t$ term and $\alpha_t$ term proportional to $M_N^2$  occur together, and can therefore not be determined separately experimentally from measurements at fixed $B$. (The same occurs for the $M_s$, $M_N$-independent contribution, that contains jointly the $\chi_s$ term and $\alpha_s$ term in eq.\,(\ref{eq:terms independent of proj qu numbers}).)
Such a separation would be possible by first determining $B$ experimentally and combining it with ab initio values of $\chi_t$. But since $\chi_{t}\,\alpha^2\Tilde{B}^2$ is seven orders larger than the $\alpha_t$ term under the present experimental conditions, a high-accuracy value for $\chi_{t}$ would be necessary; so far it is not available. An alternative determination procedure would be measurements at different magnetic fields.

It appears that RF spectroscopy would be better suited compared to vibrational spectroscopy, simply because the linewidth of a RF signal source is much smaller (in absolute terms) than that of a laser. Therefore, splittings between spin states having equal $M_s$ would be more easily measured (shorter integration time) with 0.1-mHz resolution using RF spectroscopy compared to laser spectroscopy. 
We refer to \cite{Bollinger1991} for an example of high-resolution RF spectroscopy in a \PMT{}.  However, such splittings only contain the very small tensor ($\alpha_t$) Stark shift contribution, eq.\,(\ref{eq:Stark tensor contribution}), approx.~$-3$\,mHz for a 1-$\mu$m cyclotron radius in $(2,2)$. Thus the sensitivity is intrinsically small. That only the field combination $\Tilde{E}$ occurs is not a limitation, since the contribution from $E_z$ is negligible.
Therefore, we point out another option: to additionally perform measurements with \Htwoplus{} in a high vibrational level. Such levels are also metastable, but their polarizability is much larger. For example, in $(8,1)$ $\alpha_t$ is approximately 30~times larger and $\alpha_s$ five times larger. With the help of these enhancement factors, the tensor Stark effect should then be precisely measurable using RF spectroscopy, at least for cyclotron orbital radii at the $\mu$m level or larger. A measurement of the ratio of two  vibrational transition frequencies, one between two low-lying levels and another between two high-lying ones would allow measuring the scalar Stark shift with high sensitivity.
Yet another option would be to use a HD$^+$ ion as a probe: its polarizabilities are larger by an additional order or magnitude.

Accessing the contributions in eqs.\,(\ref{eq:Stark tensor contribution},\ref{eq:terms independent of proj qu numbers}) that depend on the electric field might be of utility for a CPTI test: $B$ being independently measurable at the $1\times10^{-9}$ fractional level implies that  (temporal) \emph{deviations} of $\alpha_t \Tilde{E}^2$ and/or $\alpha_s \mathbf{E}^2$ 
can be identified at the level of the spectroscopic resolution employed, assuming no other systematic shifts (the light shift) are varying. This means that one would have a way to ensure that the electric-field conditions during  measurements on \Htwoplus{} and on \antiHtwoplus{} are sufficiently equal. This could be done even if the measurements occur in different traps at different times. Moreover, the second-order Doppler shift associated with transverse motion could be deduced from the transverse Stark shift (see appendix~\ref{app:Stark shift and QDS correlation}).

For a CPTI test, one could proceed as follows:  $B$ is measured with appropriate accuracy by cyclotron resonance (to a level compatible with the goal experimental vibrational frequency resolution), and set appropriately close to the same value during the experiments on both species. In addition, at least one RF transition in a favorable vibrational level is measured. Alternatively two suitable vibrational transitions are measured, with maximum optical resolution, possibly requiring a substantial effort. After analyzing the results one could adjust the trap electrode voltages to set $\Tilde{E}$ (or $\mathbf{E}$) to the same value before performing the experiments on both species. 

We mention that such a procedure assumes that the transitions used for identifying the Stark shift are not affected by a CPT invariance violation.

In the ideal case, when the cooling of the ion is so good that the residual r.m.s.~electric fields are small (as assumed in Table\,\ref{tab:summary of the shifts}), this procedure will not be necessary.

\section{Correlations between QDS and Stark shift}
\label{app:Stark shift and QDS correlation}

Note that the transverse motion's velocity gives a instantaneous contribution to the QDS that is correlated with the d.c.~Stark shift contribution originating from the induced electric field, eq.\,(\ref{eq:E_r,motion as a fct of vx, vy}):
\begin{eqnarray}
    \Delta f_\mathrm{QDS,transv}&=&-\frac{ v_x(t)^2+v_y(t)^2}{2\,c^2}\,f_0\ ;\nonumber\\
    \Delta f_\mathrm{dc-Stark,transv}&=&-\frac{1}{2\,h}
 \Delta\alpha
 (v_x(t)^2+v_y(t)^2)B^2\ .
\label{eq:comparison transverse QDS and dc-Stark}
\end{eqnarray}
with an appropriate polarizability difference $\Delta\alpha=\Delta\alpha(v,N,M_N;v',N',M_N')$. For the \Htwoplus{} transitions considered here, $\Delta\alpha$ is positive and so the two shifts have the same sign and therefore cannot be made to offset or cancel each other, as is considered for atomic ion frequency standards \cite{Berkeland1998a,Arnold2015}. The transverse QDS is about three orders larger than the d.c.~Stark shift.

A cancellation of the transverse effects would occur at a ``magic" magnetic field \cite{Arnold2015}
\begin{equation}
    B_\mathrm{magic}=\sqrt{\frac{h\,f_0}{(-\Delta\alpha)c^2}}\ .
\end{equation}
In the heteronuclear \HDplus{} (and presumably in other heteronuclear MHI as well) the  polarizability difference $\Delta\alpha$ can be  negative \cite{Schiller2014a}. Feasible magic magnetic field values occur for rotational transitions, e.g. $(v=0,N=0)\rightarrow(0,1)$ and for $\Delta v=1$ vibrational transitions between excited levels, e.g. $(v=4,N=0)\rightarrow(5,1)$. 

A cancellation of the axial effects (eqs.~(\ref{eq:mean squared axial electric field}) and (\ref{eq:mean QDS})) via a ``magic axial frequency" is not possible, the required negative polarizability is too large in magnitude.


\section{Line shapes and determination of unperturbed line center}
\label{app:Determination of line center in presence of QDS}

We begin by noting the different regimes for axial and transverse optical spectroscopy, assumed to be implemented with $\lambda=2.4\,\mu$m wavelength. At e.g.\,$T_\mathrm{z}=0.2\,$K the thermal r.m.s.~displacement  for axial motion is $\langle z^2 \rangle^{1/2}\simeq4.5\,\mu$m. Since this exceeds $\lambda$, the spectroscopy with axial wave propagation will exhibit a spectrum with many sidebands. 

For transverse spectroscopy, we consider both cyclotron and magnetron motion.
If the cyclotron mode was thermally excited to the same temperature, $\langle \rho_\mathrm{cyc}^2 \rangle^{1/2}\simeq0.21\,\mu$m, leading to only weak first-order sidebands. However, the mode can be cooled to mK-temperature, reducing the radius even further (Sec.~\ref{sec:Estimate of the QDS shift}). 
The magnetron mode can also be reproducibly and measureably cooled to 1-mK-level, corresponding to  approximately 20\,$\mu$m radius.
Thus, the spectrum in transverse direction will consist of essentially only the cyclotron carrier and a substantial number of magnetron sidebands.
It will be simple to experimentally determine which transition is the carrier transition: since the magnetron frequency depends on the quadrupole potential the carrier transition is the one that does not shift when the quadrupole potential is modified.

Myers points out the advantage of performing spectroscopy with transverse laser beam propagation. This direction is also consistent with the here proposed preferred types of transitions, $\Delta M_F=0$.

According to Myers, the QDS leads to a finite linewidth of the transition because the Boltzmann energy distribution also defines a probability distribution of the QDS values. It has a width comparable to the mean QDS,
$\sigma(f_\mathrm{QDS})\simeq|\Delta f_\mathrm{QDS}|$.
This issue does not seem to have been discussed previously. 
In the standard theoretical treatments of thermal motion of an ion in a RF trap (e.g.\,\cite{Fisk1997,Berkeland1998a}), one usually only considers that the carrier transition frequency is shifted by the mean QDS \ref{eq:mean QDS}.
The thermal character of the motion determines the strengths of the sidebands, but not the width.
Analogously, also the Stark shift -- which also exhibits a quadratic dependence on the dynamic variables in a \PMT{} -- would lead to a line broadening, apart from a shift. In Ref.~\cite{Fisk1997,Berkeland1998a} it was treated on the same footing as the QDS, without considering line broadening.

To the best of our knowledge, a QDS-broadened line has not been observed experimentally yet.
One reason could be that in experiments to date, thanks to laser cooling and the usually relatively high ion mass, the purported width, with numerical value similar to the value after eq.\,(\ref{eq:mean QDS}), was smaller than the experimental linewidth caused by finite interrogation time, magnetic field inhomogeneities, linewidth of the spectroscopy wave, etc. This holds both for ions in RF traps and in \PMT{}s (see Appendix~\ref{app:experimental studies in PMTs with narrow linewidth}). 
In ref.~\cite{Fisk1997} a microwave ion clock based on buffer-gas cooled Yb$^+$ ions is described. The ions were trapped at relatively high temperature, approximately 400\,K, so that the mean second-order Doppler shift was relatively large. A modification of the line shape of the transition was not reported, perhaps because the spectral resolution was insufficient.

However, analogs have been observed: optical and RF transitions in inhomogeneous magnetic traps. Ensembles of particles that over time cover a volume that includes a magnetic field minimum and an approximately quadratic spatial variation and that experience a linear Zeeman effect will exhibit such line shapes \cite{brown1986geonium}.
This is clearly observed even for a single particle in a \PMT{} \cite{nagahama2017sixfold}. 

We may model the shape of a spectroscopic line 
in presence of QDS as a a convolution of a Lorentzian and a one-sided exponential. The exponential describes the Boltzmann distribution of QDSs. The Lorentzian describes the combined effect of finite linewidth of the laser and the finite interrogation time, resulting in an ``interrogation linewidth" $\delta_\mathrm{int}$.
 We may neglect the finite lifetime of the upper spectroscopy level and other broadening mechanisms.

Realistically, a minimum laser linewidth of 0.05\,Hz and minimum interrogation-duration-related linewidth of 0.1\,Hz may be considered, giving a smallest combined linewidth of approximately 0.12\,Hz. Considering the value $f_0$ of the reference transition, the interrogation linewidth is $\delta_{\mathrm{int}}=1\times10^{-15}$ fractionally.

The result of a simple calculation of the line shape in the case of a QDS of a particle at finite temperature is shown in Fig.\,\ref{fig:line shapes}. 
We recognize that there are two limiting cases.
\begin{figure}
    \centering
\includegraphics[width=0.9\linewidth]{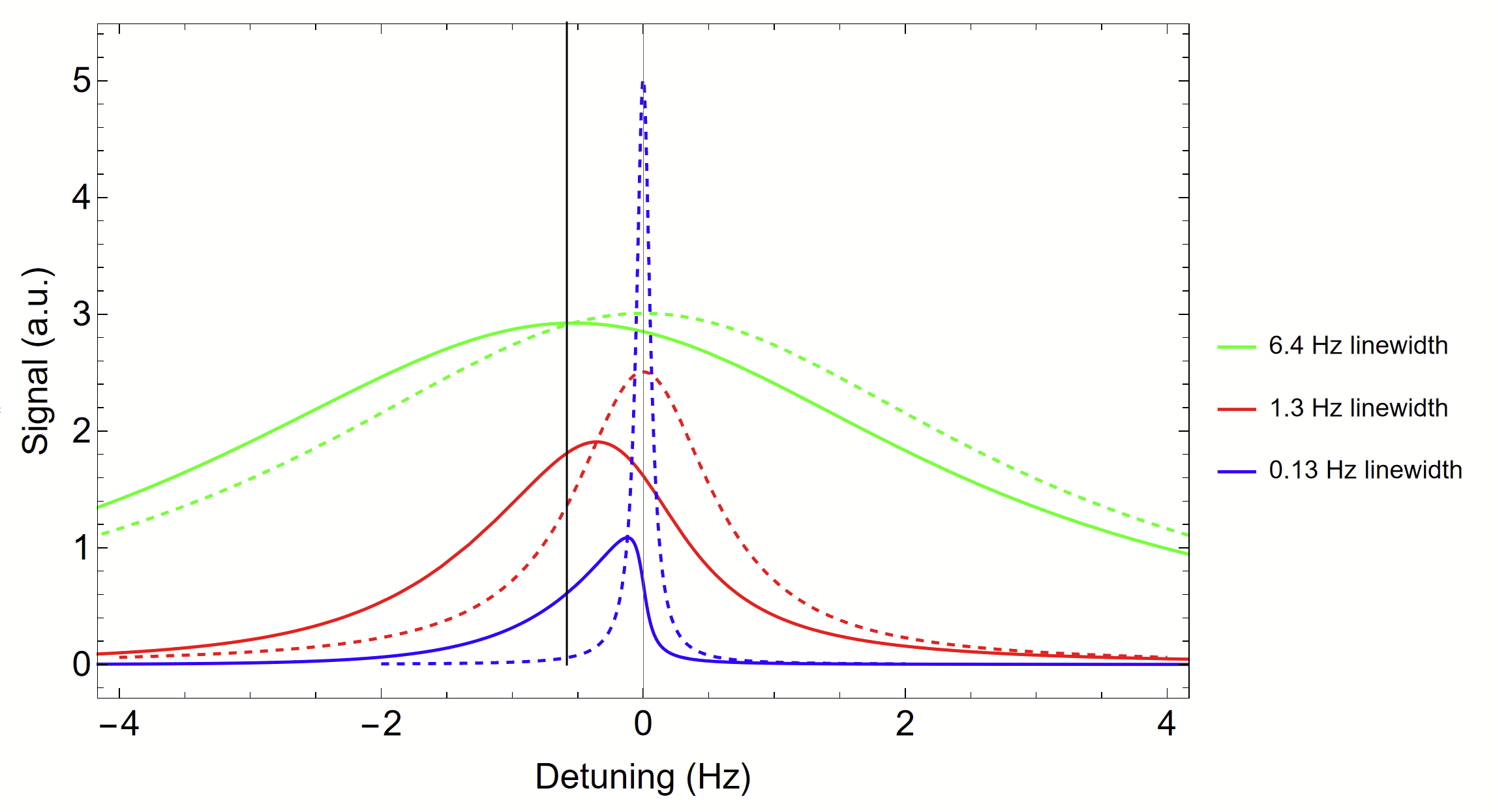}
    \caption{ \textbf{Line shapes in presence of QDS (full lines) and without QDS (dashed).} The vertical \textcolor{black}{\textbf{black}} line shows the mean QDS, assuming an axial temperature $T_z=0.2\,$K and negligible radial temperatures. The transition frequency is $f_0\simeq130\,\mathrm{THz}$. The colors correspond to three different values of the interrogation linewidth $\delta_{\rm int}$, defined as the total linewidth in absence of QDS, i.e.~stemming from the laser linewidth and the finite interrogation duration. Note that the maxima of the full lines occur at the mean QDS only in the limit of  $\delta_\mathrm{int}$ large compared to the mean QDS (\textcolor{green}{\textbf{green}} full line). 
    }
    \label{fig:line shapes}
\end{figure}

\underline{Case CSGE-PMT:} If QDS cancellation (App.\,\ref{sec:QDS cancellation}) is not implemented, the QDS linewidth (similar to eq.\,(\ref{eq:Delta fQDS for 0.4 K})) will be larger than the interrogation linewidth $\delta_\mathrm{int}$ if a laser of ultra-narrow linewidth is available. 
The convolution line shape then resembles the QDS Boltzmann line shape, and it is asymmetric. The low-frequency side shows a sharp rise, on a frequency scale given by the interrogation linewidth. This side is nearly ``centered" at the unperturbed frequency.

This is a very favorable situation: it appears that resolving the unperturbed frequency is possible without requiring an independent and precise value of the temperature, and moreover, the resolving ``power" of the spectroscopy is given by the interrogation linewidth rather than by the QDS linewidth.

The goal is to determine the ``center" of the low-frequency side of the line to a small fraction of the interrogation linewidth. In practice, one would measure also the broad side of the line - possibly with fewer data points - and then fit the unperturbed frequency, using all information, including independent temperature data obtained as discussed above, if available. 
We consider feasible to achieve a  30-fold ``splitting" of the sharp low-frequency side of the line. For an interrogation linewidth of $1\times10^{-15}$ this implies a goal statistical uncertainty of $u_{\rm stat}(\Delta f^{\mathrm{CSGE}}_\mathrm{QDS})=3\times10^{-17}$. The systematic uncertainty is difficult to analyze at this point; in the framework of the present simple model it could well be small in comparison, since one fits for the unperturbed frequency and there is no need to apply the numerical correction $\Delta f_\mathrm{QDS}$.
The above argument holds if the radial temperatures are sufficiently small. This is approximately satisfied if they are less than e.g.\,4\,mK each, the associated QDS broadening then being $4\times10^{-16}$, smaller than $\delta_\mathrm{int}$.

Achieving the above $u_{\rm stat}(\Delta f^\mathrm{CSGE}_\mathrm{QDS})$ will require on the order of several 1000 interrogations of the ion. A single CSGE interrogation requires on the order of  10~min, bringing the total measurement time to several weeks. 

We consider this feasible, because it has already been demonstrated that  a high-duty-cycle and month-long ESR ion interrogation in a \PMT{} is possible \cite{Koenig2025}, and a month-long laser frequency lock and frequency comb lock are possible.

\underline{Case QLS-\PMT{}:} A detailed treatment of this case is beyond the scope of this work. Qualitatively, we expect that because of the very low ion temperature (the ion is nearly in the ground state of motion)  the  QDS linewidth ($\simeq5\times10^{-17}$, see Sec.\,\ref{The QDS in a QLS-PT}) is much smaller than the interrogation linewidth. 
Then, the line shape is essentially a Lorentzian of width determined by the interrogation linewidth, and the peak is at a detuning corresponding to the mean QDS. 
The green line in the figure attempts to show this case. For display purposes the axial temperature is set to 0.2\,K  and the interrogation linewidth is scaled up to $50\,\delta_\mathrm{int}$.

The statistical uncertainty of a line center determination, $u_{\rm stat}(\Delta f^{\mathrm{QLS}}_{\rm QDS})$, is of order $\delta_\mathrm{int}\times \sqrt{\tau_c/\tau}$, where $\tau$ is the averaging time, and $\tau_c$ is the cycle time for excitation and interrogation of the single ion. We assume $\tau_c$ to be dominated by the ion interrogation time, as discussed in Sec.\,\ref{sec:QLS-PT: State preparation and time consumption}. Thus, $\tau_c\simeq1/f_0\,\delta_\mathrm{int}\approx10\,$s. We can thus expect  $u_{\rm stat}(\Delta f^{\mathrm{QLS}}_{\rm QDS})\simeq1\times10^{-17}$ after $\tau\simeq2\,$days.

\section{A magnetic bottle for the cancellation of the QDS?}
\label{sec:QDS cancellation}

In some \PMT{}s it is common practice to tune the relativistic shifts of the cyclotron motion (see beginning of Sec.\,\ref{sec:Quadratic Doppler shift}) to zero by counteracting them with additionally applied magnetic field inhomogeneities. 

Similarly, we now consider the possibility of tuning the axial QDS to zero by implementing a magnetic bottle, in which the rovibrational transition frequency becomes spatially dependent. We assume the ion motion to be classical, so the following treatment applies to the CSGE-\PMT{}.

The magnetic field will always follow the multipolar expansion of the Maxwell equations. Let $z=0$ be the reference position from which the magnetic field is expanded. In lowest order, a magnetic bottle is described by \begin{equation}
    \mathbf{B}(z,r)=[B_0+B_2(z^2-r^2/2)]\,\mathbf{e}_z-B_2(z\,r)\,\mathbf{e}_r\ ,
    \label{eq:bottle magnetic field}
\end{equation}
with the distance from the trap axis $r$ and the expansion coefficient $B_2$.  
\subsection{Cancellation of the axial QDS}

First, let us only consider the axial motion  and assume that $r$ is negligible. Let the equilibrium ion position be $z_0$, the value being determined by the electric potentials on the trap electrodes. The  shifted transition frequency is
\begin{equation}
    f_\mathrm{tot}(t)=f(B_0)+\beta 
    B_2 z(t)^2 
     - \frac{v_z(t)^2}{2 c^2} f(B_0)\ .
\end{equation}
 The axial harmonic oscillation is described by $z(t)=z_0+A \cos{\omega_z t}$, $v_z=-\omega_z A \sin{\omega_z t}$, where the amplitude $A$ varies stochastically in time, but on a time scale much longer than the oscillation period. 

The instantaneous frequency can be written as
\begin{eqnarray}
        f_\mathrm{tot}(t)&=&f(B_0)+
        \beta B_2  z_0^2 \nonumber \\ 
    &&+2\beta
     z_0 B_2 A \cos{\omega_z t}\nonumber \\
    &&+ \beta B_2 A^2 (1+\cos2{\omega_z t})/2 \nonumber \\
    &&- (A^2/2)(1-\cos2{\omega_z t})\,\omega_z^2 f(B_0)/{2 c^2}
    \ .
    \label{eq:instantaneous QDS - fundamental equation v2}
\end{eqnarray}
The average over one axial oscillation period is
\begin{eqnarray}
    \langle f_\mathrm{tot}(t) \rangle&=&f(B_0)+
    \beta B_2  z_0^2\nonumber \\
    && + 
    {\small\frac{A^2}{2}}(\beta B_2 - \omega_z^2 f(B_0)/{2 c^2})\ ,
    \label{eq:QDS axial final}
\end{eqnarray}
where we have used $\langle \cos{\omega_z t}\rangle=0$, $\langle \cos{2\omega_z t}\rangle=0$. 
The choice
\begin{equation}
    B_2=\beta^{-1}\frac{f(B_0)}{2 c^2}(2\pi \nu_z)^2 \ 
    \label{eq:magic B_2 v2}
\end{equation}
makes the period-averaged frequency independent of $A$. The statistical variations of the axial energy will thus not affect the transition frequency any more. Possible drifts of 
$B_2$ and $z_0$, as well as imperfect electric-potential polarity switching between \Htwoplus{} and \antiHtwoplus{} trapping,  are deemed to cause a negligible uncertainty. 

\subsection{Cancellation of the radial QDS}
\label{sec:QDS cancellation - radial case}

In the same vein as above, we now consider the impact of the magnetic bottle on the radial QDS. Since we consider the situation where the axial magnetic field $B_0$ is large, we may altogether ignore the transverse component of the magnetic field in eq.\,\ref{eq:bottle magnetic field}.
The motion orthogonal to the trap axis contributes the frequency shifts 
\begin{equation}
    \delta f_\mathrm{tot}(t)=-\frac{\small 1}{\small 2}\beta B_2 r(t)^2 
     - \frac{v_r(t)^2}{2 c^2} f(B_0)\ ,
     \label{eq:radial QDS formula}
\end{equation}
We decompose $r$ and the transverse velocity $v_r$ into magnetron $(-)$ and cyclotron $(+)$ contributions, 
\begin{eqnarray}
r(t)^2&=&[x_-(t)+x_+(t)]^2+[y_-(t)+y_+(t)]^2\\
    v_r(t)^2&=&[v_{x,-}(t)+v_{x,+}(t)]^2+[v_{y,-}(t)+v_{y,+}(t)]^2\ .
\end{eqnarray}
Averaging over the much faster cyclotron motion, the cross terms average to zero and we obtain
\begin{equation}
    \langle \delta f_\mathrm{tot}(t)\rangle_\mathrm{sec}=
    -\frac{1}{2}\beta B_2 (\langle r_-(t)^2\rangle_\mathrm{sec}+\langle r_+(t)^2\rangle_\mathrm{sec}) 
     - \frac{\langle v_-(t)^2\rangle_\mathrm{sec}+\langle v_+(t)^2\rangle_\mathrm{sec}}{2 c^2} f(B_0)\ ,
\end{equation}
We have the relationships $(2\pi\nu_-) r_-(t)=v_-(t)$, $(2\pi\nu_+) r_+(t)=v_+(t)$. Therefore
\begin{equation}
    \langle \delta f_\mathrm{tot}(t)\rangle_\mathrm{sec}=
    \left(\frac{-\beta B_2}{2(2\pi\nu_-)^2}-\frac{f(B_0)}{2 c^2}\right)\langle v_-(t)^2\rangle_\mathrm{sec}+
    \left(\frac{-\beta B_2}{2(2\pi\nu_+)^2}-\frac{f(B_0)}{2 c^2}\right)\langle v_+(t)^2\rangle_\mathrm{sec} \ .
    \label{eq:QDS radial final}
\end{equation}
Comparing eqs.\,(\ref{eq:QDS axial final}) and (\ref{eq:QDS radial final}) we see that if QDS cancellation of the axial QDS is implemented, an increase of the radial-motion-related shift magnitude will occur, and vice-versa. 

Usually $\nu_-\ll\nu_z\ll \nu_+$, so that it is possible to null the magnetron QDS -- by choosing the negative value $\beta B_2=-f(B_0) (2\pi\nu_-)^2/c^2$ -- while keeping the increase of the shift magnitudes related to the cyclotron and axial motion small. However, this does not seem useful, since as discussed in Sec.\,\ref{sec:The cyclotron mode}, in a state-of-the-art CSGE-\PMT{} the kinetic energy of cyclotron motion ($E_+\propto v_+^2$) is much larger than the kinetic energy of magnetron motion. Furthermore, both $E_-$, $E_+$ can be reduced by feedback cooling to a level much smaller than the axial energy. 

A possible operating scenario could instead be a (unusually) small $B_0\simeq0.2\,$T and large $\nu_z$, so that $\nu_-$ and $\nu_+$ can be made comparable to $\nu_z$. $\beta B_2$ would be chosen according to condition (\ref{eq:magic B_2 v2}). The axial QDS would be nulled, while the relative increase of the radial shift would be approximately a factor 2. This may well be tolerable when $E_-$, $E_+$ are small, and may altogether lead to a substantial reduction of the total shift. We may indeed expect to be able to cool the axial mode to $T_z\approx 1\,K$,  the magnetron motion to the 10\,mK level after an appropriate time (noting that this mode experiences negligible reheating once it has been cooled), and the cyclotron motion to the same level, too. For the latter one would use a cooling trap. The small cyclotron frequency for the above small $B_0$ implies that an electric resonator of large inductance can be implemented, leading to a favorably short time constant.

Experimentally, it is easier to implement $B_2>0$. Therefore, to satisfy eq.\,(\ref{eq:magic B_2 v2}) we must select a transition having $\beta>0$. We consider all cases reported in Table\,\ref{tab:sensitvities beta}. The $(v=0,N=2,M_s=+1/2,M_N=0)\rightarrow$ $(2,2,1/2,0)$ transition
represents one option, requiring
\begin{equation}
    B_2\simeq9.0\times10^2\,(\nu_z/1\,\mathrm{MHz})^2\,\mathrm{kT/m}^2\ .\label{eq:B2 value 1 for QDS cancellation}
\end{equation}
In order to keep $B_2$ at realizable values ($<250\,$kT/m$^2$), the axial frequency should be 0.55\,MHz or less. 

Instead, for the 
$(v=0,N=2,M_s=\pm1/2,M_N\ne0)\rightarrow$ $(2,2,M_s'=M_s,M_N'=M_N)$ transitions, whose magnetic sensitivities are larger by an approximate factor $5 M_N$, a correspondingly smaller $B_2$ is obtained
\begin{equation}
    B_2\approx1.6\times10^2\,\frac{(\nu_z/1\,\mathrm{MHz})^2}{M_N}\,\mathrm{kT/m}^2\ .
    \label{eq:B2 value 2 for QDS cancellation}
\end{equation}
In such strong magnetic bottles the measurement of the magnetic field via the cyclotron frequency required the application of cyclotron resonance spectroscopy via the CSGE, as e.g.~described in \cite{nagahama2017sixfold}. A fractional resolution of  $0.5\times10^{-6}$ is possible, but takes several hours of measurement time.
In combination with the value $\beta/f(B_0)\simeq3\times10^{-10}/$T assumed in eq.\,(\ref{eq:B2 value 1 for QDS cancellation}) or $1\times 10^{-9}$/T in eq.\,(\ref{eq:B2 value 2 for QDS cancellation}) one obtains measurement uncertainties as high as $2\times10^{-15}$.
However, the magnetic field could also be  monitored via ESR spectroscopy, or by cyclotron frequency measurement in a neighboring homogeneous trap. Both approaches would yield much lower B-field uncertainties, below the $1\times 10^{-9}$ level, as discussed earlier (Sec.\,\ref{Summary of magnetic effects}). We therefore see no obstacle to achieve, in particular in the case of interleaved measurements on \Htwoplus{}/\antiHtwoplus{}, a low uncertainty,  $u_\mathrm{sys}(\Delta f_\mathrm{mag,diff})\simeq 3\times10^{-18}$.

How can requirement eq.\,(\ref{eq:magic B_2 v2}) be implemented in practice, given that $B_2$ is fixed once the \PMT{} has been built? Possibly, $B_2$ could be measured by studying the shift and sidebands of a ``probe" transition having large $\beta_\mathrm{probe}$ (for which the QDS does not cancel). Then, the required value of $\nu_z$ for studying the actual transition of interest is set to a value computed from eq.\,(\ref{eq:magic B_2 v2}) using the measured $B_2$ and the ab initio theory values of $\beta$, $\beta_\mathrm{probe}$. 

Summarizing, it appears possible to cancel the axial QDS. How well is a matter of experimental procedures. Possibly, a factor $100$ can be reached in the CSGE-\PMT{}. The uncertainty of the residual axial QDS arising from $\Delta T_z$(\Htwoplus{}-\antiHtwoplus{}), $u^{(1)}$, would then also be reduced by the same factor, and could be averaged down to a level comparable to the contribution from magnetron and cyclotron motion. 
We note that the magnetic bottle for axial QDS cancellation could also serve for CSGE detection and radial cooling.

The cancellation approach will also remove the axial QDS line broadening and asymmetry (App.\,\ref{app:Determination of line center in presence of QDS}) This does, however, not lead to higher resolution: because the effective linewidth of the transition is then equal to the interrogation linewidth, assumed to be $\delta_\mathrm{int}=1\times10^{-15}$. A 30-fold splitting of this line appears realistic. Therefore, the statistical uncertainty for the CSGE-\PMT{} would remain similar to the value discussed in the main text.

However, since a \PMT{} operating with the required small $B_0$ is an untested concept, we shall not pursue it in the analysis in the main text.

\section{High-resolution spectroscopy in \PMT{}s}
\label{app:experimental studies in PMTs with narrow linewidth}
We mention some spectroscopy studies in \PMT{}s that have achieved high resolution. 
Mavadia \textit{et al.}~\cite{Mavadia2014} observed a carrier (and magnetron sidebands) of linewidth 1\,kHz on a $^{40}\mathrm{Ca}^+$ transition (729\,nm) at a magnetron temperature 0.042\,K, an axial temperature on the order of 1\,mK and a cyclotron temperature on the order of 7\,mK.
Goodwin \textit{et al.}~\cite{Goodwin2016} observed 150\,kHz linewidth on the same ion species in laser spectroscopy at 0.45\,K.

Von Boehn \textit{et al.}~\cite{Boehn2025}~observed $0.1\,$MHz-level linewidths for a Raman transition between ground-state electron-spin states on a single laser-cooled Be$^+$ ion. Axial sidebands were measured and evaluated to determine a 1.6(2)~mK axial temperature.

K\"onig \textit{et al.} \cite{Koenig2025b} found a fractional linewidth of $3\times10^{-9}$ in ESR of a HD$^+$ molecule at 4\,K. 
Bollinger \textit{et al.} \cite{Bollinger1991}  in RF spectroscopy observed a fractional linewidth of $3\times10^{-12}$ for $^9\mathrm{Be}^+$ at 0.25\,K.

\vfill\eject

\end{document}